\documentclass[12pt,a4paper]{article}

\usepackage{amsmath}
\usepackage{amssymb}
\usepackage{amsfonts}
\usepackage{amsthm}
\usepackage{graphicx}
\usepackage{mathrsfs}
\usepackage[mathscr]{euscript}
\usepackage[mathscr]{euscript}
\usepackage{epsfig}

\usepackage[T1]{fontenc}

\newcommand{\bqa}{\begin{eqnarray*}}
\newcommand{\eqa}{\end{eqnarray*}}
\newcommand{\bqan}{\begin{eqnarray}}
\newcommand{\eqan}{\end{eqnarray}}
\newcommand{\bqt}{\begin{quote}}
\newcommand{\eqt}{\end{quote}}
\newcommand{\bt}{\begin{tabbing}}
\newcommand{\et}{\end{tabbing}}
\newcommand{\bit}{\begin{itemize}}
\newcommand{\eit}{\end{itemize}}
\newcommand{\ben}{\begin{enumerate}}
\newcommand{\een}{\end{enumerate}}
\newcommand{\beq}{\begin{equation}}
\newcommand{\eeq}{\end{equation}}
\newcommand{\bdefi}{\begin{definition}}
\newcommand{\edefi}{\end{definition}}
\newcommand{\bpro}{\begin{proposition}}
\newcommand{\epro}{\end{proposition}}
\newcommand{\blem}{\begin{lemma}}
\newcommand{\elem}{\end{lemma}}
\newcommand{\bth}{\begin{theorem}}
\newcommand{\bco}{\begin{corollary}}
\newcommand{\eco}{\end{corollary}}
\newcommand{\bdes}{\begin{description}}
\newcommand{\edes}{\end{description}}
\newcommand{\bre}{\begin{remark}}
\newcommand{\ere}{\end{remark}}

\newtheorem{definition}{Definition}
\newtheorem{proposition}{Proposition}
\newtheorem{lemma}{Lemma}
\newtheorem{theorem}{Theorem}
\newtheorem{corollary}{Corollary}
\newtheorem{remark}{Remark}
\newtheorem{assumption}{Assumption}

\def\1{{\mathbf 1}}

\begin{document}

\title{Eigenvalue tests for the number of latent factors \\ in short panels\thanks{This paper underlies the Halbert White Jr. Memorial JFEC invited lecture given by Patrick Gagliardini at the Annual Society for Financial Econometrics Conference on June 25th 2022 at the University of Cambridge. We thank the JFEC Editors Allan Timmermann and Fabio Trojani for the invitation, the discussants Alexei Onatski and Markus Pelger for very insightful and constructive comments, as well as G. Genoni, L. Mancini and participants at the Annual SoFiE conference 2022 and at seminars at the Universities of Geneva and Warwick  for helpful remarks.}}
\author{Alain-Philippe Fortin\thanks{University of Geneva and Swiss Finance Institute.}, Patrick Gagliardini\thanks{Universit\`a della Svizzera italiana (USI, Lugano) and Swiss Finance Institute. E-mail address: patrick.gagliardini@usi.ch.}, Olivier Scaillet\thanks{University of Geneva and Swiss Finance Institute.}}
\date{First Version: June 2022. This Version: October 2022}
\maketitle

\abstract{
This paper studies new tests for the number of latent factors in a large cross-sectional factor model with small time dimension. These tests are based on the eigenvalues of variance-covariance matrices of (possibly weighted) asset returns, and rely on either an assumption of spherical errors, or instrumental variables for factor betas. We establish the asymptotic distributional results using expansion theorems based on perturbation theory for symmetric matrices. Our framework accommodates semi-strong factors in the systematic components. We propose a novel statistical test for weak factors against strong or semi-strong factors.  We provide an empirical application to US equity data. Evidence for a different number of  latent factors according to market downturns and market upturns, is statistically ambiguous in the considered subperiods. In particular, our results contradicts the common wisdom of a single factor model in bear markets. 
}


\newpage

\section{Introduction}

A central and practical issue in applied work with unobservable (i.e. latent) factors is to determine the number of factors. For models with latent factors only, Connor and Korajczyk (1993) are the first to develop a statistical test for the number of factors for large balanced panels of individual stock returns in time-invariant models under covariance stationarity and homoskedasticity. Unobservable factors are estimated by the method of asymptotic principal components developed by Connor and Korajczyk (1986) (see also Stock and Watson (2002)). For heteroskedastic settings, the recent literature on large balanced panels with static factors has extended the toolkit available to researchers. 
A first strand of that literature focuses on consistent estimation procedures for the number of factors. Bai and Ng (2002) introduce a penalized least-squares strategy to estimate the number of factors, at least one. Ando and Bai (2015) extend that approach when explanatory variables are present in the linear specification (see Bai (2009) for homogeneous regression coefficients). Onatski (2010) looks at the behavior of the adjacent eigenvalues to determine the number of factors when the cross-sectional dimension ($n$) and the time-series dimension ($T$) are both large and comparable. Ahn and Horenstein (2013) opt for the same strategy and cover the possibility of zero factors through specifying a mock eigenvalue whose functional form vanishes too. Caner and Han (2014) propose an estimator with a group bridge penalization to determine the number of unobservable factors. Based on the framework of Gagliardini, Ossola and Scaillet (2016), Gagliardini, Ossola and Scaillet (2019) build a simple diagnostic criterion for approximate factor structures in large panel datasets. Given observable factors, the criterion checks whether the errors are weakly cross-sectionally correlated or share at least one unobservable common factor (interactive effects). A general version allows to determine the number of omitted common factors also for time-varying structures (see Gagliardini, Ossola and Scaillet (2020) for a survey of estimation of large dimensional conditional factor models in finance). A second strand of that literature develops inference procedures for hypotheses on the number of latent factors. Onatski (2009) deploys a characterization of the largest eigenvalues of a Wishart-distributed covariance matrix with large dimensions in terms of the Tracy-Widom Law. To get a Wishart distribution, Onatski (2009) assumes either Gaussian errors or $T$ much larger than  $n$. Kapetanios (2010) uses subsampling to estimate the limit distribution of the adjacent eigenvalues.

This paper aims at complementing the above literature by considering a large cross-sectional dimension but a fixed  time series dimension, i.e., a short panel. We develop new tests for the number of latent factors with statistics based on the  eigenvalues, and spacings thereof, of variance-covariance matrices. The key idea is that, under assumptions on the error terms detailed in the paper, the eigenvalues of some finite-dimensional variance-covariance matrices constructed from returns feature a flat pattern (possibly equal to zero) for orders larger than $k$ when ranked in decreasing order, where $k$ is the number of latent factors. By establishing the asymptotic distributions of the small eigenvalues of estimated variance-covariance matrices we develop testing procedures on the number of latent factors $k$. 

In a short panel setting, Zaffaroni (2019) considers a methodology for inference on conditional asset pricing models linear in latent risk factors, valid when the number of assets diverges but the time series dimension is fixed, possibly very small. He shows that the no-arbitrage condition permits to identify the risk premia as the expectation of the latent risk factors. This result paves the way to an inferential procedure for the factor risk premia and for the stochastic discount factor, spanned by the latent risk factors. Raponi,  Robotti and Zaffaroni (2020) has recently developed tests of beta-pricing models and  a two-pass methodology  to estimate the ex-post risk premia (Shanken (1992)) associated to observable factors. Kim and Skoulakis (2018) deals with the error-in-variable problem of the two-pass methodology with small $T$ by regression-calibration. The small $T$ perspective yields an effective approach to capture general forms of time-variation in factor betas, risk premia and number of factors by performing the factor analysis in short subperiods (either non-overlapping, or rolling windows) of the sample of interest.

The recent literature has extended asymptotic principal component methods to accommodate more general factor models. Fan, Liao and Wang (2016) extend the characteristic-based modeling in Connor and Linton (2007) and Connor, Hagmann and Linton (2012) by allowing the betas to include unknown asset-specific additive constants. They propose a so-called Projected Principal Component Analysis  to estimate this specification with time-invariant loadings, and show that their factor estimates are consistent even if $T$ is finite. Pelger and Xiong (2022) instead let the factor loadings be functions of an observable state variable. Their estimation under large $n$ and $T$ relies on minimizing a local version of the least-squares criterion underlying PCA, where localization is implemented by kernel smoothing. Gu, Kelly and Xiu (2021) consider the setting where the loadings are a nonparametric function of a large dimensional vector of characteristics, and use an autoencoder to estimate this relationship. Among the parametric approaches, Kelly, Pruitt and Su (2017, 2019) model the coefficients as linear functions of characteristics plus some noise term, while Chen, Roussanov and Wang (2022) opt for semi-nonparametric nonlinear modeling of nonlinear betas. Gagliardini and Ma (2019) study the problem of conducting inference on the conditional factor space, including its dimension. The adopted nonparametric framework is general regarding the beta dynamics and encompasses the aforementioned linear and nonlinear beta specifications. Finally, let us mention that there is also work on inference for large dimensional models with unobservable factors with high frequency data (Ait-Sahalia and Xiu (2017), Pelger (2019, 2020), Cheng, Liao and Yang (2021)). None of these papers considers the problem of testing for the number of latent factors.

The outline of the paper is as follows. In Section 2, we present the static factor model for asset (excess) returns, and discuss identification either via instrumental variables, or a sphericity assumption for the variance-covariance matrix of returns. We study the (in)consistency of the PCA factor estimator, as well as interpretation in terms of Error-in-Variable and in terms of incidental parameters.
Section 3 develops the eigenvalue test statistics based on instrumental variables and based on eigenvalues of the return variance-covariance. Section 4 characterizes the  asymptotic distributions of the test statistics. To do so under large $n$ and fixed $T$, we establish a new second-order uniform asymptotic expansion of the small eigenvalues of a symmetric matrix via perturbation theory. We indicate how to achieve feasible statistics  by providing adequate estimators of the characteristics of the asymptotic distribution. We dedicate Section 5 to extending our analysis to cover inference within a more general framework including weak factors. We analyze testing for (semi-)strong factors vs vanishing factors, power under local alternative hypotheses, and testing for weak factors. In Section 6, we provide the results of Monte Carlo experiments to investigate the finite-sample properties of the considered test statistics. Section 7 presents the findings of our empirical analysis in short subpanels of stock returns in the US market. The concluding remarks are given in Section 8. 

\section{An eigenvalue testing problem}


We develop our inferential theory for the number of latent factors under a static model:
\begin{equation} \label{eq:M1}
y_{i,t} = \beta_i'f_t + \varepsilon_{i,t},
\end{equation}
where $i=1,...,n$ is the index for ``individuals" (e.g., assets) and $t=1,...,T$ for time periods (e.g., months), $f_t$ is a $k$-dimensional vector of unobservable factors and $\varepsilon_{i,t}$ is the idiosyncratic error term. We introduce below some high-level conditions on latent factors and error terms underlying our analysis, while we refrain from detailing the specific regularity conditions. \footnote{Those are given in Fortin, Gagliardini, Scaillet (2022a) for the approach based on factor analysis, which generalizes the approach based on the PCA of variance-covariance matrix of returns considered in this paper.} In asset pricing applications, variables $y_{i,t}$ denote asset (excess) returns and the components of vector $f_t$ represent pervasive risk factors in the economy. 
We assume that the time series dimension $T$ is fixed, i.e., we face short panels, while the cross-sectional dimension $n$ tends to infinity in our asymptotics. We rewrite the model in matrix notation as:
\begin{equation} \label{eq:M2}
y_i = F \beta_i + \varepsilon_i,
\end{equation}
where $y_i$ and $\varepsilon_i$ are $T \times 1$ vectors and $F$ is a $T \times k$ matrix. We work conditionally on a given realization of the factor path, i.e., we treat $F$ as an unknown matrix parameter. Our focus is on inference on the number of latent factors $k$.

In this section, we develop the framework with strong factors, namely matrix $\Sigma_{\beta}:= \underset{n \rightarrow \infty}{\lim}~\frac{1}{n}\sum_{i=1}^n \beta_i \beta_i'$ is positive definite. We consider the setting with semi-strong and weak factors in a later section. Next, we present two approaches for identification of the unknown number $k$ of factors.

\subsection{Identification by instrumental variables}

We start by assuming the existence of an overidentified vector of instrumental variables. In a large $T$ framework, identification with instrumental variables is considered by Gagliardini and Gourieroux (2017) in static (i.e., unconditional) factor models, and by Kelly, Pruitt and Su (2017, 2019) and Gagliardini and Ma (2019) in dynamic (i.e., conditional, or time-varying beta) models.

\begin{assumption} \label{ass:IV}
There exists a $K$-dimensional vector of instrumental variables $z_i$, for $K > k$, such that:
\begin{itemize}
\item[(i)] $\underset{n \rightarrow \infty}{\text{plim}} ~ \frac{1}{n} \sum_{i=1}^n z_i \varepsilon_{i}' = E[z_i \varepsilon_{i}'] = 0$, 

\item[(ii)] The $K \times k$ matrix $\Gamma = \underset{n \rightarrow \infty}{plim} ~ \frac{1}{n} \sum_{i=1}^n z_i \beta_i'$ has full column rank.
\end{itemize}
\end{assumption}

Instrumental variables are cross-sectionally uncorrelated with error terms at all dates $t=1,...,T$, and full-rank correlated with the betas. We can take assets characteristics measured at $t=0$, or their time average in a period previous to sample dates, as candidates for instrumental variables.  

Following Gagliardini, Gourieroux (2017) and Gagliardini, Ma (2019), let us define the limit cross-sectional average:
\begin{equation} \label{xit}
\xi_t = \underset{n \rightarrow \infty}{\text{plim}} ~ \frac{1}{n} \sum_{i=1}^n z_i y_{i,t}, 
\end{equation}
for any $t$, i.e., the vector of returns of asymptotic static portfolios with weights proportional to the characteristics that are the elements of $z_i$. In a general setting with dynamic betas and with large $T$, Gagliardini and Ma (2019) use time-varying characteristics $z_{i,t}$ in Equation (\ref{xit}), which yield managed portfolios returns, but it make inferential theory on the number of factors very challenging, and is not considered in our small $T$ setting.  Under Assumption \ref{ass:IV}, we get from Equations (\ref{eq:M1}) and (\ref{xit}):
\begin{equation} \label{xiGamma}
\xi_t = \Gamma f_t,  \qquad t=1,...,T,
\end{equation}
or equivalently in matrix notation $
\Xi = F \Gamma',$
where $\Xi$ is a $T \times K$ matrix, i.e., a rank-$k$ exact matrix factorization. Hence, the asymptotic portfolio returns $\xi_t$ have a singular factor structure without error terms. 
The sample second-moment  matrix of $\xi_t$ for $t=1,...,T$ is given by:
\begin{equation} \label{eq:SVD}
V_{\xi} = \Gamma \tilde{\Sigma}_f \Gamma',
\end{equation}
where $\tilde{\Sigma}_f = \frac{1}{T} \sum_{t=1}^T f_t f_t'$ is the sample second-moment matrix of the latent factor. We can normalize the latent factors vector such that $\tilde{\Sigma}_f$ is diagonal, with diagonal elements ranked in decreasing order, and the columns of matrix $\Gamma$ are orthonormal vectors. Then, Equation (\ref{eq:SVD}) corresponds to the spectral decomposition of $V_{\xi}$ with diagonal matrix of eigenvalues $\tilde{\Sigma}_f$ and matrix of normalized eigenvectors $\Gamma$. 

We assume that the factor path is such that $\tilde{\Sigma}_f$ is invertible, which requires $T \geq k$, with distinct eigenvalues. From Equation (\ref{eq:SVD}), the $K\times K$ variance matrix $V_{\xi}$ is reduced rank, with rank $k$. Thus, by denoting $\delta_j(\cdot)$ the $j$th largest eigenvalue of a symmetric matrix, the number of latent factors is identifiable by the property:
$
\delta_j( V_{\xi}) = 0, \quad j=k+1,...,K,$
while these eigenvalues are strictly larger than zero for $j\leq k$. Moreover, under the proposed normalization the latent factor values are identifiable as well (up to sign changes), namely
$
f_t = \Gamma' \xi_t
$ is the $k$-dimensional (population) Principal Components (PC) of vector $\xi_t$.

The inference on the number of factors $k$ coincides with testing on the rank of symmetric matrix $V_{\xi}$. While there is an extensive literature on testing for the rank of a matrix, it is known that standard procedures may not apply in the case of symmetric matrices. In our setting, if the $\xi_t$ were observed in sample without estimation error, the testing problem would be ``degenerate" since the components of vector $\xi_t$ feature deterministic relationships allowing for exact determination of $k$. Since the $\xi_t$ have to be estimated by cross-sectional averaging, the corresponding estimation error drives the distributional properties of the test. It explains the nonstandard setting of the testing problem at hand, and its similarity with the problem of inference on the number of unit canonical correlations among two principal components vectors estimated from large panels as studied by Andreou, Gagliardini, Ghysels, and Rubin (2019). 

\subsection{Identification by the variance-covariance matrix of returns}

As a second approach to identification of the latent factor space with fixed $T$, let us consider the cross-sectional 
second-moment matrix of returns
\begin{equation} \label{Vy}
V_y = \underset{n\rightarrow\infty}{\text{plim}} ~ \frac{1}{n} \sum_{i=1}^n y_i y_i'.
\end{equation}
In alternative to the availability of instruments (Assumption \ref{ass:IV}), here we assume asymptotic unconditional homoschedasticity and no serial correlation of the errors, namely spherical error terms.
\begin{assumption} \label{ass:var:returns}
We have 
$
V_{\varepsilon}:= \underset{n\rightarrow\infty}{\text{plim}} ~ \frac{1}{n} \sum_{i=1}^n \varepsilon_i \varepsilon_i' = \bar{\sigma}^2 I_T,
$
where $\bar{\sigma}^2>0$ is a constant.
\end{assumption}
\noindent Assumption \ref{ass:var:returns} allows e.g. for idiosyncratic conditional heteroscedasticity in the individual error processes, as in DGP 3 considered for the Monte Carlo experiments (Section 6). It excludes e.g. a strong factor in idiosyncratic return volatilities; see Renault, Van Der Heijden and Werker (2022) for an arbitrage pricing theory with idiosyncratic variance factors. The assumption of spherical error terms underlies the testing methodology of Connor, Korajczyk (1993) and the analysis of Zaffaroni (2019).

From Assumption \ref{ass:var:returns}, Equation (\ref{Vy}), and $ \underset{n\rightarrow\infty}{\text{plim}} ~ \frac{1}{n} \sum_{i=1}^n \beta_i \varepsilon_i' = 0$, we have:
$
V_y = F {\Sigma}_{\beta} F' + \bar{\sigma}^2 I_T.  
$
We can work with the factor normalization such that $\tilde{\Sigma}_f = I_k$ and matrix ${\Sigma}_{\beta} = diag(\sigma_{\beta,j}^2)$ is diagonal.
Then, the first $k$ eigenvalues of $V_y$ are $T \sigma_{\beta,j}^2 + \bar{\sigma}^2$, for $j=1,...,k$, associated with eigenvectors that are the columns of matrix $\frac{1}{\sqrt{T}}F$, while the $T-k$ smallest eigenvalues of $V_y$ are equal to $\bar{\sigma}^2$.  
We assume that the eigenvalues $\sigma_{\beta,j}^2$ are distinct. Then, the number of factors $k$ is identifiable under Assumption \ref{ass:var:returns} since the eigenvalue difference is such that 
$
\delta_{j}(V_y) -\delta_{j+1}(V_y)=0$, for $j = k+1,...,T-1,
$
  while this difference is strictly larger than $0$, for $j \leq k$.

\subsection{PCA, Error-in-Variable and incidental parameters}

In large panels, the standard estimator for the latent factor space is based on Principal Component Analysis (PCA). It consists in the (normalized) eigenvectors of the sample analogue of matrix $V_y$ associated with the $k$ largest eigenvalues. Bai and Ng (2002), Bai (2003), Stock and Watson (2002) provide pioneering work for the study of the large sample properties with both $n$ and $T$ large, including a consistent selection procedure for the number of latent factors.  Forni et al.\ (2000) consider identification and estimation of generalized dynamic-factor models with large $n,T$. Theorem 4 in Bai (2003) shows that the PCA estimator is consistent with fixed $T$ and $n$ large if, and only if, Assumption \ref{ass:var:returns} holds (see also Connor and Korajczyk (1987)). Zaffaroni (2019) establishes the asymptotic normality of the PCA estimator and consistent selection of the number of latent factors in that setting. Essentially consistency of PCA estimators holds because the columns of $F$ are eigenvectors of $V_y$ associated with the $k$ largest eigenvalues under Assumption \ref{ass:var:returns}.  

In this section, we first derive the result in Theorem 4 of Bai (2003) using different arguments. \footnote{In particular, a proof not using Lemma D.1 in Bai (2003).} Then, we illustrate that result and Assumptions \ref{ass:IV} and \ref{ass:var:returns} from the view point of the Error-in-Variable (EIV) problem and the incidental parameter problem. 

\subsubsection{(In)consistency of the PCA factor estimator}

For a generic matrix $V_{\varepsilon}$, the limit variance-covariance matrix of returns is $V_y = F \Sigma_{\beta} F' + V_{\varepsilon}$. To get the spectral decomposition of this matrix, let us define the orthogonal matrix $R = [ \frac{1}{\sqrt{T}} F ~ : ~Q ]$, where $Q$ is a $T \times (T-k)$ matrix whose columns are an orthonormal basis of the complement of the range of $F$ (with $F'F/T=I_k$ in our normalization). Then:
\begin{eqnarray*}
R' V_y R = \left( 
\begin{array}{cc}
T \Sigma_{\beta} + \frac{1}{T} F'V_{\varepsilon} F & \frac{1}{\sqrt{T}} F' V_{\varepsilon} Q \\
\frac{1}{\sqrt{T}} Q' V_{\varepsilon} F & Q' V_{\varepsilon} Q
\end{array}
\right).
\end{eqnarray*}
The first $k$ eigenvectors of matrix $V_y$ are the columns of $F$ up to a rotation if, and only if, (1) the out-of-diagonal block $\frac{1}{\sqrt{T}} F' V_{\varepsilon} Q$ vanishes, and (2) the eigenvalues of the upper-left block $T \Sigma_{\beta} + \frac{1}{T} F'V_{\varepsilon} F$ are larger than any eigenvalue of the lower-right block $Q' V_{\varepsilon} Q$. Using that $M_F = I_T - \frac{1}{T} F F'$ is the orthogonal projection onto the orthogonal complement of the range of $F$, we get the next result. 

\begin{proposition} \label{prop:PCA:consist}
The PCA estimator is consistent for given $F$ with fixed $T$ and $n\rightarrow\infty$, i.e., $\underset{n\rightarrow\infty}{\text{plim}}~ \hat{F} = F$ up to a rotation, if and only if, 
\begin{equation} \label{cond:MFV}
\left\{ \begin{array}{l}
M_F V_{\varepsilon} F = 0, \\
\delta_k( T \Sigma_{\beta} + \frac{1}{T} F'V_{\varepsilon}F ) > \delta_1 ( M_F V_{\varepsilon} M_F ).
\end{array}
\right.
\end{equation}
\end{proposition} 

The condition $M_F V_{\varepsilon} F = 0$ is equivalent to $V_{\varepsilon} F = F A$ for a $k \times k$ (symmetric, non-singular) matrix $A$, i.e., the range of $F$ is an invariant subspace of $V_{\varepsilon}$. \footnote{The condition $M_F V_{\varepsilon} F = 0$ is the condition for OLS and GLS estimators to coincide in a regression with design matrix $F$ and variance-covariance matrix of the errors $V_{\varepsilon}$.} As in Theorem 4 of Bai (2003), let us require consistency of the PCA estimator for any $F$, and not just for a given $F$. Then, for a given matrix $V_{\varepsilon}$ independent of $F$, the condition $M_F V_{\varepsilon} F = 0$ holds for every $F$ if, and only if, $V_{\varepsilon} = \bar{\sigma}^2 I_T$ for a constant $\bar{\sigma}^2>0$, namely Assumption \ref{ass:var:returns} holds. In that case, the second condition in System (\ref{cond:MFV}) is met as well. Hence, we deduce that Assumption \ref{ass:var:returns} is sufficient and necessary for the consistency of the PCA estimator for any factor path $F$ as stated  in Theorem 4 of Bai (2003).

\subsubsection{Interpretation in terms of Error-in-Variable}

In this subsection, we provide an interpretation for the conditions in Proposition \ref{prop:PCA:consist} in terms of an EIV problem. For this purpose, recall that the PCA estimator solves the first-order conditions of the Least-Square (LS) problem for factor values and loadings:
\begin{eqnarray*}
\hat{F} = \left( \frac{1}{n} \sum_{i=1}^n y_i \hat{\beta}_i' \right) \left( \frac{1}{n} \sum_{i=1}^n \hat{\beta}_i \hat{\beta}_i' \right)^{-1}, \qquad 
\hat{\beta}_i = \frac{1}{T} \hat{F}'y_i, \quad i=1,...,n,
\end{eqnarray*}
with the normalization $\hat{F}'\hat{F}/T=I_k$ and $\frac{1}{n} \sum_{i=1}^n \hat{\beta}_i \hat{\beta}_i'$ diagonal. Hence, the factor values result from a multivariate cross-sectional regression of returns onto estimated loadings. With fixed $T$, the estimation error in the latter does not vanish asymptotically, and it originates an EIV problem. In fact, the LS problem can be seen as a multivariate regression 
\begin{equation} \label{endo:reg}
y_i = F \hat{\beta}_i + v_i, \quad i=1,...,n,
\end{equation}
 with matrix parameter $F$, error $v_i = \varepsilon_i - F ( \hat{\beta}_i - \beta_i)$ and endogenous regressor $\hat{\beta}_i$. The OLS estimator $\hat{F}$ in the regression (\ref{endo:reg}) \footnote{This regression is unfeasible because the $\hat{\beta}_i$ need to be estimated at the same time as $\hat{F}$, and not sequentially.} is consistent if, and only if,
\begin{equation} \label{orth:EIV}
\underset{n\rightarrow\infty}{\text{plim}}~\frac{1}{n} \sum_{i=1}^n v_i \hat{\beta}_i' = 0.
\end{equation}
To understand how this orthogonality condition of the error and the estimated regressor is linked to the conditions of Proposition \ref{prop:PCA:consist} on the consistency of the PCA estimator, suppose indeed $\text{plim} ~ \hat{F} = F$. Then, using $\hat{\beta}_i = \beta_i - \frac{1}{T}\hat{F}'(\hat{F}-F) \beta_i + \frac{1}{T} \hat{F}'\varepsilon_i$ and $\underset{n\rightarrow\infty}{\text{plim}}~\frac{1}{n} \sum_{i=1}^n \varepsilon_i \beta_i' = 0$, we have:
\begin{equation*}
\underset{n\rightarrow\infty}{\text{plim}}~\frac{1}{n} \sum_{i=1}^n v_i \hat{\beta}_i' = \frac{1}{T}
\underset{n\rightarrow\infty}{\text{plim}}~\frac{1}{n} \sum_{i=1}^n \left( \varepsilon_i - \frac{1}{T}F F' \varepsilon_i\right) \varepsilon_i' F = \frac{1}{T} M_F V_{\varepsilon} F.
\end{equation*}
Thus, (\ref{orth:EIV}) holds if $M_F V_{\varepsilon} F = 0$, that is the first condition in (\ref{cond:MFV}). 
This condition is necessary for the consistency of the PCA estimator with fixed $T$ and to eliminate the endogeneity issue from the EIV problem.

The EIV framework is also useful to interpret the IV condition in Assumption \ref{ass:IV}. Indeed, the variables $z_i$ can be seen as instruments for the endogenous regressors $\hat{\beta}_i$ in regression (\ref{endo:reg}). However, in contrast to the standard IV framework, regression (\ref{endo:reg}) is infeasible since the $\hat{\beta}_i$ have to be obtained at the same time as the estimate of $F$. Moreover, while Equation (\ref{xiGamma}) corresponds to the ``population normal equation" for IV in the multivariate cross-sectional regression at date $t$, we cannot identify matrix $\Gamma $ by the sample cross-moments of the $z_i$ and ${\beta}_i$, the true betas being unknown. Instead, matrices $\Gamma$ and $F$ have to be identified jointly by the spectral decomposition of the sample second-moment $V_{\xi}$ under the normalization of the latent factors to have $\tilde{\Sigma}_f = F'F/T$ diagonal and $\Gamma'\Gamma = I_{k}$. \footnote{Shanken (1992), Kim and Skoulakis (2018), and Raponi, Robotti and Zaffaroni (2020) use IV approaches (or similar) with fixed  $T$  to estimate ex-post risk premia of observed factors in the second pass cross-sectional regression. That use of IV differs from ours because betas are obtained in the first pass by regressing returns onto observed factors across time.} 

\subsubsection{Interpretation in terms of incidental parameters}

We can also analyse the (in)consistency of the PCA estimator with fixed $T$ from the vantage point of the well-known incidental parameter problem of the panel data literature (Neyman and Scott (1948), Lancaster (2000)). \footnote{Based on a discussion of Zaffaroni (2019) by P.\ Gagliardini, possibly in some form in a new version of Zaffaroni (2019).} Indeed, in a setting with fixed $T$, we can see matrix $F$ as a common parameter in the panel model (\ref{eq:M2}), while the loadings $\beta_i$ play the role of incidental parameters. The number of incidental parameters grows with the cross-sectional sample size $n$, so that the information to estimate $F$ does not necessarily accumulate, leading potentially to inconsistency of the estimator for $F$. 

The PCA estimator minimizes the LS criterion (i.e., minus the Gaussian pseudo log-likelihood) given by $\mathcal{L}_n(\beta,F) = \frac{1}{n} \sum_{i=1}^n (y_i - F\beta_i)'(y_i - F \beta_i)$ with the normalization $F'F/T = I_k$. The minimizer of $\beta_i$ for a given $F$ is $\hat{\beta}_i(F) = \frac{1}{T} F'y_i$. The concentrated LS criterion becomes:
\begin{eqnarray} \label{concentrated}
\mathcal{L}^c_n(F) &:=& \mathcal{L}_n(\hat{\beta}(F),F) = \frac{1}{n} \sum_{i=1}^n y_i' M_F y_i .
\end{eqnarray}
Using that $y_i = F^0 \beta_i + \varepsilon_i$, where $F^0$ is the matrix of true factor values, and the properties of the trace, we get from (\ref{concentrated}):
\begin{eqnarray} \label{lim}
 \underset{n \rightarrow \infty}{\text{plim}} ~ \mathcal{L}^c_n(F) = Tr[ F^{0\prime} M_F F^0 \Sigma_{\beta}] + Tr[ M_F V_{\varepsilon}]  = - \frac{1}{T} Tr[ F' ( F^0 \Sigma_{\beta} F^{0\prime} + V_{\varepsilon}) F ],
\end{eqnarray}
up to terms that do not dependent on $F$. The minimizer $F^{*}$ of (\ref{lim}) is the matrix of the standardized eigenvectors of $V_y = F^0 \Sigma_{\beta} F^{0\prime} + V_{\varepsilon}$ associated with the $k$ largest eigenvalues. Under the conditions of Proposition \ref{prop:PCA:consist}, we get $F^{*}=F^0$. The population concentrated criterion (\ref{lim}) being minimized at the true value implies the consistency of the PCA estimator.

An interesting perspective on factor estimation with instrumental variables (Assumption \ref{ass:IV}) from the view point of the incidental parameter problem is suggested by Section 4 in Chamberlain (1992). The idea is to construct orthogonality restrictions that get rid of the incidental parameters and identify the common parameter in a panel model with random effects. Suppose we have ``exogeneity" of the instrumental variables such that $E[ \varepsilon_i \vert z_i] = 0$, which is a stronger condition than Assumption \ref{ass:IV}. Then, considering the loadings $\beta_i$ as random, the setting of Chamberlain (1992) Section 4 applies. Indeed, we have $E[ y_i \vert z_i, \beta_i] = F(\theta) \beta_i$, where $F(\theta)$ denotes the matrix of factor values once we impose the normalization restriction that the lower $k\times k$ block is the identity $I_k$, and denote the vec of the upper $(T-k)\times k$ block of this matrix as the parameter $\theta$, i.e., $F(\theta) = ( \tilde{\theta}' : I_k)'$ and $\theta = vec(\tilde{\theta})$. By the Law of Iterated Expectation, we get $E[ y_i \vert z_i ] = F(\theta) h(z_i)$, where $h(z_i):= E[ \beta_i \vert z_i]$. Thus, we end up with a conditional moment restriction model with a finite-dimensional parameter $\theta$ and a functional parameter $h(\cdot)$. Chamberlain (1992) shows how to design a method of moment estimator for $\theta$ that achieves the semiparametric efficiency bound with $n\rightarrow  \infty$ and fixed $T$. Specifically, we have the conditional moment restriction $E[ M(z_i,\theta) y_i \vert z_i ] = 0$, where $M(z,\theta) = I_T -  F(\theta) [ F(\theta)' \Omega (z)^{-1} F(\theta)]^{-1} F(\theta)' \Omega(z)^{-1}$ is the oblique projection on the orthogonal complement of the range of $F(\theta)$  associated with the scalar product corresponding to the positive definite matrix $\Omega(z)^{-1}$. With optimal instruments $A(z,\theta)$, we get an orthogonality restriction $E[ A(z_i,\theta)' M(z_i,\theta) y_i ] = 0$ for semi-parametric efficient estimation of $\theta$. As  we focus mainly on the inference on the number of factors in this paper, we do not explore further this route. We conjecture however that a test on the number of latent factors can be designed as a specification test for the conditional moment restriction. 

Finally, we note that Fan, Liao and Wang (2016) also follow a random effects approach and assume that $E[ \beta_i \vert z_i] = h( z_i)$ for an unknown function $h(\cdot)$ which they estimate by a Sieve approach using the variance-covariance matrix of returns projected onto the instruments $z_i$. They develop an estimation method for the number of latent factors in the vein of Ahn and Horenstein (2013).

\section{Eigenvalue test statistics}

We develop statistics to test the null hypothesis $H_0(k)$ of $k$ latent common factors against the alternative $H_1(k)$ of more than $k$ latent factors in short panels, namely $T$ kept fixed.  

\subsection{Test statistic based on instrumental variables}

Let us estimate vector $\xi_t$ by the cross-sectional average of the observations times the instruments:
\begin{equation} \label{hatxit}
\hat{\xi}_t = \frac{1}{n} \sum_{i=1}^n z_i y_{i,t},
\end{equation}
for any $t$. From (\ref{eq:M1}) and (\ref{hatxit}), these aggregate measurements satisfy:
\begin{equation} \label{eq:Mtilde}
\hat{\xi}_t = \tilde{\Gamma} f_t + \frac{1}{\sqrt{n}} u_t,
\end{equation}
where
$
\tilde{\Gamma} = \frac{1}{n}\sum_{i=1}^n z_i \beta_i'$ and $u_t = \frac{1}{\sqrt{n}} \sum_{i=1}^n z_i \varepsilon_{i,t}.$
The symbol tilde is used instead of the hat since $\tilde{\Gamma}$ is an infeasible (yet consistent) estimator of the matrix $\Gamma$. From the CLT and Assumption \ref{ass:IV} (i),  vector $u_t$ for any $t$ is asymptotically Gaussian as $n \rightarrow \infty$ (see below). From Equation (\ref{eq:Mtilde}), the vectors $\hat{\xi}_t$ obey a ``small'' latent factor model, with latent factors $f_t$ and idiosyncratic noise scaled with $1/\sqrt{n}$. 

The estimator of the second-moment matrix $V_{\xi}$ is the sample second-moment 
\begin{equation}  \label{hatvxi}
\hat{V}_{\xi} = \frac{1}{T}\sum_{t=1}^T  \hat{\xi}_t \hat{\xi}_t'.
\end{equation}
The first test statistic is based on the sum of the $K-k$ smallest eigenvalues of matrix $\hat{V}_{\xi}$ in (\ref{hatvxi}), i.e.,
\begin{equation} \label{calTk}
\mathscr{T}(k) = \sum_{j= k+1}^K  \delta_j( \hat{V}_{\xi} ).
\end{equation}
Under the regularity conditions detailed below, $\mathscr{T}(k)$ in (\ref{calTk}) converges to the sum of the $K-k$ smallest eigenvalues of matrix $V_{\xi}$, namely $0$ under the null hypothesis $H_0(k)$, and a strictly positive constant under the alternative $H_1(k)$. Thus, values of the test statistic $\mathscr{T}(k)$ above a well-chosen threshold imply rejection of $H_0(k)$ in favor of $H_1(k)$. To determine the threshold for the rejection region, we obtain the asymptotic distribution of $\mathscr{T}(k)$ and show that, after suitable rescaling, this statistic is asymptotically distributed as a weighted sum of independent chi-square variates under $H_0(k)$ as $n \rightarrow \infty$ and $T$ is fixed. 

Following the literature on rank testing (see, e.g., Robin and Smith (2000)) the statistic can be generalized considering the family $\mathscr{T}(k) = \sum_{j= k+1}^K  \phi\left(\delta_j( \hat{V}_{\xi} )\right)$, where function $\phi(\cdot)$ is such that $\phi(0)=0$, $\phi(u)>0$ for $u>0$, and $\phi'(0)=1$. By the delta method, the asymptotic distribution of the test statistic under the null does not depend on the choice of the function $\phi(\cdot)$. The latter has an impact on the power properties. 

To study the large sample properties of the test statistic $\mathscr{T}(k)$, let us note that:
\begin{eqnarray}  \label{eq:VhatVtilde}
\hat{V}_{\xi} = \tilde{V}_{\xi} + \hat{\Psi},
\end{eqnarray}
where $\tilde{V}_{\xi} = \tilde{\Gamma} \tilde{\Sigma}_f \tilde{\Gamma}',$ $\tilde{\Sigma}_f = \frac{1}{T} \sum_{t=1}^T f_t f_t',$
 and
\begin{equation}  \label{Psihat}
\hat{\Psi} = \frac{1}{\sqrt{n}} \tilde{\Gamma} \left( \frac{1}{T} \sum_{t=1}^T f_t u_t' \right) 
+ \frac{1}{\sqrt{n}} \left( \frac{1}{T} \sum_{t=1}^T u_t f_t' \right) \tilde{\Gamma}' +  \frac{1}{n} \left( \frac{1}{T} \sum_{t=1}^T u_t u_t' \right).
\end{equation}
It is convenient to assume the normalization of the latent factor vector such that the columns of matrix $\tilde{\Gamma}$ are orthonormal and matrix $\tilde{\Sigma}_f$ is diagonal. This normalization is sample dependent (because it involves $\tilde{\Gamma}$ instead of $\Gamma$), and is coherent with the normalization adopted in the previous section for identification. It yields the spectral decomposition of $\tilde{V}_{\xi}$ with diagonal eigenvalues matrix $\tilde{\Sigma}_f$ and matrix of standardized eigenvectors $\tilde{\Gamma}$. 

In Equation (\ref{eq:VhatVtilde}), the matrix $\hat{V}_{\xi}$ is written as the sum of a reduced rank matrix $\tilde{V}_{\xi}$, with rank $k$ in sample, and a ``small perturbation'' $\hat{\Psi}$ given by (\ref{Psihat}). The distributional properties of the test are driven by the perturbation 
$\hat{\Psi}$, which is affecting the $K-k$ smallest eigenvalues of $\hat{V}_{\xi}$. The perturbation $\hat{\Psi}$ has a term at order $O_p(\frac{1}{\sqrt{n}})$ and a term at order $O_p(\frac{1}{n})$. The first term is the dominant one in probability order. However, the joint distribution of its elements is degenerate because it involves reduced rank matrices. It means that the term at order $O_p(1/n)$ dominates for certain linear combinations of the elements of $\hat{\Psi}$. In particular, the asymptotic distribution of small eigenvalues of $\hat{V}_{\xi}$ involve second-order effects (see below).

\subsection{Test statistics based on eigenvalues of the return variance-covariance}

We estimate $V_y$ by the sample second-moment matrix $\hat{V}_y = \frac{1}{n} \sum_{i=1}^n y_i y_i'$. A test statistic based on the  eigenvalue difference is:
\begin{equation} \label{calSk}
\mathscr{S}(k) =  \delta_{k+1}( \hat{V}_y) - \delta_{T}( \hat{V}_y).
\end{equation}
The statistic $\mathscr{S}(k)$ in (\ref{calSk}) converges in probability to $0$ under the null $H_0(k)$, and to a positive constant under the alternative $H_1(k)$. The differencing has the purpose to eliminate the term $\bar{\sigma}^2$ that is common across all eigenvalues of $V_y$. Note that statistic $\mathscr{S}(k)$ equals the telescope sum of eigenvalue differences $\delta_{j}( \hat{V}_y) - \delta_{j+1}( \hat{V}_y)$ from $j=k+1$ to $j=T-1$.

Other eigenvalue differences can be considered, and different functional forms can be used to aggregate those differences. In the vein of Onatski (2009), we can consider the statistic built by the maximal ratio of consecutive eigenvalue differences 
\begin{equation} \label{calS*k}
\mathscr{S}^*(k) = \underset{j=k+1,...,k^*}{\max} \frac{\delta_{j}( \hat{V}_y) - \delta_{j+1}( \hat{V}_y)}{\delta_{j+1}( \hat{V}_y) - \delta_{j+2}( \hat{V}_y)},
\end{equation}
with $k+1 \leq k^* \leq T-2$. 
Under the alternative of more than $k$ (but less than $k^*+2$) factors, the statistic $\mathscr{S}^*(k)$ in (\ref{calS*k}) diverges because there is a ratio between a strictly positive numerator and an asymptotically vanishing denominator. In the large $T$ setting of Onatski (2009), the statistic's denominator is the difference between two asymptotically vanishing quantities, while this is not the case with finite $T$.

We use the expansion
\begin{equation} \label{hatVy}
\hat{V}_y =  \tilde{V}_y +  \hat{\Phi},
\end{equation}
where
\begin{eqnarray}
\tilde{V}_y &=& F \tilde{\Sigma}_{\beta} F'  + \tilde{\sigma}^2 I_T , \qquad \tilde{\Sigma}_{\beta} = \frac{1}{n} \sum_{i=1}^n \beta_i \beta_i' , \nonumber \\
\hat{\Phi} &=& \frac{1}{\sqrt{n}} \left[  \left( \frac{1}{\sqrt{n}} \sum_{i=1}^n \varepsilon_i \beta_i' \right)F' + F \left( \frac{1}{\sqrt{n}} \sum_{i=1}^n  \beta_i \varepsilon_i' \right) +  \frac{1}{\sqrt{n}} \sum_{i=1}^n \left( \varepsilon_i \varepsilon_i' - {\sigma}_i^2 I_T \right) \right], \qquad  \label{exp:Phihat}
\end{eqnarray}
where $\tilde{\sigma}^2 = \frac{1}{n}\sum_{i=1}^n \sigma_i^2$ and the $\sigma_i^2>0$ are positive constants. 
We can normalize the latent factors such that $\tilde{\Sigma}_f = I_k$ and the matrix $\tilde{\Sigma}_{\beta} = diag ( \tilde{\sigma}_{\beta,j}^2)$ is diagonal, with diagonal elements ranked in decreasing order. This normalization  is sample-dependent, and  coherent with that considered in Section 2.2 in the population. In fact, under this normalization, the matrix $F = F_n$ may be sample dependent, but we omit index $n$ for expository purpose. 
Then we have $\delta_j( \tilde{V}_y ) =  T \tilde{\sigma}_{\beta,j}^2 + \tilde{\sigma}^2$, for $j=1,...,k$, and $\delta_j( \tilde{V}_y ) = \tilde{\sigma}^2$, for $j =k+1,...,T$. The eigenvectors of $\tilde{V}_y$ to the first $k$ eigenvalues are the columns of the matrix $\frac{1}{\sqrt{T}} F$. The perturbation matrix $\hat{\Phi}$ is of probability order $O_p(1/\sqrt{n})$ under Assumption \ref{ass:var:returns} and regularity conditions. 

\section{Asymptotic distributions of the test statistics}

Let $T$ be kept fixed in the asymptotics, and $n \rightarrow \infty$. To derive the asymptotic distribution of the statistic $\mathscr{T}(k)$, resp.\ the statistics $\mathscr{S}(k)$ and $\mathscr{S}^*(k)$, we use a second-order, resp.\ first-order, expansion for the small eigenvalues of matrices $\hat{V}_{\xi}$ and $\hat{V}_y$. We start with a general result which covers matrix perturbations as in Equations (\ref{eq:VhatVtilde}) and (\ref{hatVy}).

\subsection{Asymptotic expansion of the small eigenvalues via perturbation theory}

Let $A = U D U'$ be a symmetric $K \times K$ matrix of rank $k$, where $D$ is the diagonal matrix of the $k$ non-zero eigenvalues, and $U$ is the $K \times k$ matrix of the associated orthonormal eigenvectors. Let 
\begin{equation}  \label{eq:Ahat}
\hat{A} = A + \hat{\Psi},
\end{equation}
be an estimator of matrix $A$, where symmetric matrix $\hat{\Psi}$ is the estimation error (i.e., a ``small perturbation''). We want to derive an asymptotic expansion for the $K-k$ smallest eigenvalues of $\hat{A}$, namely $\delta_j ( \hat{A})$ for $j=k+1,...,K$, as a power series of $\hat{\Psi}$. 

We have the following result that is proved in Appendix. 
\begin{theorem} \label{prop:expansion:eigenvalues}
Let $A = U D U'$ be a symmetric $K \times K$ matrix of rank $k$, where $D$ is the diagonal matrix of the $k$ non-zero eigenvalues, and $U$ is the $K \times k$ matrix of the associated orthonormal eigenvectors. Let $\hat{A} = A + \hat{\Psi}$, where $\hat{\Psi}$ is a symmetric ``small perturbation'' matrix such that $\Vert \hat{\Psi} \Vert \leq \frac{1}{3 \Vert D^{-1} \Vert (K+1)^{3/2}}$. Then:
\begin{eqnarray}  \label{eq:eigv:Ahat}
\ \delta_{k+j} (\hat{A}) =  \delta_j \left( Q ' \hat{\Psi} Q -  Q'\hat{\Psi} U D^{-1} U' \hat{\Psi} Q \right) 
+ O \left(    \Vert D^{-1} \Vert^2 K^{4} \Vert \hat{\Psi} \Vert^3  \right),
\end{eqnarray}
for $j=1,...,K-k$, 
where $Q$ is a $K\times (K-k)$ matrix whose orthonormal columns span the null space of $A$, and the remainder term is uniform. 
\end{theorem}

The eigenvalue in the RHS of (\ref{eq:eigv:Ahat}) is invariant to the choice of matrix $Q$ whose range spans the orthogonal complement of the range of $F$. 
The uniformity of the remainder term in Theorem  \ref{prop:expansion:eigenvalues} is in the sense that its norm is upper bounded by $C \Vert D^{-1} \Vert^2 K^{4} \Vert \hat{\Psi} \Vert^3$ for a universal constant $C$ that is independent of $\hat{A}$ and $A$. The remainder term in (\ref{eq:eigv:Ahat}) is of third-order in perturbation $\hat{\Psi}$ and depends on matrix $A$ solely via its dimension $K$ and the squared Frobenius norm of its generalized inverse $\Vert D^{-1} \Vert^2 = \sum_{j=1}^k \delta_j(A)^{-2}$. Hence, when the matrix $A$ has small eigenvalues among the first $k$, the remainder term in the expansion gets larger, other things being equal. Accounting for these effects is important when considering semi-strong or weak factors (see Section 5). Also, we highlight the effect of the matrix dimension $K$, which is finite in the applications under fixed $T$ in this paper, but allows to cover cases with $T$ growing in the double asymptotics  case. It is because of the asymptotic expansion holding under a representation with a uniform remainder term. This representation applies with both deterministic, and random matrices, in which case the bound is almost sure in probability.

\subsection{Asymptotic distribution of $\mathscr{T}(k)$ with fixed $T$}

\subsubsection{Asymptotic characterization}

We apply Theorem \ref{prop:expansion:eigenvalues} to the statistic $\mathscr{T}(k)$, i.e., the sum of the $K-k$ smallest eigenvalues of the matrix $\hat{V}_{\xi}$ which satisfies Equation (\ref{eq:VhatVtilde}). We have $\Vert \tilde{\Sigma}_f^{-1}\Vert$ finite and $\hat{\Psi} = O_p(\frac{1}{\sqrt{n}})$, so that $\Vert \hat{\Psi} \Vert \leq \frac{1}{3 \Vert \tilde{\Sigma}_f^{-1} \Vert (K+1)^{3/2}}$ w.p.a.\ $1$. Then, by using that the sum of the eigenvalues of a matrix corresponds to its trace, we have
\begin{equation}  \label{asy:exp:1}
\mathscr{T}(k) =  Tr \left[ \tilde{\Pi} ' \hat{\Psi} \tilde{\Pi} - \tilde{\Pi}'\hat{\Psi} \tilde{\Gamma} \tilde{\Sigma}_f^{-1} \tilde{\Gamma}' \hat{\Psi} \tilde{\Pi} \right]
+ O_p \left(   \frac{1}{n^{3/2}}  \right),
\end{equation} 
where $\tilde{\Pi}$ is a $K \times (K-k)$ matrix whose orthonormal columns span the orthogonal complement of the range of matrix $\tilde{\Gamma}$, and $\hat{\Psi}$ is given in (\ref{Psihat}). Hence, we get:
\begin{eqnarray*}
\tilde{\Pi} ' \hat{\Psi} \tilde{\Pi} &=& \frac{1}{n} \tilde{\Pi}' \left( \frac{1}{T} \sum_{t=1}^T u_t u_t' \right) \tilde{\Pi}, \\
\tilde{\Pi}' \hat{\Psi} \tilde{\Gamma} &=& \frac{1}{\sqrt{n}} \tilde{\Pi}' \left(  \frac{1}{{T}} \sum_{t=1}^T u_t f_t' \right) + \frac{1}{n} \tilde{\Pi}' \left( \frac{1}{T} \sum_{t=1}^T u_t u_t' \right) \tilde{\Gamma}.
\end{eqnarray*}
In particular, in the ``first-order'' term $\tilde{\Pi} ' \hat{\Psi} \tilde{\Pi}$, the components of $\hat{\Psi}$ scaled by $\frac{1}{\sqrt{n}}$ in (\ref{Psihat}) yield no contribution because $\tilde{\Pi}'\tilde{\Gamma}=0$. From (\ref{asy:exp:1}), and using $\tilde{\Gamma} = \Gamma + O_p(\frac{1}{\sqrt{n}})$, we get:
\begin{equation}  \label{asy:exp:2}
\mathscr{T}(k) =  \frac{1}{n} Tr \left[    {\Pi}' \left\{ \frac{1}{T} \sum_{t=1}^T u_t u_t' - \left(  \frac{1}{{T}} \sum_{t=1}^T u_t f_t' \right)   \tilde{\Sigma}_f^{-1} 
 \left(  \frac{1}{T} \sum_{t=1}^T f_t u_t' \right)
\right\}  {\Pi} \right] 
+ O_p \left(    \frac{1}{n^{3/2}} \right).
\end{equation} 
On the RHS within the curly brackets, we have the residual matrix of the multivariate regression of $u_t$ onto $f_t$, for $t=1,...,T$. 

Let $U = [ u_1 ~: ~ ...~ :~u_T]'$ and $F = [ f_1 ~: ~ ...~ :~f_T]'$. Then we can write:
\begin{eqnarray*}
&& Tr \left[  {\Pi}' \left\{ \frac{1}{T} \sum_{t=1}^T u_t u_t' - \left(  \frac{1}{{T}} \sum_{t=1}^T u_t f_t' \right)   \tilde{\Sigma}_f^{-1} 
 \left(  \frac{1}{T} \sum_{t=1}^T f_t u_t' \right)
\right\} {\Pi} \right]   \\
&=& \frac{1}{T} Tr \left[ \Pi' ( U'U - U' F ( F 'F)^{-1} F' U) \Pi \right] 
= \frac{1}{T} Tr \left[ U' M_F U M_{\Gamma} \right]  \\
&=& \frac{1}{T} vec [ U']'( M_F \otimes M_{\Gamma}) vec[U'],
\end{eqnarray*}
where $M_F = I_T - F ( F ' F)^{-1} F'$ and $M_{\Gamma} = \Pi \Pi' = I_K - \Gamma (\Gamma'\Gamma)^{-1}\Gamma'$ are idempotent matrices of rank $T-k$ and $K-k$. Thus, matrix $M_F \otimes M_{\Gamma}$ is idempotent with rank $(T-k)(K-k)$. Moreover, under regularity conditions, the next assumption is implied by a CLT.

\begin{assumption} \label{ass:CLT:instr}
We have $ \displaystyle 
vec[U'] =  \frac{1}{\sqrt{n}} \sum_{i=1}^n \varepsilon_i \otimes z_i \Rightarrow N( 0, \Sigma_U),
$ as $n\rightarrow\infty$, where $\Sigma_U$ is a $KT \times KT$ matrix. 
\end{assumption}

By the result on the distribution of idempotent quadratic forms of Gaussian vectors, we get the next result.

\begin{proposition} \label{prop2}
Under Assumptions \ref{ass:IV} and \ref{ass:CLT:instr}, regularity conditions and the null hypothesis $H_0(k)$ of $k$ latent factors, as $n \rightarrow \infty$ and $T$ is fixed, we have:
\begin{equation*}
n  \mathscr{T}(k)  ~ \Rightarrow ~ \frac{1}{T} \sum_{j=1}^{(T-k)(K-k)} \lambda_j \chi_j^2,
\end{equation*}
where the $\chi_j^2$ are independent chi-square variables with one degree of freedom, and the $\lambda_j$ are the $(T-k)(K-k)$ non-zero eigenvalues of matrix $\Lambda:=  ( M_F \otimes M_{\Gamma}) \Sigma_U ( M_F \otimes M_{\Gamma})$. Under the alternative hypothesis $H_1(k)$ that we have more than $k$ strong factors, $n  \mathscr{T}(k)$ diverges to infinity in probability at order $O_p(n)$. 
\end{proposition}

In Proposition \ref{prop2}, the asymptotic distribution under the null is a weighted average of independent chi-square distributions. The divergence of the statistic under the alternative $H_1(k)$ ensures a consistent test. Robin and Smith (2000) consider tests for the rank of a matrix. They also use statistics based on sums of (functions of) the small eigenvalues, and show that they are distributed asymptotically as weighted sums of chi-square distributions. However, their Theorem 3.2 does not apply for the test statistic $\mathscr{T}(k)$ because their Assumption 2.4 is not met here. 

When the vectors $u_t$ are asymptotically independent across time and homoschedastic, we have $\Sigma_U = \bar{\sigma}^2(I_T\otimes Q_{zz})$, where $Q_{zz} = \lim~\frac{1}{n} \sum_{i=1}^n E[ z_i z_i']$. \footnote{More precisely, we have $ \Sigma_U = \lim~\frac{1}{n}\sum_{i=1}^n E[ \varepsilon_i \varepsilon_i' \otimes z_i z_i'] = \lim \frac{1}{n} \sum_{i=1}^n E[\varepsilon_i \varepsilon_i'] \otimes E[z_i z_i']= $ \\ $V_{\varepsilon} \otimes Q_{zz}$ with $V_{\varepsilon} = \bar{\sigma}^2 I_T$.} Then, $\Lambda = \bar{\sigma}^2 \left(M_F \otimes ( M_{\Gamma} Q_{zz} M_{\Gamma} )\right)$, and its non-zero eigenvalues are equal to the eigenvalues of the matrix $\bar{\sigma}^2 \Pi'Q_{zz} \Pi$, each with multiplicity $T-k$. Thus, we have
\begin{equation} \label{asy:dist:tauk:simple}
n  \mathscr{T}(k)  ~ \Rightarrow ~ \frac{\bar{\sigma}^2}{T} \sum_{j=1}^{K-k} \delta_j( \Pi'Q_{zz} \Pi) \chi_j^2(T-k),
\end{equation}
where the $\chi_j^2(T-k)$ are independent chi-square variables with $T-k$ degrees of freedom.

\subsubsection{Feasible statistic}

 We can compute the critical values associated to the weighted sum of chi-square variables in (\ref{asy:dist:tauk:simple}) by simulations after estimating the eigenvalues of the matrix $\Pi'Q_{zz} \Pi$ through their empirical counterparts. In particular, we estimate $\hat{\Pi}$ from the orthogonal complement of the range of $\hat{\Gamma}$, i.e., the eigenvector matrix of $\hat{V}_{\xi}$ associated with the $k$ largest eigenvalues. To estimate $\bar{\sigma}^2$, we use the residuals $\hat{\varepsilon}_i = M_{\hat{F}}y_i$ for $\hat{F} = \hat{\Xi}\hat{\Gamma}$. By using that $\hat{F} = F + o_p( 1)$, we have
\begin{equation*}
\underset{n\rightarrow \infty}{\text{plim}}~\frac{1}{n} \sum_{i=1}^n \hat{\varepsilon}_i \hat{\varepsilon}_i ' =  M_F \left( \underset{n\rightarrow \infty}{\text{plim}} \frac{1}{n} \sum_{i=1}^n {\varepsilon}_i {\varepsilon}_i ' \right) M_F = \bar{\sigma}^2 M_F.
\end{equation*}
By the properties of the trace and $Tr[M_F]=T-k$, we deduce that a consistent estimator for $\bar{\sigma}^2$ is
\begin{equation} \label{hatsig2}
\hat{\sigma}^2 = \frac{1}{n (T-k)}\sum_{i=1}^n \sum_{t=1}^T \hat{\varepsilon}_{i,t}^2.
\end{equation}
With fixed $T$, we need a correction for the degrees of freedom  in (\ref{hatsig2}). 

To get a feasible statistic in the more general setting of Proposition \ref{prop2}, let us define $\hat{\Sigma}_U = \frac{1}{n}\sum_{i=1}^n ( \hat{\varepsilon}_i \hat{\varepsilon}_i') \otimes (z_i z_i')$ and let
\begin{equation*}
\hat{\Lambda} =  ( M_{\hat{F}} \otimes \hat{M}_{\Gamma} ) \hat{\Sigma}_U ( M_{\hat{F}} \otimes \hat{M}_{\Gamma} ),
\end{equation*}
with $\hat{M}_{\Gamma} = I_K - \hat{\Gamma}\hat{\Gamma}'$. 
 Under regularity conditions, we have $\underset{n\rightarrow\infty}{\text{plim}} ~ \hat{\Sigma}_U = (M_F \otimes I_K) \Sigma_U (M_F \otimes I_K)$. Then, the matrix $\hat{\Lambda}$ is a consistent estimator of $\Lambda$ for $n\rightarrow\infty$, and we can use its eigenvalues $\hat{\lambda}_j$ to weight the chi-square distributions and simulate the critical values of the statistic. The projection matrix $M_F \otimes M_{\Gamma}$ in $\Lambda$ implies the consistency of $\hat{\Lambda}$ despite the fact that $\hat{\Sigma}_U$ is inconsistent for $\Sigma_U$ with fixed $T$. 

Under  the alternative $H_1(k)$, the critical value of the simulated distribution with estimated quantities converges to a finite constant as well. This fact, together with the divergence of the test statistics under $H_1(k)$, guarantees the consistency of the test based on the feasible statistics. 

\subsection{Asymptotic distributions of $\mathscr{S}(k)$ and $\mathscr{S}^*(k)$ with fixed $T$}

\subsubsection{Asymptotic characterization}

To get the asymptotic distribution of the small eigenvalues of $\hat{V}_y$, we apply Theorem \ref{prop:expansion:eigenvalues} to the matrix $\hat{V}_y - \tilde{\sigma}^2 I_T$ using expansion (\ref{hatVy}). Indeed, the matrix $\tilde{V}_y - \tilde{\sigma}^2I_T = F \tilde{\Sigma}_{\beta} F'$ has reduced rank $k$. Then:
\begin{eqnarray*}
\delta_{k+j}( \hat{V}_y ) &=& \tilde{\sigma}^2 +  \delta_{k+j}( \hat{V}_y - \tilde{\sigma}^2 I_T ) 
= \tilde{\sigma}^2 +  \delta_{k+j}( \tilde{V}_y - \tilde{\sigma}^2 I_T  + \hat{\Phi}) \\
&=& \tilde{\sigma}^2 + \frac{1}{\sqrt{n}} \delta_j \left(Q' \left(\frac{1}{\sqrt{n}} \sum_{i=1}^n ( \varepsilon_i \varepsilon_i' - {\sigma}^2_i I_T) \right) Q \right) + O_p \left( \frac{1}{n} \right),
\end{eqnarray*}
for $j=1,...,T-k$, where $Q$ is a $T \times (T-k)$ matrix whose columns are orthonormal vectors spanning the orthogonal complement of the range of $F$. \footnote{Theorem \ref{prop:expansion:eigenvalues} implies that the asymptotic distribution involves the orthogonal complement to the range of $F_n$, i.e., the rotation of $F$ ensuring the sample-dependent normalization such that $\tilde{\Sigma}_{\beta}$ is diagonal. However, the ranges of $F_n$ and $F$ coincide, which explains why we use matrix $Q$.} Stopping the expansion at first-order is enough to characterize the asymptotic distribution. \footnote{Because second-order terms are negligible asymptotically for statistics $\mathscr{S}(k)$ and $\mathscr{S}^*(k)$, their large sample distributions can be established by simpler methods than Theorem \ref{prop:expansion:eigenvalues}. The second-order expansion in Theorem \ref{prop:expansion:eigenvalues} is needed for statistic $\mathscr{T}(k)$.}
\begin{assumption} \label{ass:CLT:epseps}
As $n \rightarrow \infty$, we have $\frac{1}{\sqrt{n}} \sum_{i=1}^n ( \varepsilon_i \varepsilon_i' - {\sigma}^2_i I_T) \Rightarrow Z$, where $Z$ is a $T \times T$ Gaussian matrix. 
\end{assumption}

Then, $\sqrt{n} [ \delta_{k+j}( \hat{V}_y ) - \delta_{k+j+1}( \hat{V}_y )] \Rightarrow \delta_j(Q'ZQ) - \delta_{j+1}(Q'ZQ)$ jointly for $j=1,...,T-k-1$. By the Continuous Mapping Theorem, we get the next result.

\begin{proposition} \label{prop:asy:SSstar}
Under Assumptions \ref{ass:var:returns} and \ref{ass:CLT:epseps}, regularity conditions and the null hypothesis $H_0(k)$ of $k$ latent factors, as $n \rightarrow \infty$ and $T$ is fixed, we have:
\begin{eqnarray*}
\sqrt{n}  \mathscr{S}(k)  & \Rightarrow &   \delta_{1}( Z^* ) - \delta_{T-k}( Z^* ), \\
\mathscr{S}^*(k) & \Rightarrow & \underset{j=1,...,k^*-k}{\max}~ \frac{\delta_{j}( Z^*) - \delta_{j+1}( Z^*)}{\delta_{j+1}( Z^*) - \delta_{j+2}( Z^*)},
\end{eqnarray*} 
where $Z^* = Q'ZQ$. Under the alternative hypothesis $H_1(k)$ of more than $k$ factors, $\sqrt{n}  \mathscr{S}(k)$ and $\mathscr{S}^*(k)$ diverge in probability to infinity at order $O_p(\sqrt{n})$.
\end{proposition}

Suppose the error terms are independent across $i$ and $t$, and stationary across $t$, i.e., a setting implying Assumption \ref{ass:var:returns}. Then, by the standard CLT, the symmetric random matrix $Z = (z_{ij})$ in Assumption \ref{ass:CLT:epseps} is such that its elements on and above the main diagonal are mutually independent with $z_{ii} \sim N(0,\eta)$ and $z_{ij} \sim N(0,q)$ for $i\neq j$, where $\eta = \lim \frac{1}{n}\sum_{i=1}^n \eta_i$ and $q = \lim \frac{1}{n} \sum_{i=1}^n \sigma_i^4$, for $\sigma_i^2 = V[\varepsilon_{i,t}]$ and $\eta_i = V[\varepsilon_{i,t}^2]$. Further, if the error terms are normal, we have $\eta = 2 q$, and the random matrix $Z/\sqrt{q}$ is in the Gaussian Orthogonal Ensemble GOE($T$) for dimension $T \times T$, see e.g. Tao (2012). \footnote{The $T\times T$ symmetric random matrix $Z=(z_{i,j})$ is in the GOE(T) if $z_{ii}\sim N(0,2)$, and $z_{i,j}\sim N(0,1)$,  \\ for $i\neq j$, and the elements on and above the diagonal are mutually independent.}  Moreover, because $Q'\varepsilon_i \sim N(0,\sigma_i^2 I_{T-k})$, the matrix $Z^{*} / \sqrt{q}$ is in GOE($T-k$). It means that, under Gaussian innovations,  the limiting distributions for large $n$ and fixed $T$ do not depend on the matrix $Q$  underlying $Z^{*}  = Q' ZQ$, and thus are independent of the specific realized path of the factor in the time window of size $T$. Our Monte Carlo results under a Gaussian error design in Section 6 corroborate that theoretical statement.

Onatski (2009) considers the large $T$ (and large $n$) setting and establishes the asymptotic distribution of the eigenvalues of the sample second-moment matrix using random matrix theory. To make a bridge between Proposition \ref{prop:asy:SSstar} and the results in Onatski (2009), we see  that $V[Q'\varepsilon_i]=\sigma_i^2 I_{T-k}$ and, by the CLT, any finite-dimensional block of $ Q'\varepsilon_i =  \sum_t q_t \varepsilon_{i,t}$ tends to a standard Gaussian distribution as $T\rightarrow \infty$, where $q_t'$ is the $t$-th row vector of matrix $Q$. This suggests that, for large $T$, the asymptotic distributions in Proposition \ref{prop:asy:SSstar} are as if the error terms were normal, and $\frac{1}{\sqrt{q}}Z^{*}$ is asymptotically in the GOE($T-k$). The distribution of the largest eigenvalues of a matrix in the GOE is $TW$, i.e., the Tracy-Widom law, when the matrix dimension is large (see e.g. Johnstone (2001) concerning the first eigenvalue). This parallels the analysis developed in Onatski (2009) and suggests that, with $n,T \rightarrow \infty$ and $k^*$ fixed,  $\mathscr{S}^*(k) \Rightarrow  \underset{j=1,...,k^*-k}{\max} \frac{\mu_{j} - \mu_{j+1}}{\mu_{j+1} - \mu_{j+2}}$, where the $\mu_j$ follow a joint $TW$ distribution. \footnote{Here, our goal is to provide a heuristic argument to show the link between the results with fixed $T$ and those with $T\rightarrow \infty$, and not to give another formal proof of the results in Onatski (2009). For instance, we deliberately overlook the difference between the double asymptotics with $n,T \rightarrow\infty$ jointly and a sequential asymptotics with first $n\rightarrow\infty$ and then $T\rightarrow\infty$.} In fact, Onatski (2009) finds a Tracy-Widom distribution of Type 2 for his statistic because it involves the eigenvalues of a Wishart matrix based on complex-valued variates. 

\begin{figure}[!ht]
	\footnotesize
	\begin{center}
		\caption{Level curves for the joint pdf of eigenvalue spacings $s_1=\delta_1(Z^*)-\delta_2(Z^*)$ and $s_2=\delta_2(Z^*)- \delta_3(Z^*)$ for random matrix $Z^*$ in GOE($3$).}
		\label{figure:GOE3:jointpdf}
			\includegraphics[width=\textwidth]{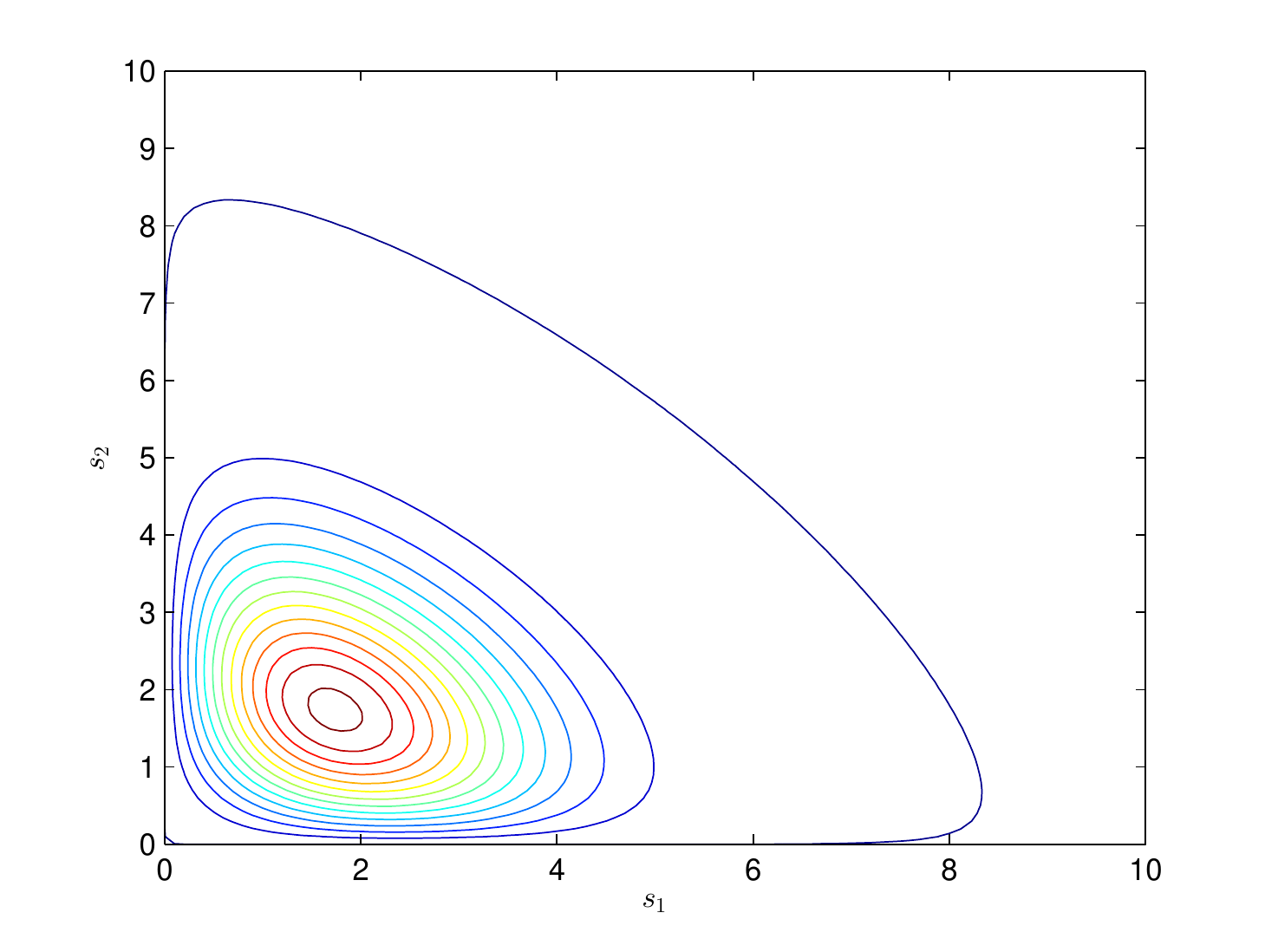}
		
		\end{center}
\end{figure}

\begin{figure}[!ht]
	\footnotesize
	\begin{center}
		\caption{The pdf $f_3(s)$ of eigenvalue spacing $s=\delta_1(Z^*)-\delta_3(Z^*)$ and the pdf $g_3(r)$ of eigenvalue spacings' ratio $r=[\delta_1(Z^*)-\delta_2(Z^*)]/[\delta_2(Z^*)-\delta_3(Z^*)]$ for random matrix $Z^*$ in GOE($3$).}
		\label{figure:GOE3:fg}
			\includegraphics[width=\textwidth]{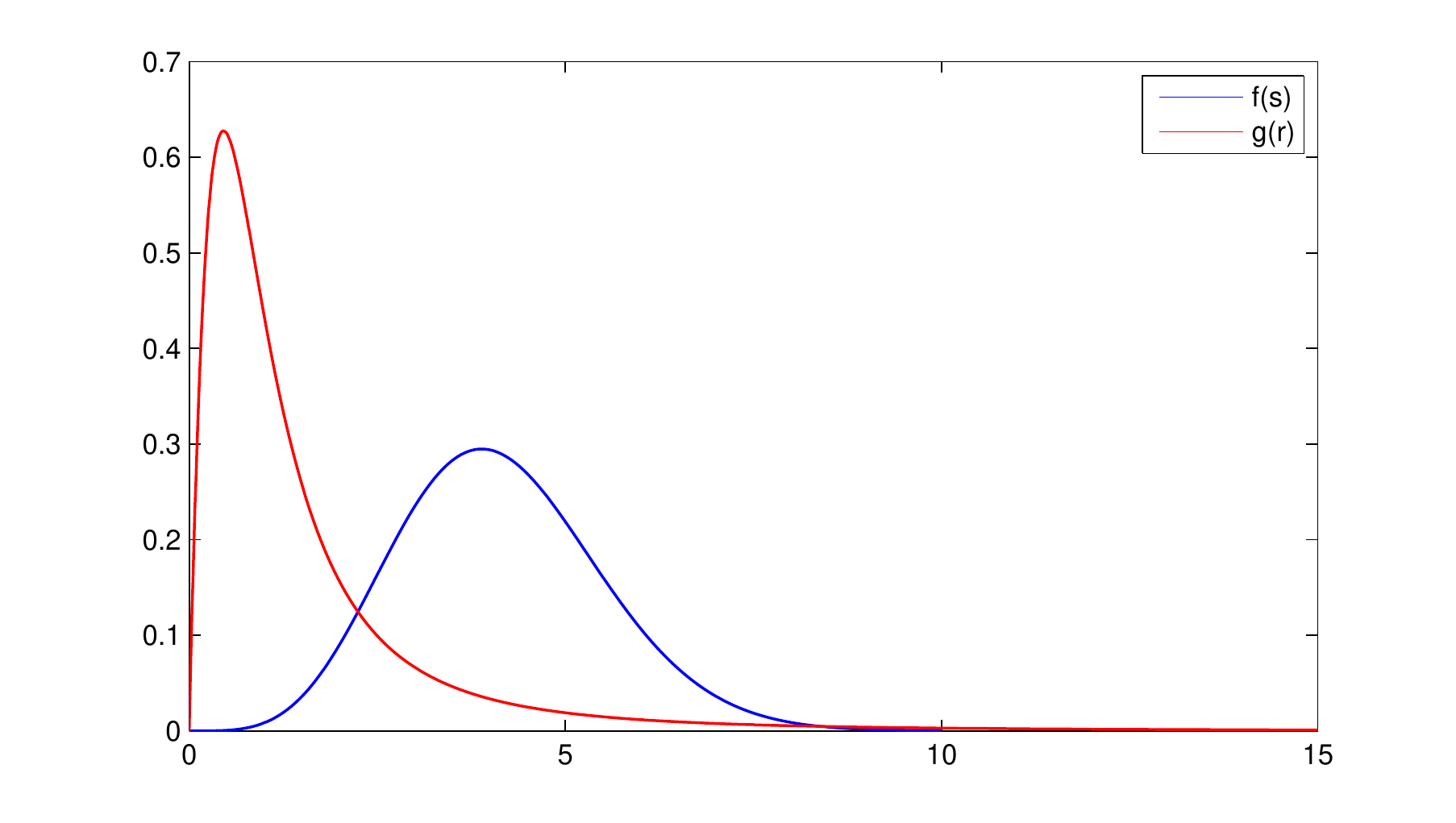}
		
		\end{center}
\end{figure}

Under fixed $T$, the asymptotic distributions in Proposition \ref{prop:asy:SSstar} do not have known analytical characterizations in terms of closed-form expressions for pdf or cdf, except in some cases - mainly in the setting with Gaussian errors (see next subsection for a discussion on how to implement feasible statistics in a general setting). For a matrix $Z^*$ in the GOE($2$), the eigenvalue difference $s=\delta_1( Z^*)- \delta_2(Z^*)$ follows the Wigner-Dyson distribution with pdf $f_2(s) = (s/4) e^{-s^2/8}$, for $s\geq0$, that is the distribution of twice the square root of a $\chi^2(2)$ variable. \footnote{The Wigner-Dyson distribution is sometimes defined with a different normalization, e.g. to have mean equal to $1$ (e.g. Rao (2020)), and is often referred to as ``Wigner surmize".} Hence, it provides the asymptotic distribution of $\sqrt{\frac{n}{q}} \mathscr{S}(k)$ with $T-k=2$ and Gaussian errors $\varepsilon_{i,t} \sim N(0,\sigma_i^2)$ independent across $i$ and $t$. In Section 5.2, we reconsider this result and expand it to local alternatives and non-Gaussian errors. For $Z^*$ in GOE($3$), using the joint distribution of eigenvalues (Ginibre formula, see e.g. Tao (2012)) and building on Rao (2020), in the companion paper Fortin, Gagliardini and Scaillet (2022b) we show that the joint distribution of the eigenvalue spacings $s_1 =\delta_1( Z^*)- \delta_2(Z^*)$ and $s_2 =\delta_2( Z^*)- \delta_3(Z^*)$ is:
\begin{equation*}
\ell(s_1,s_2 ) = \frac{1}{4\sqrt{6\pi}} \exp\left\{ - \frac{1}{6} ( s_1^2 + s_2^2 + s_1 s_2)\right\} s_1 s_2 (s_1 + s_2) 1_{s_1\geq 0, s_2 \geq 0}.
\end{equation*}
The level curves of this distribution are displayed in Figure \ref{figure:GOE3:jointpdf}. The distribution is symmetric and shows negative association between $s_1$ and $s_2$ for large values. 
By marginalization, the distribution of  $s=\delta_1( Z^*)- \delta_3(Z^*)=s_1+s_2$ has pdf $$f_3(s) = \frac{s}{4} e^{-\frac{1}{8}s^2} \left[ \left( 2 \Phi( \frac{s}{2\sqrt{3}}) - 1\right) \left( \frac{s^2}{4} - 3\right) + \frac{3s}{\sqrt{6\pi}} e^{-\frac{1}{24}s^2} \right],$$ for $s\geq0$. Furthermore, the eigenvalue spacings'  ratio $r = \frac{\delta_1(Z^*)-\delta_2(Z^*)}{\delta_2(Z^*) -\delta_3(Z^*)} = s_1/s_2$  for $ Z^*$ in GOE($3$) follows a distribution with pdf $g_3(r) =\frac{27}{8}\frac{r+r^2}{(1+r+r^2)^{5/2}}$, $r\geq 0$ (see Atas et al. (2013) for the result up to the normalizing constant). The pdfs $f_3(s)$ and $g_3(r)$ are displayed in Figure \ref{figure:GOE3:fg}. The pdf of the eigenvalue spacings ratio is skewed and  features Pareto tail. These results yield the asymptotic distributions of $\sqrt{\frac{n}{q}} \mathscr{S}(k)$ and $\mathscr{S}^*(k)$ for $T-k=3$ and Gaussian errors and allow to get easily the critical values.  

\subsubsection{Feasible statistics}

To get feasible statistics in  a general setting, we have to determine the critical value from simulations of the asymptotic distributions in Proposition \ref{prop:asy:SSstar}, which needs draws of $Z^* = Q'ZQ$. 

\medskip

\noindent \textit{i) Independent errors}

\medskip

To start with, let us consider the setting where the errors are independent across $i$ and $t$, with generic distribution admitting finite fourth-order moment. First, we need to simulate matrix  $Z$ from i.i.d.\ draws of  $z_{ii} \sim N(0,\eta)$ and $z_{ij} \sim N(0,q)$ for $i\neq j$. Let us provide consistent estimators of $q= \lim ~\frac{1}{n}\sum_{i=1}^n \sigma_i^4$ and $\eta = \lim~\frac{1}{n}\sum_{i=1}^n \eta_i$, where $\sigma_i^2 = V[\varepsilon_{i,t}]$ and $\eta_i = V[\varepsilon_{i,t}^2]$. Let $\hat{\varepsilon}_i = M_{\hat{F}} \varepsilon_i$ be the vector of residuals, and $\tilde{\varepsilon}_i = M_{{F}}\varepsilon_i$. Define the estimators:
\begin{eqnarray*}
\hat{m}_1 = \frac{1}{n} \sum_{i=1}^n \left( \sum_{t=1}^T  \hat{\varepsilon}_{i,t}^2 \right)^2 , \qquad
\hat{m}_2 = \frac{1}{n} \sum_{i=1}^n  \sum_{t=1}^T   \hat{\varepsilon}_{i,t}^4.
\end{eqnarray*}
To compute their probability limits as $n\rightarrow \infty$, we use that, for $\varepsilon_{i,t}\sim (0,\sigma_i^2)$ independent across $i$ and $t$,  we have $\Omega_{\varepsilon} := \lim~ \frac{1}{n} \sum_{i=1}^n E[(\varepsilon_i \varepsilon_i') \otimes (\varepsilon_i \varepsilon_i')] = \eta A_1 + q A_2$ where $A_1 = diag[\mathcal{K}_T]$ and $A_2= I_{T^2} + \mathcal{K}_T - 2 diag [\mathcal{K}_T] + vec[I_T] vec[I_T]' $, and $\mathcal{K}_T$ is the commutation matrix for order $T$ (see Magnus and Neudecker (2007)).
From the consistency of $\hat{F}$ and using $\sum_{t=1}^T \tilde{\varepsilon}_{i,t}^2 = \varepsilon_i' M_F \varepsilon_i = vec[M_F]'( \varepsilon_i \otimes \varepsilon_i)$,  we get:
\begin{eqnarray}
\underset{n\rightarrow\infty}{plim} ~ \hat{m}_1 &=& \underset{n\rightarrow\infty}{plim} ~ \frac{1}{n} \sum_{i=1}^n \left( \sum_{t=1}^T \tilde{\varepsilon}_{i,t}^2 \right)^2 = 
\underset{n\rightarrow\infty}{lim}~ \frac{1}{n} \sum_{i=1}^n E \left[\left(  \sum_{t=1}^T \tilde{\varepsilon}_{i,t}^2 \right)^2 \right] \notag  \\
&=& vec[M_F]'\Omega_{\varepsilon} vec[M_F]
= \eta a + q b, \label{eq1}
\end{eqnarray} 
where  $a = vec[M_F]'diag[\mathcal{K}_T] vec[M_F] = \sum_{t=1}^T [(M_{F})_{tt}]^2$ and $b=vec[M_F]'A_2 vec[M_F] 
= 2( T- k - a) + (T-k)^2$. Moreover, by using $\tilde{\varepsilon}_{i,t}^2 = (e_t \otimes e_t)'(M_F \otimes M_F) ( \varepsilon_i \otimes \varepsilon_i)$, where $e_t$ is the $t$-th unit vector in $\mathbb{R}^T$, and $\sum_{t=1}^T \tilde{\varepsilon}_{i,t}^4 = Tr[ (M_F \otimes M_F)  [(\varepsilon_i \varepsilon_i') \otimes (\varepsilon_i \varepsilon_i')] (M_F \otimes M_F) diag [ \mathcal{K}_T)]$, we have:
\begin{eqnarray}
\underset{n\rightarrow\infty}{plim}~\hat{m}_2 &=& \underset{n\rightarrow\infty}{plim} ~ \frac{1}{n} \sum_{i=1}^n \sum_{t=1}^T \tilde{\varepsilon}_{i,t}^4 
= \underset{n\rightarrow\infty}{lim} ~ \frac{1}{n} \sum_{i=1}^n E\left[ \sum_{t=1}^T \tilde{\varepsilon}_{i,t}^4 \right]  \notag \\
&=& Tr[ (M_F \otimes M_F)  \Omega_{\varepsilon} (M_F \otimes M_F) diag [ \mathcal{K}_T]] =
\eta c  + q d, \label{eq2}
\end{eqnarray} 
where $c = Tr[ (M_F \otimes M_F) diag [ \mathcal{K}_T]  (M_F \otimes M_F) diag [ \mathcal{K}_T]] = \sum_{t=1}^T \sum_{s=1}^T [(M_F)_{t,s}]^4$ and
$d = Tr[ (M_F \otimes M_F) A_2  (M_F \otimes M_F) diag [\mathcal{K}_T]] = 3a - 2c$. If $3 a^2 \neq c (T-k)(T-k+2)$,  we have that the determinant $ad-bc \neq 0$ in the linear system defined by (\ref{eq1}) and  (\ref{eq2}), and the linear mapping from $\eta,q$ and the plims of $\hat{m_1}, \hat{m}_2$ is one-to-one. Moreover, the coefficients $a,b,c,d$ can be consistently estimated by replacing $F$ with $\hat{F}$. Hence, by solving the two linear equations (\ref{eq1}) and  (\ref{eq2}) with estimated coefficients $\hat{a}, \hat{b}, \hat{c}, \hat{d}$ and unknowns $\eta$ and $q$, we get consistent estimators $\hat{\eta}$ and $\hat{q}$ from $\hat{m}_1$ and $\hat{m}_2$. 

Second, once we have i.i.d.\ Gaussian draws for $Z$ based on the estimates $\hat{\eta}$ and $\hat{q}$, we need to estimate $Q$ to build the draws for  $Z^* = Q'ZQ$. We can consistently estimate the orthogonal complement of the range of $F$ by $M_{\hat F}$. Since the eigenvalues of $Z^*$ are invariant by rotation of the columns of $Q$,  we can pick any $T-k$ columns of $M_{\hat F}$ and orthogonalize them to build $\hat{Q}$. In practice, we take the first $T-k$ ones. 

\medskip

\noindent \textit{ii) General case: parametric variance structure}

\medskip

In the general case, the errors are idiosyncratic martingale difference sequences but may feature some form of time dependence. Suppose that we have a parametric specification $V[vec[Z]] = \Omega(\theta)$ for the variance of the Gaussian matrix $Z$ in Assumption \ref{ass:CLT:epseps}, with an unknown vector parameter $\theta \in \mathbb{R}^p$ (an element of which is $q$). We derive this parametric specification for a  model with ARCH(1) errors that we consider in our Monte Carlo experiments in Section 6. Now, we use $V[vec[Z^*]] = (Q'\otimes Q') V[vec[Z]] (Q \otimes Q)$, and $V[vec[Z]] = \underset{n\rightarrow\infty}{lim} \frac{1}{n} \sum_{i=1}^n V[\varepsilon_i \otimes \varepsilon_i] = \underset{n\rightarrow\infty}{plim} \frac{1}{n} \sum_{i=1}^n (\varepsilon_i \varepsilon_i') \otimes (\varepsilon_i \varepsilon_i')  - q \cdot vec[I_T] vec[I_T]' $. Then, we can obtain a consistent estimator of parameter $\theta$ with fixed $T$ by minimum distance:
\begin{equation} \label{est:theta}
\hat{\theta} = \underset{\theta}{\arg \min} \left\Vert 
 (\hat{Q}\otimes \hat{Q})' \left( \frac{1}{n}\sum_{i=1}^n  (\hat{\varepsilon}_i \hat{\varepsilon}_i') \otimes (\hat{\varepsilon}_i \hat{\varepsilon}_i')  - {q} \cdot vec[I_{T}] vec[I_{T}]'  - \Omega(\theta) \right) (\hat{Q}\otimes \hat{Q}) \right\Vert
\end{equation}
where $\hat{Q}$ is a consistent estimators of $Q$ as in the previous subsection, and $\Vert \cdot \Vert$ denotes the Frobenius matrix norm. While the residual $\hat{\varepsilon}_{i}$ approximates $M_F \varepsilon_{i}$ and not $\varepsilon_{i}$ with fixed $T$, this fact does not affect the consistency of $\hat{\theta}$ because $Q'M_F = Q'$. Since $vec[Z^*]$ has $\frac{1}{2}(T-k)(T-k+1)$ different elements, the necessary order condition for identification is $p \leq \frac{1}{2}(T-k)(T-k+1)[\frac{1}{2}(T-k)(T-k+1)+1]/2$.

\medskip

\noindent \textit{iii) General case: nonparametric variance estimator}

\medskip 

When a parametric specification $\Omega(\theta)$ for the variance structure is not available, we can construct a feasible test statistic by using a nonparametric estimator for the variance-covariance matrix of the Gaussian matrix appearing in the limiting distribution. Indeed, the asymptotic distribution of statistic $\sqrt{n} \mathscr{S}(k)$ is the distributional limit of $\delta_1( Q'Z_n Q) - \delta_{T-k}(Q'Z_n Q)$, where $Z_n = \frac{1}{\sqrt{n}} \sum_{i=1}^n[ \varepsilon_i \varepsilon_i' - \sigma_i^2 I_T]$.  Now, because
\begin{eqnarray*}
Q'Z_n Q &=& \frac{1}{\sqrt{n}} \sum_{i=1}^n[ Q'\varepsilon_i \varepsilon_i'Q - \sigma_i^2 I_{T-k}] \\
&=& \frac{1}{\sqrt{n}} \sum_{i=1}^n[ Q'\varepsilon_i \varepsilon_i'Q - \frac{\varepsilon_i'M_F \varepsilon_i}{T-k} I_{T-k}] + \left( \frac{1}{\sqrt{n}} \sum_{i=1}^n[\frac{\varepsilon_i'M_F \varepsilon_i}{T-k} - \sigma_i^2]  \right) I_{T-k},
\end{eqnarray*}
and adding to a matrix a multiple, here $\left( \frac{1}{\sqrt{n}} \sum_{i=1}^n[\frac{\varepsilon_i'M_F \varepsilon_i}{T-k} - \sigma_i^2]  \right)$, of the identity matrix changes all eigenvalues by the same amount, we deduce that $\sqrt{n} \mathscr{S}(k) \Rightarrow \delta_1 (\bar{Z}^*) - \delta_{T-k}(\bar{Z}^*)$, where $\bar{Z}^*$ is the distributional limit of $\frac{1}{\sqrt{n}} \sum_{i=1}^n[ Q'\varepsilon_i \varepsilon_i'Q - \frac{\varepsilon_i'M_F \varepsilon_i}{T-k} I_{T-k}]$, i.e., $\bar{Z}^* = Z^* - \frac{1}{T-k} Tr[Z^*] I_{T-k}$. Besides, under the sphericality assumption of error terms, i.e., $E[ \varepsilon_i \varepsilon_i']= \sigma_i^2 I_T$, the matrix variables $Q'\varepsilon_i \varepsilon_i'Q - \frac{\varepsilon_i'M_F \varepsilon_i}{T-k} I_{T-k}$ have zero mean. Then, we can consistently estimate  the variance matrix $\bar{\Omega} := V[ vec[ \bar{Z}^*]]$  by
\begin{equation*}
\hat{\bar{\Omega}} = \frac{1}{n}\sum_{i=1}^n \left( (\hat{Q}'\hat{\varepsilon}_i) \otimes (\hat{Q}'\hat{\varepsilon}_i) -\frac{\hat{\varepsilon}_i'\hat{\varepsilon}_i}{T-k} vec[I_{T-k}] \right)
\left( (\hat{Q}'\hat{\varepsilon}_i) \otimes (\hat{Q}'\hat{\varepsilon}_i) - \frac{\hat{\varepsilon}_i'\hat{\varepsilon}_i}{T-k} vec[I_{T-k}] \right)',
\end{equation*}
as $n\rightarrow\infty$ and $T$ is fixed. Again, the equality $M_{F}Q = Q$ results in the ``inconsistency" of the residuals for fixed $T$ having no effect on the consistency of estimator $\hat{\bar{\Omega}} $. 

A direct approach based on the nonparametric estimation of the variance of $vec[Z^*]$ would be more difficult because of the unobserved $\sigma_i^2$. We avoid this difficulty by recognizing that replacing $\sigma_i^2$ with the unbiased (infeasible) estimate $\frac{\varepsilon_i' M_F \varepsilon_i}{T-k}$ does not affect the eigenvalues spacing underlying our tests. This strategy paves the way to the nonparametric variance estimator $\hat{\bar{\Omega}}$ relying on the inconsistent $\hat{\varepsilon}_i$, that we have presented in this section.

\section{Weak factors}

In this section, we extend our analysis to cover inference with weak factors and similar deviations from the dichotomy of strong factor vs no factor considered so far. From the view point of testing for the number of factors, we can see a weak factor as a local alternative hypothesis. We focus on the setting with identification via the variance-covariance matrix of the returns. We normalize the latent factors with $\tilde{\Sigma}_f = I_k$ and $\tilde{\Sigma}_{\beta} = diag ( \tilde{\sigma}_{\beta,j}^2)$ diagonal, with diagonal elements ranked in decreasing order. We assume that, as $n\rightarrow\infty$, we have:
\begin{equation} \label{c:j}
n^{\kappa_j} \tilde{\sigma}_{\beta,j}^2 \rightarrow c_j > 0,
\end{equation} 
for an exponent $\kappa_j \geq 0$ and any $j$ (the sequence of the $\kappa_j$ is non-decreasing by construction). This setting accommodates various forms of factors: a strong factor  with $\kappa_j = 0$,  a semi-strong factor with $\kappa_j \in (0,1/2)$, a weak factor with $\kappa_j = 1/2$  (Kleibergen (2009)), and  a vanishing factor with $\kappa_j > 1/2$ (Onatski (2012, 2015)). The latter ones include factors with zero loadings across all assets ($\kappa_j = + \infty$), the so-called useless or irrelevant factors; see Kan and Zhang (1999a,b) and Gospodinov, Kan and Robotti (2014). \footnote{We work here with the factor rotation such that $\tilde{\Sigma}_{\beta}$ is diagonal. However, when applied to the eigenvalues of $\tilde{\Sigma}_{\beta}$, the exponents $\kappa_j$ are invariant to the chosen factor rotation. In fact, the eigenvalues of $R'\tilde{\Sigma}_{\beta}R$ and $\tilde{\Sigma}_{\beta}$ coincide, when $R$ is an orthogonal matrix.} When $\kappa_j >0$ the eigenvalue $\tilde{\sigma}_{\beta,j}^2$ shrinks to zero at rate $O(n^{-\kappa_j})$ in the drifting DGP. This can originate e.g. from the fact that a given factor loads exclusively on firms in a sector with negligible weight compared to the entire economy, or that the loadings of a factor are very small across all stocks. For a weak factor, the magnitude of the eigenvalue is on the same scale as the estimation error, namely $O(n^{-1/2})$.

Below we use the next result, that has been proved in Carlini and Gagliardini (2022) and provides a perturbation expansion for the non-zero eigenvalues of a symmetric matrix and the associated eigenvectors. \footnote{See Proposition 6 in Carlini and Gagliardini (2022). The proof of that proposition yields the statement as in our Proposition \ref{prop:perturbation:1:k}.} It extends the results in, e.g., Izenman (1975) by providing a more accurate control of the remainder terms. 

\begin{theorem} \label{prop:perturbation:1:k}
Let $A = U D U'$ be a symmetric $K \times K$ matrix of rank $k$, where $D$ is the diagonal matrix of the $k$ distinct non-zero eigenvalues $\lambda_j = \delta_j(A)$, and $U=[U_1: \cdots : U_k]$ is the $K \times k$ matrix of the associated orthonormal eigenvectors. Let $\hat{A} = A + \hat{\Psi}$, where symmetric matrix $\hat{\Psi}$ is a ``small perturbation'', with normalized eigenvectors $\hat{U}_j$ and eigenvalues $\delta_j(\hat{A})$. Then:
\begin{eqnarray*}
 \delta_{j} (\hat{A}) = \delta_j(A) + U_j' \hat{\Psi} U_j + O \left( \rho_j(A) \Vert \hat{\Psi} \Vert^2 \right),
\end{eqnarray*}
and
\begin{equation*}
\hat{U}_j  = U_j + \sum_{\ell=0, \ell \neq j}^k \frac{1}{\lambda_j - \lambda_{\ell}} P_{\ell} \hat{\Psi} U_j + O \left( \rho_j(A)^2 \Vert \hat{\Psi} \Vert^2 \right),
\end{equation*}
for $j=1,...,k$, where $P_{j} = U_{j} U_{j}'$, $P_0 = I_K- U U' = Q Q'$, the orthonormal columns of the $K\times (K-k)$ matrix $Q$ span the null space of $A$, scalar $\lambda_0=0$ is the  null eigenvalue,
\begin{equation*}
\rho_j(A)=   \sum_{\ell=0, \ell \neq j}^k \vert \lambda_j - \lambda_{\ell} \vert^{-1}, \qquad j=1,...,k,
\end{equation*}
and the remainder terms are uniform w.r.t.\ $A$ and $\hat{\Psi}$.
\end{theorem}

The remainder terms in the expansions of the $j$-th eigenvalues and eigenvectors involve the matrix $A$ solely by means of $\rho_j(A)$, i.e., a measure of the (inverse) proximity of eigenvalues. More precisely, $\rho_j(A)$ is large if there are other eigenvalues of $A$ close to $\delta_j(A)$. Thus, Theorem \ref{prop:perturbation:1:k} covers the case where matrix $A$ has nearly overlapping eigenvalues, or some eigenvalues are nearly null. 

\subsection{Testing for (semi-)strong factors vs vanishing factors}

Let us suppose there are $k$ strong or semi-strong factors, and $T-k$ vanishing factors. Namely, we have $\kappa_j < 1/2$ and $c_j>0$ for $j=1,...,k$, and $\kappa_j > 1/2$ and $c_j \geq 0$ for $j=k+1,...,T$. We want to conduct inference on the number of (semi-)strong factors by testing the null $H_0(k)$ of $k$ (semi-)strong factors against the alternative $H_1(k)$ of more than $k$ (semi-)strong factors, for a  given integer $k$. This setting extends the analysis of Section 3 to accommodate intermediate forms of factors, namely semi-strong factors in the systematic component, and factors with asymptotically vanishing loadings in the idiosyncratic component. To simplify, we assume that the eigenvalues up to rank $k$ remain distinct asymptotically \footnote{i.e, we have either $\kappa_{j+1}> \kappa_j$ or $c_{j+1} < c_{j}$ (or both) for $j=1,...,k-1$.}. Then, we have  $\rho_j(\tilde{\Sigma}_{\beta}) = O( n^{\kappa_j})$.  From (\ref{hatVy}) and Theorem \ref{prop:perturbation:1:k}, we get:
\begin{eqnarray*}
\delta_j(\hat{V}_y) &=& \tilde{\sigma}^2 +   T \tilde{\sigma}_{\beta,j}^2 + \frac{1}{T} F_j' \hat{\Phi} F_j +  O_p( n^{\kappa_j-1})  \\
&=&  \tilde{\sigma}^2 +   T \tilde{\sigma}_{\beta,j}^2 + \frac{1}{\sqrt{n}} \left( \sqrt{T} U' W_n + \sqrt{T} W_n' U + U' Z_n U \right)_{jj} +  O_p( n^{\kappa_j-1} ),
\end{eqnarray*}
for $j=1,...,k$, where 
$
Z_n = \frac{1}{\sqrt{n}} \sum_{i=1}^n ( \varepsilon_i \varepsilon_i' - {\sigma}^2_j I_T),$ $W_n = 
\frac{1}{\sqrt{n}} \sum_{i=1}^n \varepsilon_i \beta_i'.$
Moreover, using Theorem \ref{prop:expansion:eigenvalues} with $\hat{\Psi} = \hat{\Phi} + \sum_{j=k+1}^T \tilde{\sigma}_{\beta,j}^2 F_j F_j'$ for $\hat{\Phi}$ as in (\ref{exp:Phihat}), and  $\Vert D^{-1} \Vert = O(n^{-\kappa_k})$ we get:
\begin{equation*}
\delta_{k+j}(\hat{V}_y) = \tilde{\sigma}^2 + \frac{1}{\sqrt{n}} \delta_j ( Q' Z_n Q ) +  O_p( n^{\kappa_k -1} + n^{-\kappa_{k+1}}),
\end{equation*}
for $j=1,...,T-k$. Hence, $\sqrt{n} [  \delta_j(\hat{V}_y) - \delta_{j+1}(\hat{V}_y)]$ diverges to $+ \infty$ in probability as $n\rightarrow \infty$, for $j=1,...,k$, while we have $\sqrt{n} [  \delta_{k+j}(\hat{V}_y) - \delta_{k+j+1}(\hat{V}_y)]~\Rightarrow \delta_j ( Q' Z Q ) - \delta_{j+1} ( Q' Z Q )$ for $j=1,...,T-k-1$. 
\begin{proposition}
Under the null hypothesis $H_0(k)$ of $k$ (semi-)strong factors, the asymptotic distributions of test statistics $\sqrt{n}\mathscr{S}(k)$ and $\mathcal{S}^*(k)$ for $n\rightarrow\infty$ and fixed $T$ are as in Proposition \ref{prop:asy:SSstar}. The statistics diverge in probability to infinity under the alternative hypothesis $H_1(k)$ of more than $k$ (semi-)strong factors.
\end{proposition}
Thus, the test statistics $\sqrt{n}\mathscr{S}(k)$ and $\mathcal{S}^*(k)$ are valid to conduct inference on the number of latent factors also when these factors are only semi-strong.  The divergence rate under the alternative is slower compared to the case with strong factors only. For instance, if the $(k+1)$-th factor is semi-strong, i.e., $\kappa_{k+1}<1/2$, statistics $\sqrt{n}\mathscr{S}(k)$ and $\mathscr{S}^*(k)$ diverge at rate $O_p( n^{1/2 - \kappa_{k+1}})$. 

Let us consider now the estimates of factor values. From (\ref{hatVy}) and Theorem \ref{prop:perturbation:1:k}, we get:
\begin{eqnarray} \label{asy:Uhatj}
\hat{U}_j  &=& U_j + \frac{1}{\sqrt{n}} \sum_{\ell=0, \ell \neq j}^k \frac{1}{\tilde{\sigma}_{\beta,j}^2 - \tilde{\sigma}_{\beta,\ell}^2} P_{\ell} \left( \sqrt{T} W_n U' + \sqrt{T}U W_n'  + Z_n \right) U_j + O_p( n^{2\kappa_j-1}) , \nonumber \\
\end{eqnarray}
for $j=1,...,k$, with $\tilde{\sigma}_{\beta,0}^2 = 0$, $P_0 = Q Q'$ and $P_{\ell} = U_{\ell} U_{\ell}'$. Thus, the estimated factor values are consistent for $n\rightarrow\infty$ and fixed $T$, but with a slower convergence rate for semi-strong factors: $\hat{U}_j = U_j + O_p ( n^{-1/2 + \kappa_{j}})$. \footnote{Here, we have that  $U_j = \frac{1}{\sqrt{T}} F_j$,  and $F_j$ contains a sample-dependent rotation due to the \\ normalization in the sample.} Due to consistency of factor estimates, we can follow the procedures detailed in Section 4.3.1 to define feasible statistics based on simulating the asymptotic laws. The asymptotic expansion (\ref{asy:Uhatj}) also paves the way to establish the asymptotic normality of the factor estimates.

\subsection{Power analysis under local alternative hypotheses}

 In this section, we consider a local power analysis. We test the null hypothesis $H_0(k)$ of $k$ (semi-)strong factors against the local alternative hypothesis $H_{1,loc}(k)$ in which the $(k+1)$-th factor is weak, namely $\kappa_j < 1/2$ for $j\leq k$, $\kappa_{k+1}=1/2$ with $\sqrt{n} \tilde{\sigma}_{\beta,k+1}^2 \rightarrow c_{k+1}>0$, and $\kappa_j > 1/2$ for $j>k+1$. 

We consider the test statistics $\sqrt{n}\mathscr{S}(k)$ and $\mathscr{S}^*(k)$. 
%
From Theorem \ref{prop:expansion:eigenvalues} with \break $\hat{\Psi} = \frac{1}{\sqrt{n}} (T \sqrt{n} \tilde{\sigma}_{\beta,k+1}^2 U_{k+1} U_{k+1}' +  \sqrt{T} W_n U' + \sqrt{T} U W_n' + Z_n )$, we have under $H_{1,loc}(k)$: 
\begin{equation*} 
\sqrt{n} [  \delta_{k+j}(\hat{V}_y) - \delta_{k+j+1}(\hat{V}_y)] ~\Rightarrow~\delta_j \left( T c_{k+1} \xi_{k+1} \xi_{k+1}' + Z^{*} \right) - \delta_{j+1} \left( T c_{k+1} \xi_{k+1} \xi_{k+1}' + Z^{*} \right),
\end{equation*}
for $j=1,...,T-k-1$, where $U_{k+1} = \frac{1}{\sqrt{T}}F_{k+1}$ is the normalized vector of the weak factor values and $\xi_{k+1} = Q'U_{k+1}$. For the choice of matrix $Q$ such that $U_{k+1}$ is its first column, we have $\xi_{k+1} = e_1$, i.e., the first unit vector of dimension $T-k$. Thus, under the local alternative $H_{1,loc}(k)$, we have the asymptotic distributions:
\begin{eqnarray*}
&& \sqrt{n} \mathscr{S}(k)~\Rightarrow~ \delta_1(Z_1^*) - \delta_{T-k}(Z_1^*) =: \zeta_1, \\
&& \mathscr{S}^*(k) ~\Rightarrow~ \underset{j=1,...,k^* - k}{\max}~\frac{\delta_j(Z_1^*)- \delta_{j+1}(Z_1^*)}{\delta_{j+1}(Z_1^*) - \delta_{j+2}(Z_1^*)},
\end{eqnarray*} 
as $n\rightarrow \infty$ and $T$ is fixed, where $Z_1^* = T c_{k+1} e_1 e_1' +Z^*$, i.e. random matrix $Z^*$ gets shifted by deterministic quantity $T c_{k+1}$ in the upper-left entry. 

Let us first consider the setting where errors are Gaussian such that  $\varepsilon_{i,t} \sim N(0,\sigma_i^2)$ mutually independent across $i$ and $t$. Then, $\frac{1}{\sqrt{q}} Z^*$ is a symmetric $(T-k)\times (T-k)$ random matrix in the GOE. The asymptotic distributions under both the null and the local alternative are independent of the factor path. The asymptotic local power curve for statistic $\sqrt{n}\mathscr{S}(k)$ is $\pi(a,T-k) = P[ \zeta_1 > \tau_{\alpha}]$ as a function of $a = \frac{T c_{k+1}}{\sqrt{q}}$ and $T-k$, for asymptotic size $\alpha$, where $\tau_{\alpha}$ is the $(1-\alpha)$-quantile of the asymptotic distribution under the null, i.e., $P[\zeta_0 \leq \tau_{\alpha}] = 1 - \alpha$ for $\zeta_0 = \delta_1(Z^*) - \delta_{T-k}(Z^*)$. The asymptotic power function depends on $\frac{T c_{k+1}}{\sqrt{q}}$ and $T-k$ only, because we can write  $Z_1^* = \sqrt{q} \left( \frac{T c_{k+1}}{\sqrt{q}} e_1 e_1'  + \frac{1}{\sqrt{q}} Z^* \right)$ (the distribution of $\frac{1}{\sqrt{q}} Z^*$ depends on $(T-k)$ only, and the scaling factor $\sqrt{q}$ is immaterial for power). A similar result applies for the asymptotic power of statistic $\mathscr{S}^*(k)$. By dividing $c_{k+1}/\sqrt{q}$ by the square root $\sqrt{n}$ of the cross-sectional sample size $n$, we get asymptotically the ratio between the  average squared loadings on the weak factor $\tilde{\sigma}_{\beta,k+1}^2$ and the square root of the average squared variance of errors - a kind of signal-to-noise ratio for the weak factor. Then, $T c_{k+1}/\sqrt{q}$ has an interpretation analogue to a concentration parameter in weak instrument regressions.

When $T-k=2$, we can easily characterize the asymptotic distribution of $\sqrt{n} \mathscr{S}(k) $ under the local alternative. Indeed, we have $\zeta_1 = \delta_1 (Z_1^*) - \delta_2 (Z_1^*) = \break {\sqrt{q}} \left( \delta_1 ( a e_1 e_1' +  \frac{1}{\sqrt{q}} Z^* ) -  \delta_2 ( a e_1 e_1' +  \frac{1}{\sqrt{q}} Z^* ) \right)$, and by using the formula for the two roots of a second-order polynomial, we get $\delta_1 ( a e_1 e_1' +  \frac{1}{\sqrt{q}} Z^* ) -  \delta_2 ( a e_1 e_1' +  \frac{1}{\sqrt{q}} Z^* ) 
= \sqrt{ ( z_{11}^* - z_{22}^* + a)^2 + 4 ( z_{12}^*)^2}$, 
where the $z_{i,j}^*$ are the elements of the symmetric $2 \times 2$ matrix $Z^*/ \sqrt{q}$ in the GOE. By using that $(z_{11}^* - z_{22}^*)/2$ and $z_{12}^*$ are mutually independent standard Gaussian variables, we deduce that $\frac{1}{4q} \zeta_1^2$ is distributed as non-central chi-square $\chi^2 (2, a^2/4)$ with 2 degrees of freedom and non-centrality parameter $a^2/4 = \frac{1}{4q} (T c_{k+1})^2$. Thus, under $H_{1,loc}(k)$, we have 
\begin{equation} \label{scaledchi}
\frac{n}{4 q} \mathscr{S}(k)^2 ~ \Rightarrow~ \chi^2 (2,\frac{T^2 c_{k+1}^2}{4q}).
\end{equation}
Hence, when $T-k=2$, the statistic obtained by rescaling $\mathscr{S}(k)^2$ in (\ref{scaledchi}) has a centered $\chi^2(2)$ asymptotic distribution under the null. In fact, by change of variable, we deduce that the pdf of the asymptotic distribution of $\sqrt{n} \mathscr{S}(k)$ is $f_2(s)=\frac{s}{4q} e^{-s^2/8q}$, which corresponds to ``Wigner surmise" for eigenvalue spacing in $2 \times 2$ GOE random matrices (see Section 4.3). Under the local alternatives, the asymptotic distribution  in (\ref{scaledchi}) features a non-centrality parameter (in analogy to, e.g., Hansen statistic for overidentification test and other chi-square tests). 

\begin{figure}[!ht]
	\footnotesize
	\begin{center}
		\caption{Asymptotic local power under local alternatives in a Gaussian setting: we take $T-k=3$, and nominal size $\alpha=0.05$. We use  10,000 draws of the symmetric matrix $Z = (z_{ij})$ with  $z_{ii} \sim N(0,2q)$ and $z_{ij} \sim N(0,q)$ for $i\neq j$.}
		\label{figure:local:power}
			\includegraphics[width=\textwidth]{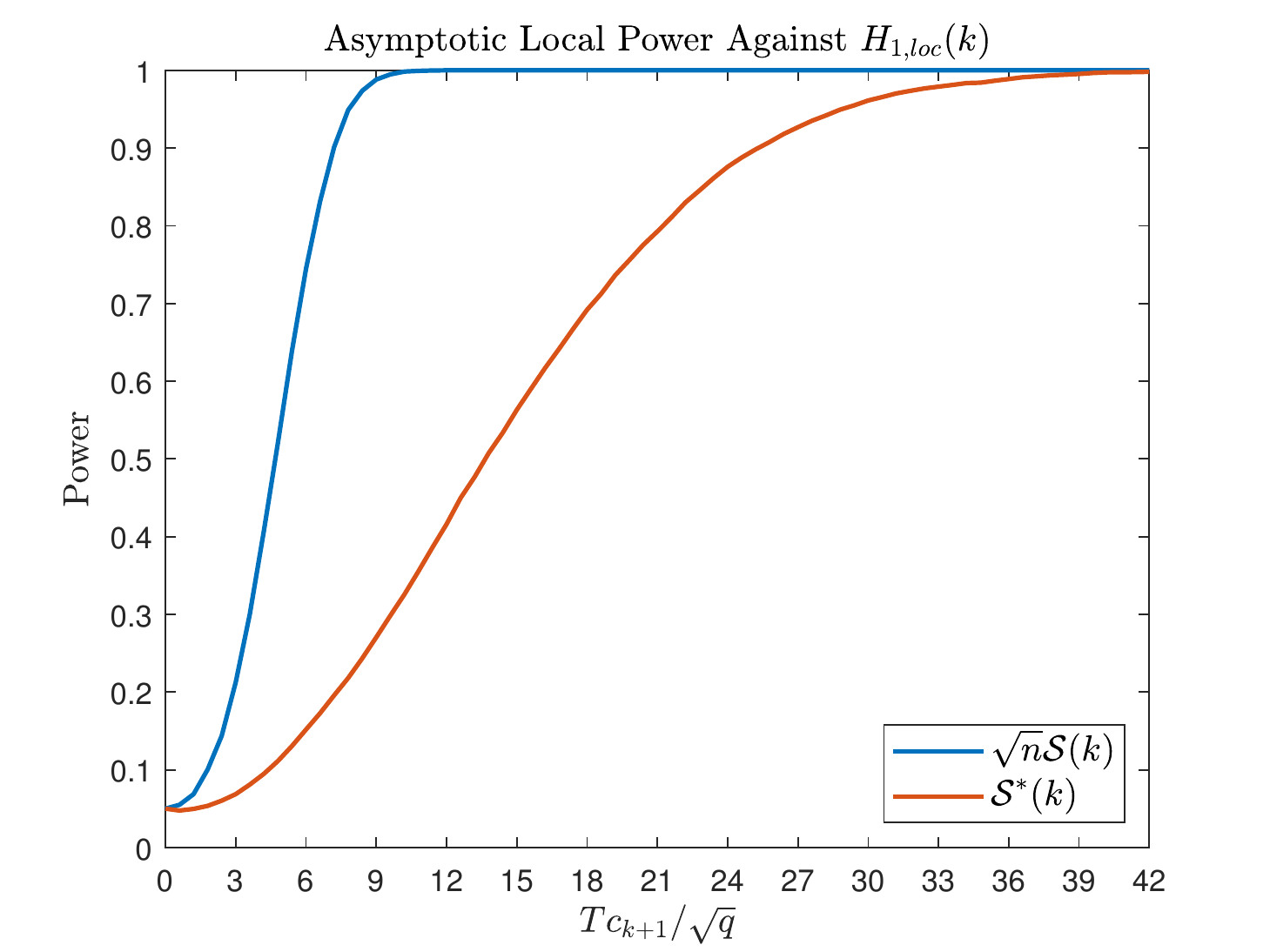}
		
		\end{center}
\end{figure}

For $T-k>2$, we can obtain  the asymptotic power curves by simulating the draws $\frac{1}{\sqrt{q}} Z^{*}$, here 10,000 draws, from the GOE in dimension $(T-k)\times(T-k)$. In Figure \ref{figure:local:power}, we display the asymptotic local power curves for statistics $\sqrt{n}\mathscr{S}(k)$ and $\mathscr{S}^*(k)$ as functions of $Tc_{k+1}/\sqrt{q}$, for $T-k=3$, and asymptotic size $\alpha=0.05$. This setting corresponds to, e.g., $T=6$ periods and $k=3$ (semi-)strong factors under the null. In this case and for a cross-sectional size of e.g. $n=1000$, the range in the horizontal axis of $c_{k+1}$ covers values of the (modified) signal-to-noise $c_{k+1}/\sqrt{q n}$ between $0$ and about $0.20$. The asymptotic local power curve for the statistic $\sqrt{n} \mathscr{S}(k)$  ramps up steeply, and the consistency of the test is achieved quickly even locally. We confirm the good power properties in our Monte Carlo study for finite samples.

In the more general case with non-Gaussian errors, the asymptotic power under local alternatives depends on the factor path via the matrix $Q$. To understand  this effect in a simple setting, consider again the case with $T-k=2$. By expanding the above arguments, it is possible to show that
\begin{equation}  \label{asy:S2:nonGaussian}
\frac{n}{4 q} \mathscr{S}(k)^2 ~ \Rightarrow~ d_1 \chi^2(1,\mu_1) +  d_2 \chi^2 (1,\mu_2),
\end{equation}
where $d_j = 1 + (\eta^*-2) \lambda_j$ and $\mu_1 = \frac{a^2}{4 d_1} \frac{\sigma_{12}^2}{\sigma_{12}^2 + (\lambda_1-\sigma_{11})^2}$ and $\mu_2 = \frac{a^2}{4 d_2} [1-\frac{\sigma_{12}^2}{\sigma_{12}^2 + (\lambda_1-\sigma_{11})^2}]$, with $\eta^* = \eta/q$, and the two non-central chi-square variables are independent. Here,  $\lambda_1 = \frac{1}{2} ( \sigma_{11} + \sigma_{22} + \sqrt{( \sigma_{11} - \sigma_{22})^2 + 4 \sigma_{12}^2 })$ and $\lambda_2 = \lambda_1 - \sqrt{( \sigma_{11} - \sigma_{22})^2 + 4 \sigma_{12}^2 }$ are the eigenvalues of the symmetric matrix $\Omega = \left( \begin{array}{cc} \sigma_{11} & \sigma_{12} \\ \sigma_{12} & \sigma_{22} \end{array} \right)$ where $\sigma_{11} = \frac{1}{4} \sum_t ( Q_{t,1}^2 - Q_{t,2}^2)^2$, $\sigma_{22} = \sum_t Q_{t,1}^2 Q_{t,2}^2$ and $\sigma_{12} = \frac{1}{2} \sum_t ( Q_{t,1}^2 - Q_{t,2}^2) Q_{t,1} Q_{t,2}$, and the $Q_{t,k}$ are the elements of the $T \times 2$ matrix $Q$. Here,  $\Omega$ is the empirical variance-covariance matrix of the zero mean vectors $\frac{1}{2} ( Q_{t,1}^2 - Q_{t,2}^2)$ and $Q_{t,1} Q_{t,2}$. For Gaussian errors, $\eta^* =2$, and hence $d_1=d_2=1$ and $\mu_1+\mu_2 = \frac{a^2}{4} = \frac{(T c_{k+1}) 2}{4q}$, which yields the asymptotic distribution in (\ref{scaledchi}). Moreover, for generic distribution of the errors but $T$ large, the elements of matrix $\Omega$ scale with $T^{-1}$, so that the eigenvalues $\lambda_j$ tend to $0$ when $T\rightarrow\infty$, the effect of the factor path vanishes and  we recover the Gaussian case. 

\begin{figure}[!ht]
	\footnotesize
	\begin{center}
		\caption{Asymptotic local power of statistic $\mathscr{S}(k)$ under local alternatives in a non Gaussian setting: we take $T=2$, $k=0$, nominal size $\alpha=0.05$, and $\eta^* = \eta/q = 5$. The parameter $\varphi=Q_{11}^2$ is the squared upper-left element of matrix $Q$.}
		\label{figure:local:power:nG}
			\includegraphics[width=\textwidth]{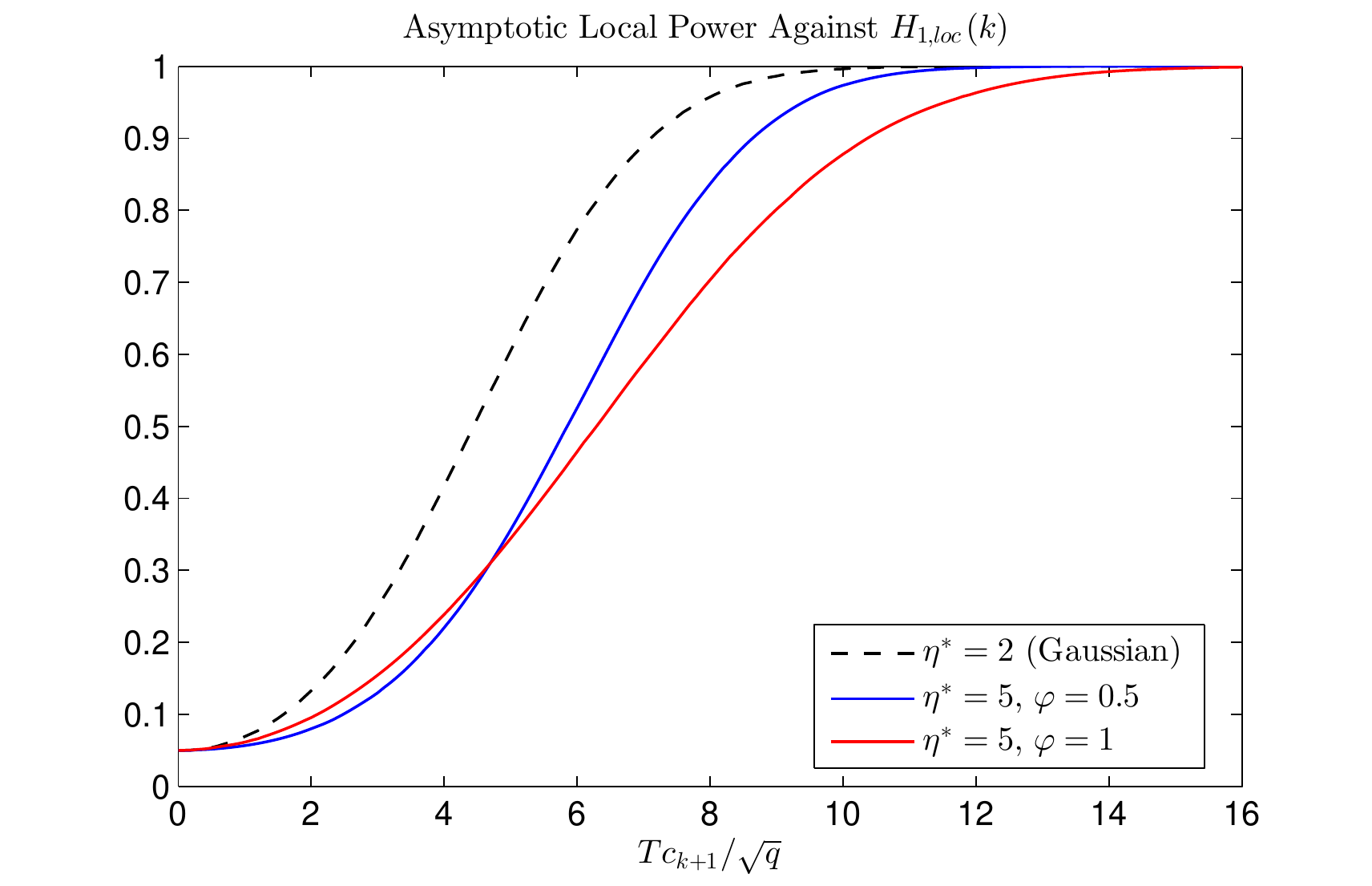}
		
		\end{center}
\end{figure}

As an illustration, let us particularize the result in the case with $T=2$ and $k=0$, namely we test the null of no factors in a large panel with two time periods and consider the local alternative of a weak factor. From (\ref{asy:S2:nonGaussian}) we get the asymptotic distribution under local alternatives $\frac{n}{4 q} \mathscr{S}(k)^2 ~ \Rightarrow~ \frac{\eta^*}{2} \chi^2(1,\frac{a^2}{2 \eta^*} ( 1 - 2\varphi)^2) +   \chi^2 (1,a^2 \varphi (1-\varphi))$, where $\varphi := Q_{11}^2$. The distribution depends on the factor path by means of $Q_{11}^2$, i.e. the squared standardized value of the weak factor in the first period. In Figure 
\ref{figure:local:power:nG} we plot the asymptotic local power curves for different values of parameters $\varphi \in [0,1]$ and $\eta^* =5$. The local power is lower than for the Gaussian design ($\eta^*=2$). The effect of the factor path on local power is not uniform. For small value of $T c_{k+1}/\sqrt{T}$, the local power is marginally larger with $\varphi=1$ (or $\varphi=0$, not displayed), i.e. when the weak factor has values that vary a lot between the two periods.  Instead, for larger values of $T c_{k+1}/\sqrt{T}$ the local power is larger for $\varphi=0.5$, i.e. when the weak factor has a more stable path.

\subsection{Testing for weak factors}

Suppose there are $k-1$ strong, or semi-strong, factors in the systematic component (maintained hypothesis). For the $k$th factor, we want to test the null hypothesis of a weak factor $H_0^k ~:~\kappa_k = 1/2$ vs the alternative of strong or semi-strong factor $H_1^k~:~\kappa_k < 1/2$. Note that, differently from the previous section, here the weak factor is under the null hypothesis. From Theorem \ref{prop:expansion:eigenvalues} with $\hat{\Psi} = \frac{1}{\sqrt{n}} (T \sqrt{n} \tilde{\sigma}_{\beta,k}^2 U_k U_k' +  \sqrt{T} W_n U' + \sqrt{T} U W_n' + Z_n )$, we have under $H_0^k$:
\begin{equation} \label{test:weak:factors}
\Delta_k := \sqrt{n} [  \delta_k(\hat{V}_y) - \delta_{k+1}(\hat{V}_y)] ~\Rightarrow~\delta_1 \left( T c_k \xi_k \xi_k' + Q' Z Q \right) - \delta_2 \left( T c_k \xi_k \xi_k' + Q' Z Q \right),
\end{equation}
where the columns of $Q$ spans the orthogonal complement of the range of $U = [U_1:\cdots:U_{k-1}]$, $\xi_k = Q'U_k$ is a vector in $\mathbb{R}^{T-k+1}$ with unit norm, and $c_k = \lim n^{1/2} \tilde{\sigma}_{\beta,k}^2 > 0$. The asymptotic distribution is invariant to rotations of the columns of $Q$ and $\xi_k$. Under the alternative hypothesis $H_1^k$, we have $\Delta_k \overset{p}{\rightarrow} +\infty$ at rate $O(n^{1/2-\kappa_k})$.

The asymptotic distribution in (\ref{test:weak:factors}) is not feasible because scalar $c_k$ and vector $\xi_k$ are not known. Under a weak factor hypothesis, the vector of factor values $U_k$ cannot be estimated consistently from PCA (matrix $Q$ instead can). In fact, the remainder term in the asymptotic expansion (\ref{asy:Uhatj}) for $j=k$ is $O_p(1)$ if $\kappa_k=1/2$. To perform the test, we can adopt a subsampling approach (see, e.g., Politis, Romano, Wolf (1999)) since the asymptotic distribution in (\ref{test:weak:factors}) is well defined.  We compute the values $\Delta_k^b$, for $b=1,...,B$, from $B$ subsamples of size $m$ (i.e., $m$-out-of-$n$ bootstrap). The critical value for the test at size $\alpha \in (0,1)$ is the $1-\alpha$ quantile of the empirical distribution of the $\Delta_k^b$. We reject the null hypothesis of weak factor $H_0^k~:~\kappa_k =1/2$ if the sample value $\Delta_k$ exceeds the critical value. 

\section{Monte Carlo analysis}

This section explores the finite sample properties of the test statistics $\sqrt{n}\mathscr{S}(k)$, $\mathscr{S}^*(k)$ and $n \mathscr{T}(k)$. We first introduce the four Data Generating Processes (DGP) that we use in our Monte Carlo analysis, and then present the results for the size and power of the statistics.  

\subsection{Data Generating Processes}
We use four DGPs. In DGP1, the betas and factor values are $\beta_i ~ \overset{i.i.d.}{\sim} ~ N(0,I_k)$ and $f_t ~ \overset{i.i.d.}{\sim} ~ N(0,I_k)$, and the error terms are $\varepsilon_{i,t} ~ \overset{i.i.d.}{\sim} ~ N(0,\sigma_i^2)$, where the variances are  uniform random draws $\sigma_i^2 ~ \overset{i.i.d.}{\sim} ~ U[a,b]$ with $a=1$ and $b=4$. All random variables are mutually independent. We generate $10,000$ panels of returns of size $n \times T$ for each of the $100$ draws of the $T\times k$ factor matrix $F$, in order to keep the factor values constant within repetitions, but also to study the potential heterogeneity of size and power results across different factor paths.  The factor betas and error variances are the same across all repetitions in all designs of the section. We use $k=3$ factors, three different	 cross-sectional sizes $n=500,1000, 5000$, and three  values of time-series dimension $T=6,12,24$. 

DGP2 accommodates various types and strengths in the third factor:
\begin{eqnarray*}
\beta_i ~ \overset{i.i.d.}{\sim} ~ N(0, \Sigma_{\beta}), \qquad \Sigma_{\beta} = \left( \begin{array}{ccc}
1 & & \\
& 1 & \\
& & c n^{-\kappa}
\end{array} \right),
\end{eqnarray*}
where the values of exponent $\kappa$ are $\kappa=0$ (strong factor), $\kappa=0.25,0.4$ (semi-strong), $\kappa=0.5$ (weak), and $\kappa=0.6,0.75,1$ (vanishing factor). The values for constant $c$ are $c=0.1$, $c=1$, and $c=10$. 
Further, $f_t ~ \overset{i.i.d.}{\sim} ~ N(0,I_3)$, $\varepsilon_{i,t} ~ \overset{i.i.d.}{\sim} ~ N(0,\sigma_i^2)$, and 
$\sigma_i^2 ~ \overset{i.i.d.}{\sim} ~ U[a,b]$ and $a=1$, $b=4$ as in DGP1. The case with $\kappa=0$ and $c=1$ corresponds to DGP1. The sample sizes are $T=6$ and $n=500,1000,5000$. We display results for one given realization of the factor path. 
\footnote{The factor path is normalized after drawing the factor values such that $\tilde{\Sigma}_f=I_k$. Consequently, $c_3 =c$ in (\ref{c:j}).}

DGP3 is aimed at covering a setting with both non-normality and idiosyncratic conditional heteroschedasticity of errors. Specifically, factor values and loadings are generated as in DGP2, but the errors now follow independent Autoregressive Conditionally Heteroschedastic dynamics of order 1, i.e., ARCH(1), see Engle (1982), namely
\begin{eqnarray*}
\varepsilon_{i,t} &=& h_{i,t}^{1/2} u_{i,t}, \qquad u_{i,t} \overset{i.i.d.}{\sim} N(0,1), \\
h_{i,t} &=& c_i + \alpha_i \varepsilon_{i,t-1}^2 , 
\end{eqnarray*}
where $c_i = \sigma_i^2 (1-\alpha_i)$, $\sigma_i^2 \overset{i.i.d.}{\sim} U[a,b]$ with $a=1$, $b=4$, and $\alpha_i \overset{i.i.d.}{\sim} U[l,u]$ with $l=0.1$ and $u=0.4$, all draws mutually independent. Here, we set $u^2 < 1/3$ to ensure existence of the fourth-order moments of errors. This specification matches the condition of sphericality of errors in Assumption \ref{ass:var:returns} with $\bar{\sigma}^2 = (a+b)/2$ (for a.e. draws of the random ARCH parameters).  The CLT condition in Assumption \ref{ass:CLT:epseps} is also met with a symmetric random matrix $Z$ such that $Z_{t,t} \sim N(0, 2 q \psi(0))$, $Z_{t,t+h} \sim N(0, q[ 1 + 2\psi(h)])$ for $h>0$, and $Cov( Z_{t,t}, Z_{t+h,t+h} ) = 2q \psi(h)$, for $h>0$, where $\psi(h) = {E}[ \frac{\alpha_i^h}{1-3\alpha_i^2}]= \int_l^u \frac{\alpha^h}{1-3\alpha^2}d\alpha$. Any pair of elements $Z_{t,s}$, $Z_{r,p}$, of which one or both are out of the diagonal, are independent.  As the econometrician may be unsure about the actual cross-sectional distribution of the $\alpha_i$ parameters, she can adopt a semi-nonparametric approach with respect to that distribution. It yields a parametric specification $\Omega(\theta)$ for the variance-covariance matrix of $vec(Z)$, with parameter vector $\theta = (q,\psi(0),\psi(1),...,\psi(T-1))' \in \mathbb{R}^{T+1}$. This specification nests the one with time independence for $\psi(1)=...=\psi(T-1)=0$ (then, $\eta = 2 q \psi(0)$), as well as Gaussian independent errors when additionally $\psi(0)=1$. In the simulations, we use $T=12$ and $n=500, 1000, 5000$. We estimate the parameter vector $\theta$ using (\ref{est:theta}). The order condition is met as long as $k \leq 9$ (here $k=2$).   

Finally, DGP4 involves instruments correlated with the factor loadings according to the model
	\begin{equation*}
	\beta_i=\Gamma'z_i+ u_i,
	\end{equation*}
where $z_i\overset{i.i.d.}\sim N(0,I_K)$ and $u_i\overset{i.i.d.}\sim N(0,I_k)$ mutually independent for $k=3$ and $K=10$. The $K \times k$ matrix $\Gamma$ is obtained from the normalized eigenvectors associated with the non-zero eigenvalues of $G G'$, where the $K\times k$ matrix $G$ has i.i.d. standard normal entries. Moreover, $f_t \overset{i.i.d.}\sim N(0,I_k)$ and $\varepsilon_i\overset{i.i.d.}\sim N(0,\sigma_i^2I_T)$ with $\sigma^2_i\overset{i.i.d.}\sim U[a,b]$ for $a=1$, $b=4$ as in DGPs 1 and 2.

\subsection{Size and power results}

We start with statistics  $\sqrt{n}\mathscr{S}(k)$ and $\mathscr{S}^*(k)$ based on the variance-covariance matrix of excess returns. The critical values are obtained from the procedure outlined in Section 4.3.1 i). 
We provide the size and power results in \% for DGP1 in Table \ref{MC:Table1}. Size is close to its nominal level $5\%$ for both statistics, with size distortions smaller than $1\%$. The impact of the factor values on size is very small, as expected from theory under Gaussian errors. The power refers to the statistics computed with $k=2$, for which DGP1 corresponds to a global alternative. It is generally larger for statistic $\sqrt{n}\mathscr{S}(k)$. It is coherent with the finding in Figure \ref{figure:local:power} when considering local alternatives. The power of both statistics varies with the factor path especially for $T=6$. It  is a finite-$n$ effect, which becomes weaker when $T$ increases. 

We provide the rejection frequencies in \%  for statistics $\sqrt{n} \mathscr{S}(k)$ and $\mathscr{S}^*(k)$ under DGP2 in Tables \ref{MC:Table2} and \ref{MC:Table3}. For statistic $\sqrt{n}\mathscr{S}(k)$, the power against omitted strong, or semi-strong, factors (i.e., $\kappa < 0.5$) is large for $c=1$ and $c=10$. As $n$ grows, the rejection frequency is expected to converge to $100\%$; the convergence is quicker for small $\kappa$ and/or large $c$.   With vanishing factors, i.e., $\kappa>0.5$, the finite sample size of the statistic is close to its nominal size $5\%$ with $c=0.1$ and $c=1$. We find a few oversize effects for $c=10$. In all cases, the rejection frequency gets closer to $5\%$ when $n$ increases, as expected. With a weak factor, i.e., $\kappa=0.5$, the power of the statistic $\sqrt{n}\mathscr{S}(k)$ increases from about $5\%$ with $c=0.1$ to $100\%$ with $c=1$. Table \ref{MC:Table3} shows that the results are qualitative similar for statistic $\mathscr{S}^*(k)$ but with smaller power and oversize effects. 

For a given sample size, the rejection rate for a semi-strong factor with small $c$ can be similar to that for a vanishing factor with larger constant $c$. For instance, for $n=1000$, the cases $\kappa=0.25,c=0.1$ and $\kappa=0.60,c=1$ yield the same rejection rate $0.066$ for statistic $\sqrt{n} \mathscr{S}(k)$ in Table \ref{MC:Table2}. In fact, in finite sample, only quantity $cn^{-\kappa}$ matters. The Pitman drifting DGP in (\ref{c:j}) is merely a mathematical tool to analyse the asymptotic behavior of the test locally around the null or the global alternative hypotheses.  Tables \ref{MC:Table2} and \ref{MC:Table3} show that the cross-sectional size $n$, for which the asymptotic regime is reached, depends on the combination of values $\kappa,c$.

Tables \ref{MC:Table4} and \ref{MC:Table5} provide the rejection frequencies in \%  for statistics $\sqrt{n} \mathscr{S}(k)$ and $\mathscr{S}^*(k)$ for DGP3 with ARCH(1) errors.	The size and power properties are good, with rejection frequencies that are rather close to those with Gaussian errors displayed in Tables \ref{MC:Table2} and \ref{MC:Table3}. It confirms that the procedure presented in Section 4.3.1 ii) 	to define a feasible test statistic works well in the setting with idiosyncratic conditional heteroschedasticity yielding non-Gaussian errors. 

Finally, we consider the statistic based on the variance-covariance matrix of instrument-weighted portfolio returns. Table \ref{MC:Table6} presents the size and power for the statistic $\mathscr{T}(k)$ with different values of sample sizes $T$ and $n$. Results are averaged over $100$ different realizations of $F$ and $\Gamma$, and standard errors in percent are given in parentheses. To simulate the critical values, in the upper panel of Table \ref{MC:Table6} we use the estimator $\hat{\Sigma}_{U,1}=\hat{\sigma}^2 [I_T\otimes \hat{Q}_{zz}]$ defined in Section 4.2.1, that is valid in the homoschedastic setting of DGP4. The lower panel uses the more general estimator $\hat{\Sigma}_{U,2}=\frac{1}{n}\sum_{i=1}^n [(\hat{\varepsilon}_{i}\hat{\varepsilon}'_{i})\otimes (z_iz'_i)]$. Overall, size distortions are rather small, except for some evidence of undersize for the statistic using the more general variance estimator $\hat{\Sigma}_{U,2}$ when $n$ is small. Power is close to $100\%$ across all combinations in our experiment.  

\section{Empirical application}

In this section, we present the results of an empirical application to testing for latent factors in short subperiods of the CRSP dataset. We consider monthly returns of individual stocks between January 1963 and December 2021. We focus on subperiods defined by the bear vs bull market classification introduced in Lunde and Timmermann (2004). We consider the three statistics $\sqrt{n}\mathscr{S}(k)$, $\mathscr{S}^*(k)$ and $n\mathscr{T}(k)$, and compute their p-values for testing different values of the number of latent factor $k$. For the third statistic,  as instruments, we use the 12 stock characteristics in Freyberger, Neuhierl, and Weber (2020) (see also Gagliardini and Ma (2019)) measured at the date prior to the subperiod start. The 12 characteristics are grouped into four categories: (i) past returns variables, which are return from 2 to 1 month (1 month horizon)  before the current period ($r_{2-1}$), return from 12 to 2 months (10 months horizon) before current period ($r_{12-2}$), return from 12 to 7 months (5 months horizon) before current period ($r_{12-7}$), and return from 36 to 13 months (23 months horizon) before current period ($r_{36-13}$); (ii) profitability-related characteristics, which are earnings per share (EPS), return on asset (ROA), return on equity (ROE); (iii) value-related characteristics, which are total assets to size (A2ME), sales to price (S2P), and (iv) trading friction variables, which include total assets (AT), price times shares outstanding (LME), and last month's volume to shares outstanding (LTurnover). We collect them from the COMPUSTAT database. Some of these instruments are recorded at frequencies lower than monthly, namely yearly, and the instrument values are considered constant within a year. For comparison purposes, for each test statistic and subperiod, we use the balanced panel of stocks with available return and instrument data at all months. Relying on short time spans mitigates the concern of survivorship bias inherent to the use of balanced panels.

\begin{figure}[!ht]
	\footnotesize
	\begin{center}
		\caption{Eigenvalue spacings of matrix matrix $\hat{V}_y$ (upper panel), their ratios (middle panel) and eigenvalues of matrix $\hat{V}_{\xi}$ (lower panel) for the period from March 1977 to February 1978. This period is classified as ``bull market" according to Lunde and Timmermann (2004) methodology.}
		\label{figure:eigv:7778}
			\includegraphics[width=\textwidth]{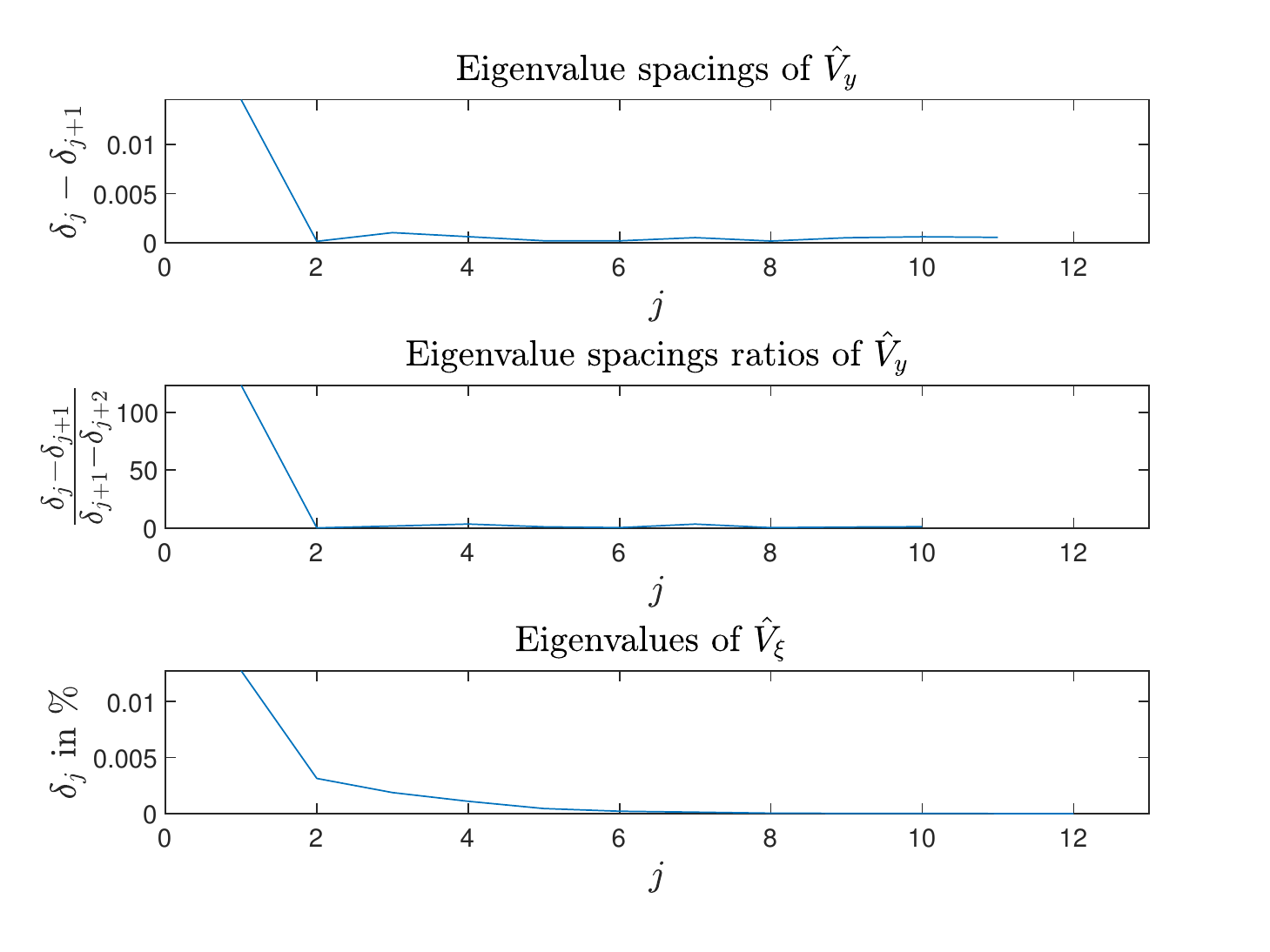}
		
		\end{center}
\end{figure}

\begin{figure}[!ht]
	\footnotesize
	\begin{center}
		\caption{The figure displays the p-values for statistics $\mathscr{S}(k)$ (upper panel), $\mathscr{S}^*(k)$ (middle panel) and $\mathscr{T}(k)$ (lower panel) for the period from March 1977 to February 1978. This period is classified as ``bull market" according to Lunde and Timmermann (2004) methodology.}
		\label{figure:pval:7778}
			\includegraphics[width=\textwidth]{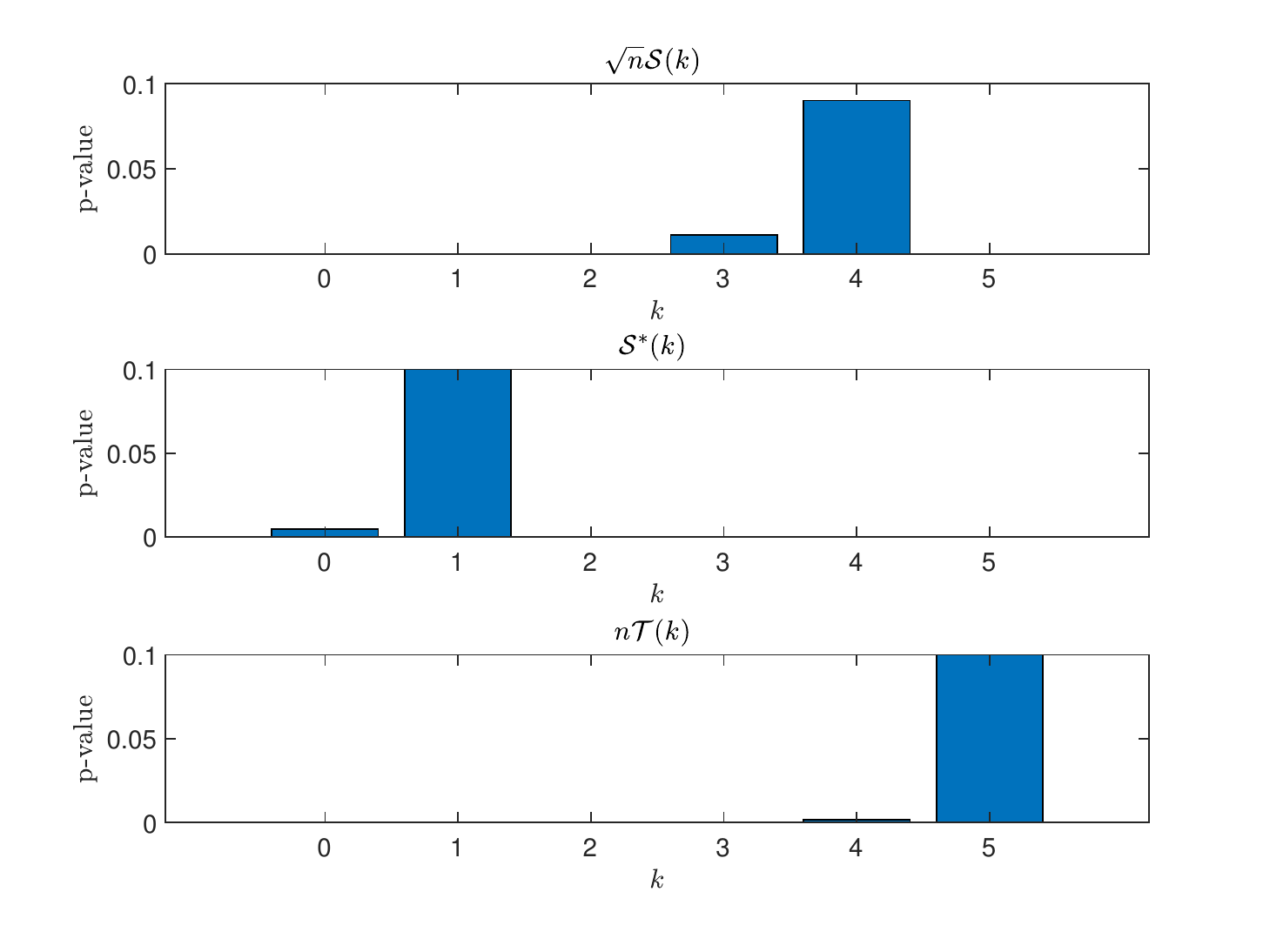}
		
		\end{center}
\end{figure}

\begin{figure}[!ht]
	\footnotesize
	\begin{center}
		\caption{Eigenvalue spacings of matrix matrix $\hat{V}_y$ (upper panel), their ratios (middle panel) and eigenvalues of matrix $\hat{V}_{\xi}$ (lower panel) for the period from July 1981 to June 1982. This period is classified as ``bear market" according to Lunde and Timmermann (2004) methodology.}
		\label{figure:eigv:8182}
			\includegraphics[width=\textwidth]{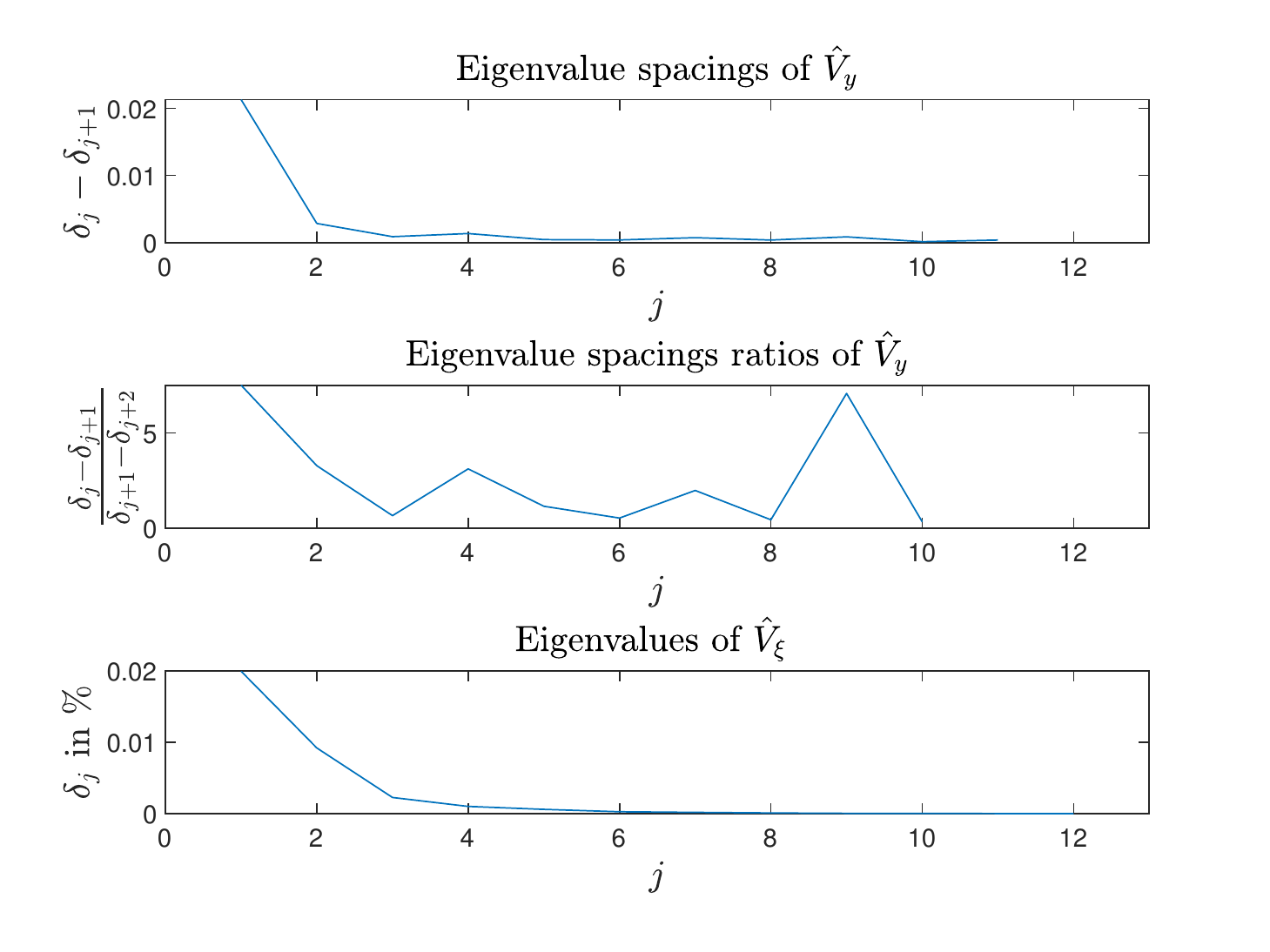}
		
		\end{center}
\end{figure}

\begin{figure}[!ht]
	\footnotesize
	\begin{center}
		\caption{The figure displays the p-values for statistics $\mathscr{S}(k)$ (upper panel), $\mathscr{S}^*(k)$ (middle panel) and $\mathscr{T}(k)$ (lower panel) for the period from  July 1981 to June 1982. This period is classified as ``bear market" according to Lunde and Timmermann (2004) methodology.}
		\label{figure:pval:8182}
			\includegraphics[width=\textwidth]{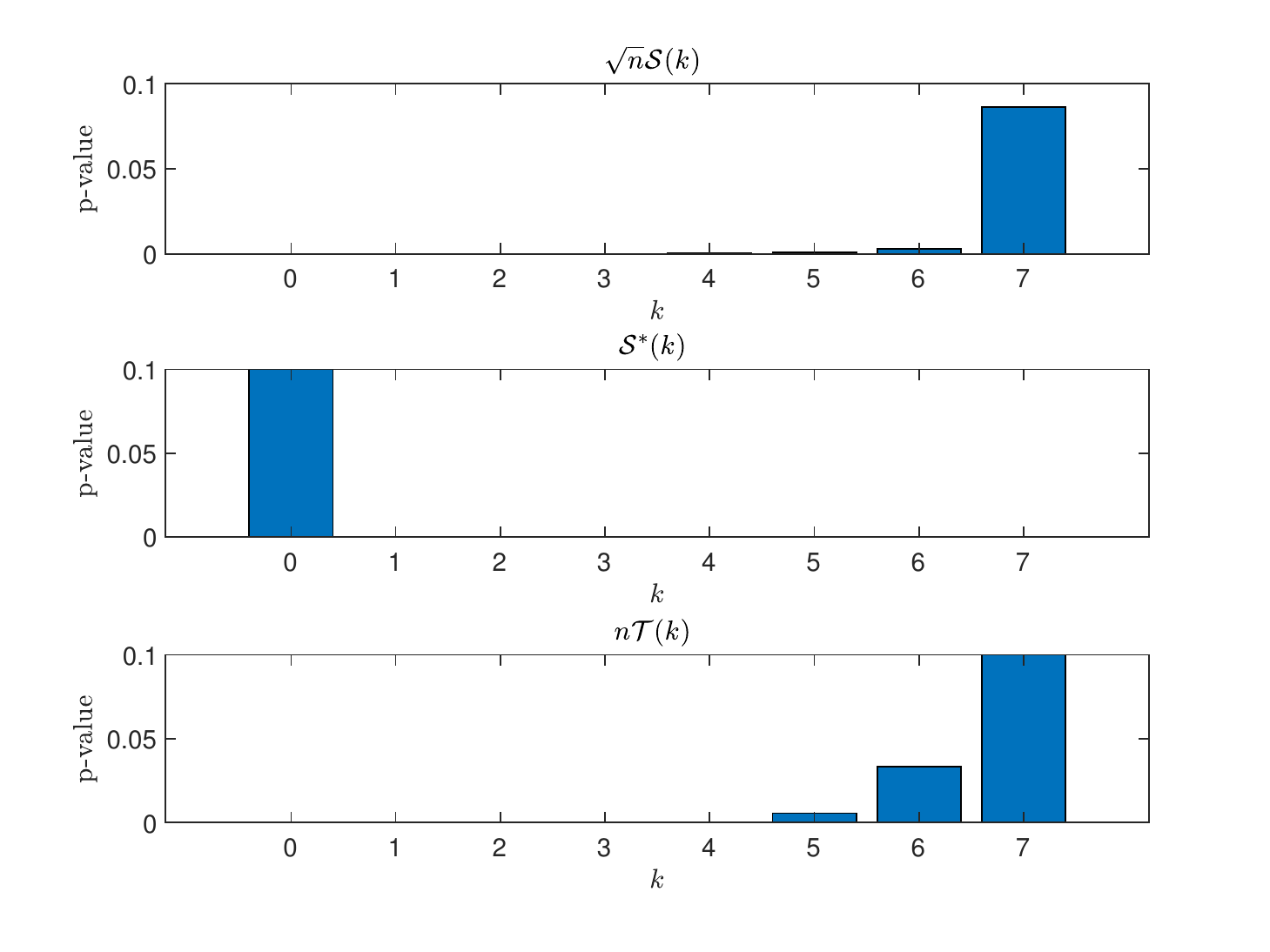}
		
		\end{center}
\end{figure}

\begin{figure}[!ht]
	\footnotesize
	\begin{center}
		\caption{Eigenvalue spacings of matrix matrix $\hat{V}_y$ (upper panel), their ratios (middle panel) and eigenvalues of matrix $\hat{V}_{\xi}$ (lower panel) for the period from December 2010 to November 2011. This period is classified as ``bull market" according to Lunde and Timmermann (2004) methodology.}
		\label{figure:eigv:1011}
			\includegraphics[width=\textwidth]{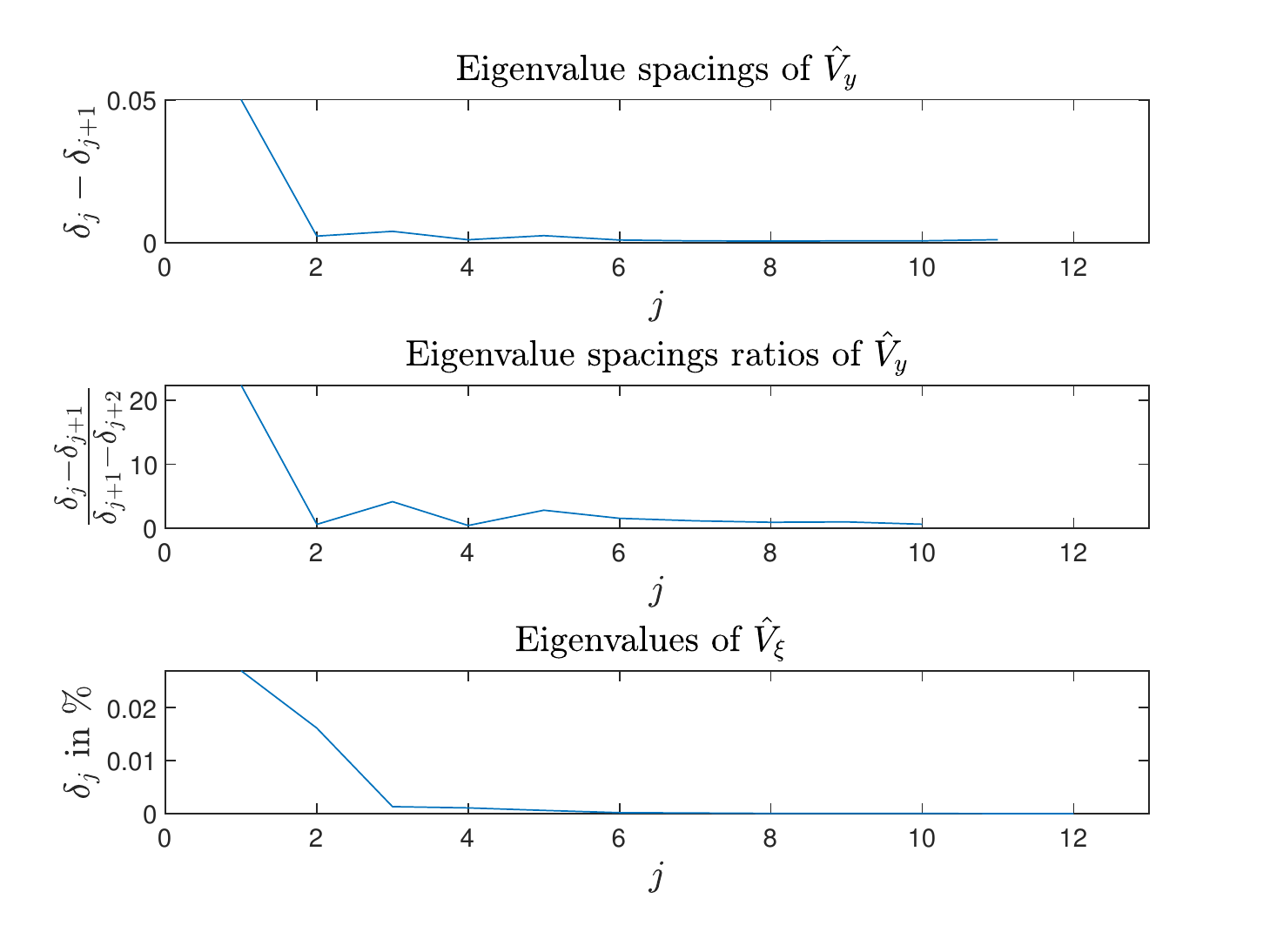}
		
		\end{center}
\end{figure}

\begin{figure}[!ht]
	\footnotesize
	\begin{center}
		\caption{The figure displays the p-values for statistics $\mathscr{S}(k)$ (upper panel), $\mathscr{S}^*(k)$ (middle panel) and $\mathscr{T}(k)$ (lower panel) for the period from December 2010 to November 2011. This period is classified as ``bull market" according to Lunde and Timmermann (2004) methodology.}
		\label{figure:pval:1011}
			\includegraphics[width=\textwidth]{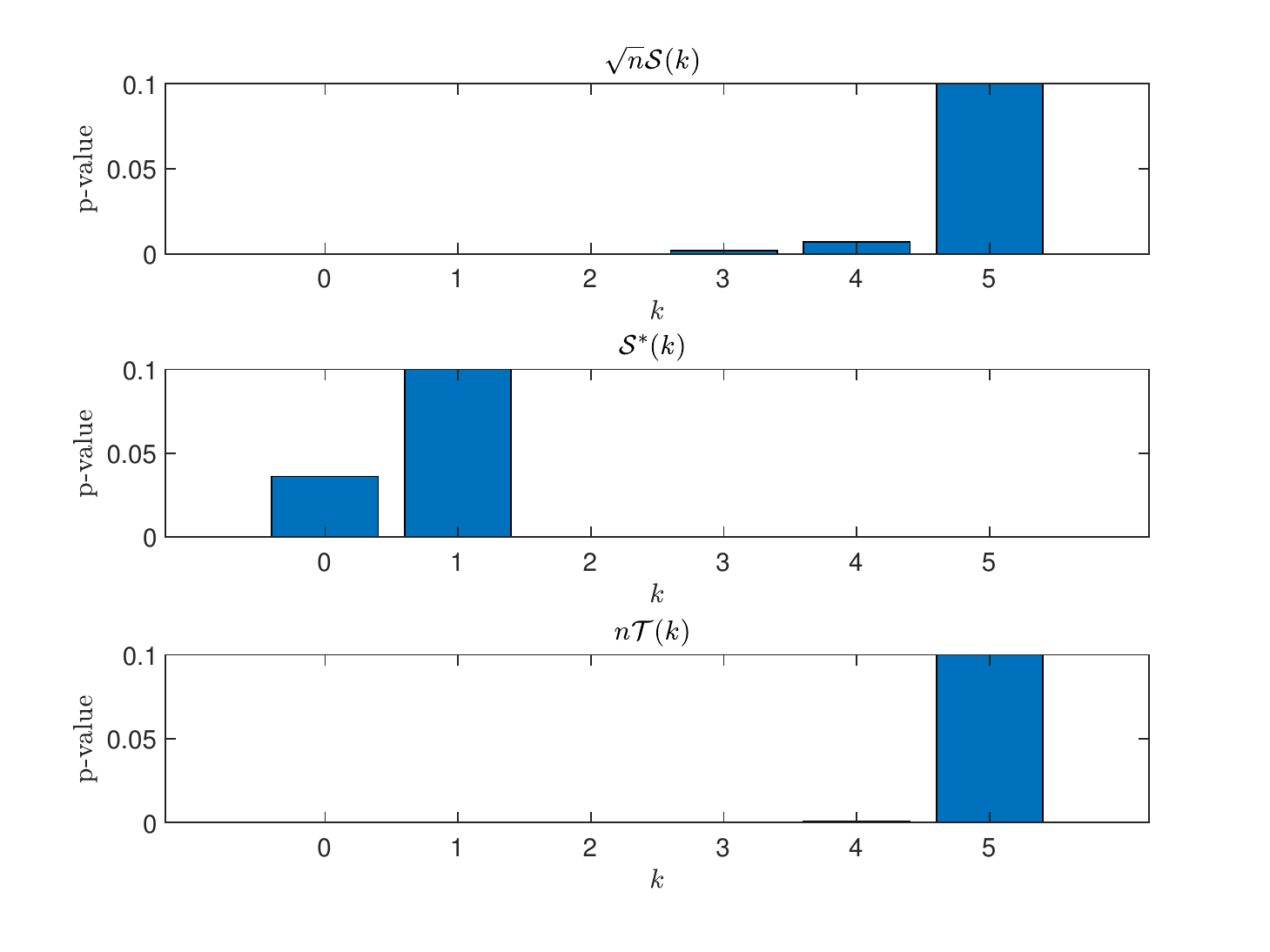}
		
		\end{center}
\end{figure}

\begin{figure}[!ht]
	\footnotesize
	\begin{center}
		\caption{Eigenvalue spacings of matrix matrix $\hat{V}_y$ (upper panel), their ratios (middle panel) and eigenvalues of matrix $\hat{V}_{\xi}$ (lower panel) for the period from January 2020 to December 2021. We associate this period in our sample to the Covid-19 pandemics.}
		\label{figure:eigv:covid}
			\includegraphics[width=\textwidth]{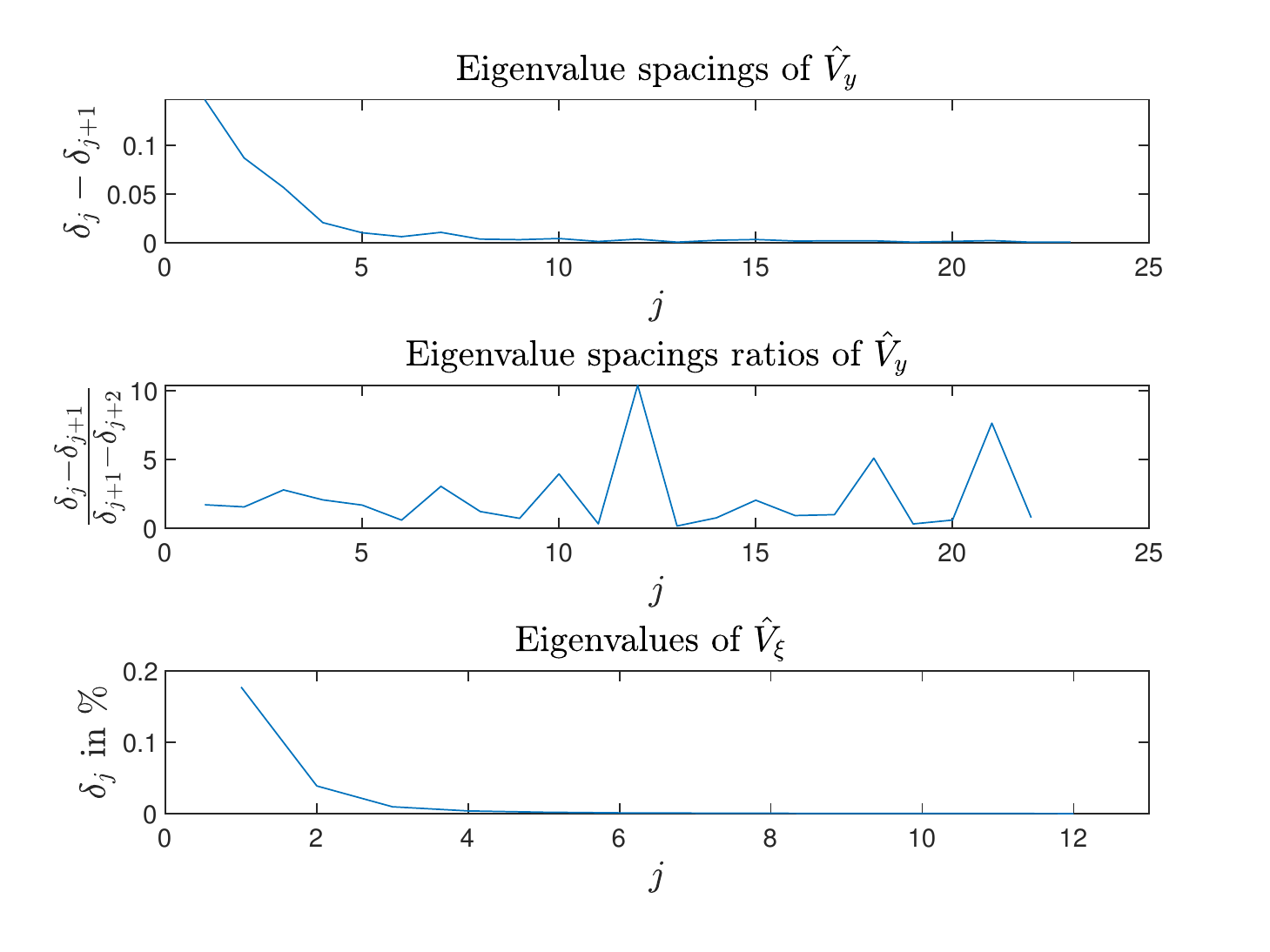}
		
		\end{center}
\end{figure}

\begin{figure}[!ht]
	\footnotesize
	\begin{center}
		\caption{The figure displays the p-values for statistics $\mathscr{S}(k)$ (upper panel), $\mathscr{S}^*(k)$ (middle panel) and $\mathscr{T}(k)$ (lower panel) for the period from January 2020 to December 2021. We associate this period in our sample to the Covid-19 pandemics.}
		\label{figure:pval:covid}
			\includegraphics[width=\textwidth]{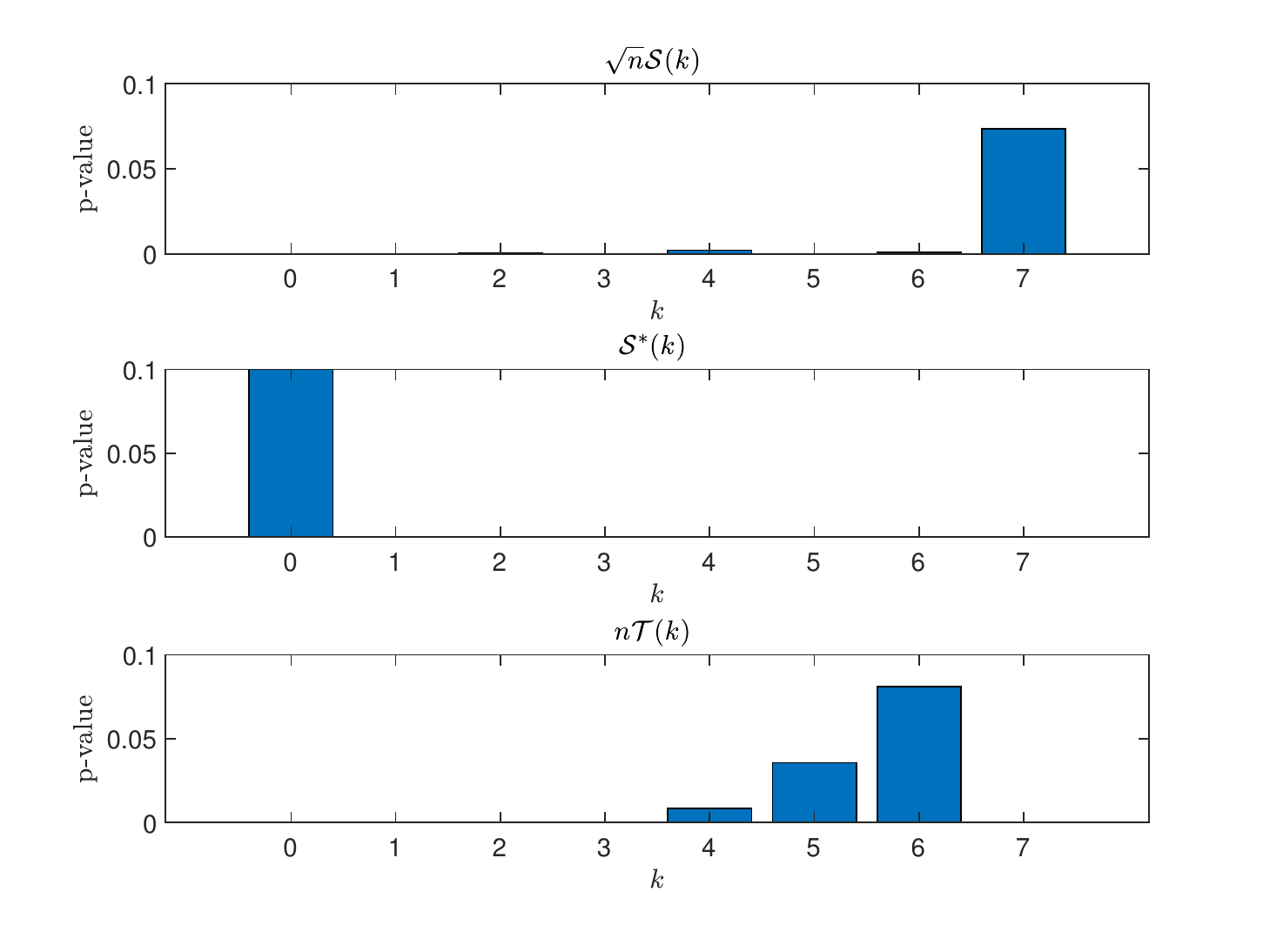}
		
		\end{center}
\end{figure}

Our empirical
 evidence based on a fixed $T$ can target particular periods. We focus on three time spans of $T=12$ months for illustrative purposes: (I) from 1977/03 to 1978/2, with $n=1781$ stocks, (II) 1981/07-1982/06, with $n=1821$, (III) 2010/12-2011/11, with $n=3129$ in the balance panels. 
We also consider   a fourth period of $T=24$ months (IV) 2020/01-2021/12, with $n=2418$.
 Periods (I) and (III) are classified as ``bull market", and periods (II) and (IV) as ``bear market", according to Lunde and Timmermann (2004). The last subperiod (IV) essentially corresponds to Covid pandemics. In Figure \ref{figure:eigv:7778}, we display the key inputs for our test statistics in subperiod (I), namely the eigenvalues' spacings and their ratios for the matrix $\hat{V}_y$ of second-order moments of returns, as well as the eigenvalues for variance matrix $\hat{V}_{\xi}$ built from instrument-based portfolios. Figure \ref{figure:pval:7778} reports the p-values of the three test statistics for the same subperiod. We display eigenvalues and test results for subperiods (II) to (IV) in Figures \ref{figure:eigv:8182} to \ref{figure:pval:covid}. 

In Figure \ref{figure:eigv:7778}, upper panel, we see that the eigenvalues differences $\delta_j(\hat{V}_y)-\delta_{j+1}(\hat{V}_y)$ are very small from order $j=5$ onward. In accordance with this feature, in Figure \ref{figure:pval:7778}, statistic $\sqrt{n}\mathscr{S}(k)$ rejects the null hypothesis $H_0(k)$ for $k=0,1,2,3$ factors at level $5\%$ (and even smaller), while the p-value for the test of $k=4$ factors is about $0.09$. The middle panel of Figure \ref{figure:eigv:7778} shows that the eigenvalue spacings ratio $\frac{\delta_j(\hat{V}_y)-\delta_{j+1}(\hat{V}_y)}{\delta_{j+1}(\hat{V}_y)-\delta_{j+2}(\hat{V}_y)}$ for $j=1$ is very large, while the other ratios are small. Then, statistic $\mathscr{S}^*(k)$ for $k=0$ is beyond the critical value and rejects the null hypothesis of no latent factor in the panel of excess returns	in subperiod 1977/03-1978/2, even at $1\%$ level, while it does not reject the null hypothesis of a single factor, see Figure \ref{figure:eigv:7778} middle panel. With test statistic $n \mathscr{T}(k)$, we reject the null $H_0(k)$ for 4 latent factors or less, while the p-value for $k=5$ factors is about $0.10$. Hence, the four largest eigenvalues of $\hat{V}_{\xi}$ in the lower panel of Figure \ref{figure:eigv:7778} are statistically significantly different from zero while the other eigenvalues are not. 

The results obtained with statistics $\sqrt{n}\mathscr{S}(k)$ and $n\mathscr{T}(k)$  are rather concordant, in each of the four subperiods under consideration. Both statistics lead to the same acceptance/rejection decisions in most cases. The discrepancies about the smallest order $k$ with p-values above $5\%$, say, is at most of one unit. For instance, this order is $k=4$ for $\sqrt{n}\mathscr{S}(k)$ and $k=5$ for $n\mathscr{T}(k)$ in 1977/3-1978/2 (Figure \ref{figure:pval:7778}), whereas $k=7$ for both statistics in 1981/7-1982/6 (Figure \ref{figure:pval:8182}). A large number of factors might point at time-varying betas with common instruments, namely scaled factors (Cochrane (2005)). With a penalisation method, Bakalli, Guerrier and Scaillet (2021) show the predominance of selected common instruments over selected stock-specific instruments in the factor loading dynamics. It is worthwhile recalling that statistics $\sqrt{n}\mathscr{S}(k)$ and $n\mathscr{T}(k)$ rely on different identification principles, i.e., errors sphericity for the former and instruments validity for the latter. Hence, concordance in the results across the two statistics provides a first robustness check for our findings vis-a-vis the identification assumptions. On the contrary, statistic $\mathscr{S}^*(k)$ based on eigenvalue spacings ratios tends to fail to reject null hypotheses with small numbers of latent factors, such as zero or one factor. We interpret this finding as a consequence of the low power of the $\mathscr{S}^*(k)$ statistic already pointed out in our numerical experiments (see Figure \ref{figure:local:power}) and Monte Carlo simulations (see Section 6).

When comparing the test results across the four subperiods, our findings seem to point to a rather similar number of latent factors in the bull and bear market periods: the test statistics $\sqrt{n}\mathscr{S}(k)$ and $n\mathscr{T}(k)$ fail to reject null hypotheses with 5 to 7 latent factors in both cases. We do not see a clear pattern relating monotonically the number of latent factors to the bear vs bull market phases, in particular we do not find a conclusive evidence for a smaller number of latent factors during bear market phases compared to market upturns, at least in those subperiods. \footnote{Besides, a selection rule for the number of latent factors based on sequential testing with our statistics would find more latent factors in the bear market periods (II) and (IV) than in bull market periods (I) and (III).}  In particular, our results contradict the common wisdom that every equity return series correlate to 1 (or at least close to) and everything boils down to a single factor model (or at least close to) in bear periods. 

\section{Concluding remarks}

In this paper, we develop new tests for the number of latent factors in short panels. Identification relies either on a sphericality assumption on the error terms, or on availability of instruments. The derivation of the asymptotic distributions for $n \rightarrow \infty$ and fixed $T$ leverages on (i) a uniform perturbation expansion for the small eigenvalues of symmetric matrices, and (ii) the distributions of eigenvalues (spacings) of Gaussian matrices of finite dimension. The setting is general enough to accommodate various forms for the factor strength, namely strong, semi-strong, weak and vanishing factors, when defining the null and alternative hypotheses. We also introduce a novel test for weak factors against (semi-)strong factors. In an empirical application for short subperiods of the CRSP panel dataset, p-values suggest a relative stability in the number of latent factors across market downturns and market upturns with 5 to 7 factors. Our findings bring evidence against the common wisdom that a (near to) single factor model with (near to) unit correlation among any pair of series prevail in bear market phases.

\newpage

\section*{References}

\noindent Ait-Sahalia, Y., and Xiu, D., 2017. Using principal component analysis to estimate a high dimensional factor model with high-frequency data. Journal of Econometrics, 201, 384-399.

\medskip

\noindent Ahn, S., and Horenstein, A.R., 2013. Eigenvalue ratio test for the number of factors. Econometrica, 81(3), 1203-1227.

\medskip

\noindent Ando, T., and  Bai, J., 2015. Asset pricing with a general multifactor structure. Journal of Financial Econometrics, 13(3), 556-604.

\medskip

\noindent  Andreou, E., Gagliardini, P., Ghysels, E., and Rubin, M., 2019. Inference in group factor models with an application to mixed frequency data. Econometrica, 87(4), 1267-1305.

\medskip

\noindent Atas, Y., Bogomolny, E., Giraud, O., and Roux, G., 2013. Distribution of the ratio of consecutive level spacings in random matrix ensembles. Physical Review Letters, 110, 084101. 

\medskip

\noindent Bai, J., 2003. Inferential theory for factor models of large dimensions. Econometrica, 71(1), 135-171.

\medskip

\noindent  Bai, J., 2009. Panel data models with interactive fixed effects. Econometrica, 77(4), 1229-1279.

\medskip

\noindent Bai, J., and Ng, S., 2002. Determining the number of factors in approximate factor models. Econometrica, 70(1),
191-221

\medskip

\noindent Bakalli, G., Guerrier, S., and Scaillet, O., 2021. A penalized two-pass regression to predict stock returns with time-varying risk premia. Working paper.

\medskip

\noindent Caner, M., and Han, X., 2014. Selecting the correct number of factors in approximate factor models: the large panel case with group bridge estimator. Journal of
Business and Economic Statistics, 32(3), 359-374.

\medskip

\noindent Carlini, F., and Gagliardini, P., 2022. Instrumental variables inference in a small-dimensional VAR model with dynamic latent factors. Forthcoming in Econometric Theory. 

\medskip

\noindent Chamberlain, G., 1992. Efficiency bounds for semi-parametric regression. Econometrica, 60, 567-596.

\medskip

\noindent Chen, Q., Roussanov, N., and Wang, X., 2022. Semiparametric conditional factor models: estimation and inference. Working paper.

\medskip

\noindent Cheng, M., Liao, Y., and Yang, X., 2021. Uniform predictive inference for factor models with
instrumental and idiosyncratic betas. Working paper.

\medskip

\noindent Cochrane, J., 2005. Asset Pricing. Princeton: Princeton University Press.

\medskip

\noindent Connor G., and Linton, O. 2007. Semiparametric estimation of a characteristic-based factor model of common
stock returns. Journal of Empirical Finance, 14, 694-717.

\medskip

\noindent Connor, G., Hagmann, M., and Linton, O., 2012. Efficient semiparametric estimation of the Fama-French model
and extensions. Econometrica, 80(2), 713-754.

\medskip

\noindent  Connor, G., and Korajczyk, R., 1986. Performance measurement with the arbitrage pricing theory: A new framework for analysis. Journal of  Financial Economics, 15(3), 373-394.

\medskip

\noindent Connor, G.,  and Korajczyk, R., 1987. Estimating pervasive economic factors with missing observations. Working
Paper No. 34, Department of Finance, Northwestern University.

\medskip

\noindent Connor, G., and Korajczyk, R., 1993. A test for the number of factors in an approximate factor model. Journal of Finance, 48(4), 1263-1291.

\medskip

\noindent Engle, R., 1982. Autoregressive Conditional Heteroscedasticity with estimates of the variance of United Kingdom inflation. Econometrica, 50, 987-1007.

\medskip

\noindent Fan, J., Liao, Y., and Wang, W., 2016.  Projected principal component analysis in factor models. Annals of Statistics, 44, 219-254.

\medskip

\noindent Forni, M., Hallin, M., Lippi, M., and Reichlin, L., 2000. The generalized dynamic-factor model: identification and estimation. Review of Economics and Statistics, 82, 540-554.

\medskip

\noindent Fortin, A.-P., Gagliardini, P., and Scaillet, O., 2022a. Latent Factor Analysis in short panels. Working paper.

\medskip

\noindent Fortin, A.-P., Gagliardini, P., and Scaillet, O., 2022b. A note on Random Matrix Theory. Working paper.

\medskip

\noindent  Freyberger, J., Neuhierl, A., and Weber, M., 2020. Dissecting characteristics nonpara- metrically. Review of Financial Studies, 33(5), 2326-2377.

\medskip

\noindent   Gagliardini, P., and Gourieroux, C., 2017.  Double instrumental variable estimation of interaction models with big data. journal of Econometrics, 201(2), 176-197.

\medskip

\noindent   Gagliardini, P., and Ma, H., 2019. Extracting statistical factors when betas are time-varying. Working Paper.

\medskip

\noindent Gagliardini, P., Ossola, E., and Scaillet, O., 2016. Time-varying risk premium in large cross-sectional equity datasets. Econometrica, 84(3), 985-1046.

\medskip

\noindent  Gagliardini, P., Ossola, E., and Scaillet, O., 2019. A diagnostic criterion for approximate factor structure. Journal of Econometrics, 212(2), 503-521.

\medskip

\noindent  Gagliardini, P., Ossola, E., and Scaillet, O., 2020. Estimation of large dimensional conditional factor models in finance, Handbook of Econometrics, Volume 7A, edited by S. Durlauf, L. Hansen, J. Heckman, and R. Matzkin, 219-282.

\medskip

\noindent Gospodinov, N., Kan, R., and Robotti, C., 2014. Misspecification-robust inference in linear asset-pricing models with irrelevant risk factors. Review of Financial Studies, 27(7), 2139-2170.

\medskip

\noindent Gu, S., Kelly, B., and Xiu, D, 2021. Autoencoder asset pricing models. Journal of Econometrics, 222(1), 429-450.

\medskip

\noindent Kan, R., and Zhang, C., 1999a. Two-pass tests of asset pricing models with useless factors. Journal of Finance, 54(1), 203-235.

\medskip

\noindent Kan, R., and Zhang, C., 1999b. GMM tests of stochastic discount factor models with useless factors. Journal of Financial Economics, 54(1), 103-127.

\medskip

\noindent  Kapetanios, G., 2010. A testing procedure for determining the number of factors in approximate factor models with large datasets. Journal of Business and Economic
Statistics, 28(3), 397-409.

\medskip

\noindent Kelly, B., Pruitt, S., and Su, Y., 2017. Instrumented Principal Component Analysis. Working
Paper.

\medskip

\noindent Kelly, B., Pruitt, S., and Su, Y., 2019. Characteristics are covariances: A unified model of risk
and return. Journal of Financial Economics, 134, 501-524.

\medskip

\noindent Kim, S., and Skoulakis, G., 2018. Ex-post risk premia estimation and asset pricing tests using large cross-sections: the regression-calibration approach. Journal of Econometrics, 204(2), 159-188.

\medskip 

\noindent Kleibergen, F., 2009. Test of risk premia in linear factor models. Journal of Econometrics, 149(2), 149-173.

\medskip

\noindent Izenman, A. J., 1975. Reduced-rank regression for the multivariate linear model. Journal of Multivariate
Analysis, 5(2), 248-264.

\medskip

\noindent Johnstone, I. M.,  2001. On the distribution of the largest eigenvalue in principal component
analysis. Annals of Statistics, 29(2), 295-327.

\medskip

\noindent Lancaster, T., 2000. The incidental parameter problem since 1948. Journal of Econometrics, 95(2), 391-413.

\medskip

\noindent Lunde, A., and Timmermann, A., 2004. Duration dependence in stock prices: An analysis of bull and bear markets. Journal of Business and Economic Statistics, 22, 253-273.

\medskip

\noindent Magnus, J. R., and Neudecker, H., 2007. Matrix differential calculus with applications in statistics and econometrics, third edition. New York, Wiley. 

\medskip

\noindent Neyman, J., and Scott, E.L., 1948. Consistent estimation from partially consistent observations. Econometrica, 16(1), 1-32. 

\medskip

\noindent Onatski, A.,  2009. Testing hypotheses about the number of factors in large factor models. Econometrica, 77, 1447-1479.

\medskip

\noindent Onatski, A., 2010. Determining the number of factors from empirical distribution of eigenvalues. Review of Economics and Statistics, 92(4), 1004-1016.

\medskip

\noindent Onatski, A, 2012. Asymptotics of the principal components estimator of large factor models with weakly influential factors. Journal of Econometrics, 168(2), 244-258.

\medskip

\noindent Onatski, A, 2015. Asymptotic analysis of the squared estimation error in misspecified factor models. Journal of Econometrics, 186(2), 388-406.

\medskip

\noindent  Pelger, M., 2019. Large-dimensional factor modeling based on high-frequency observations. Journal of Econometrics, 208, 23-42.

\medskip

\noindent   Pelger, M., 2020. Understanding systematic risk: A high-frequency approach. Journal of Finance, 75(4), 2179-2220.

\medskip

\noindent Pelger, M., and Xiong, R., 2022. State-varying factor models of large dimensions. Journal of Business and Economic Statistics, 40(3), 1315-1333.

\medskip

\noindent Politis, D., Romano, J., and Wolf, M., 1999. Subsampling. Springer Series in Statistics.

\medskip

\noindent Rao, W.-J., 2020. Higher-order level spacings in random matrix theory based on Wigner's conjecture, Working paper. 

\medskip

\noindent Renault, E., Van Der Heijden, T., and Werker, B., 2022. Arbitrage pricing theory for idiosyncratic variance factors. Journal of Financial Econometrics, forthcoming.

\medskip

\noindent Robin, J.-M., and Smith, R., 2000. Tests of rank. Econometric Theory, 16(2), 151-175.

\medskip

\noindent Shanken, J., 1992. On the estimation of beta-pricing models. Review of Financial Studies, 5, 1-33.

\medskip

\noindent Stock, J., and Watson, M., 2002. Macroeconomic forecasting using diffusion indexes. Journal of Business and Economics Statistics, 20(2), 147-162.

\medskip

\noindent Stock, J., and Watson, M., 2002. Forecasting using principal components from a large number of predictors.
Journal of the American Statistical Association, 97(460), 1167-1179.

\medskip

\noindent  Tao, T., 2012. Topics in random matrix theory. Graduate Studies in Mathematics, Volume 132, American Mathematical Society.

\medskip

\noindent Raponi, V., Robotti, C., and Zaffaroni, P., 2020. Testing beta-pricing models using large cross-sections. Review of Financial Studies, 33(6), 2796-2842.

\medskip

\noindent Zaffaroni, P., 2019. Factor models for conditional asset pricing. Working Paper. 

\newpage

\section*{Appendix: Proof of Theorem \ref{prop:expansion:eigenvalues}}
\setcounter{equation}{0}\def\theequation{A.\arabic{equation}}

Let $\hat{W}$ be the $K \times (K-k)$ matrix of the standardized eigenvectors of $\hat{A}$ associated with the $K-k$ smallest eigenvalues, and $\hat{\Lambda} = diag \left( \delta_j(\hat{A}), j=k+1,...,K\right)$ the diagonal matrix with these eigenvalues along the diagonal. Then:
\begin{equation}  \label{eq:EVEV}
\hat{A} \hat{W} = \hat{W} \hat{\Lambda}.
\end{equation} 
Let $Q$ be a $K \times (K-k)$ matrix whose columns are an orthonormal basis of the null space of matrix $A$ so that $Q' U$ and $Q'Q$ give the null and identity matrices. Since the columns of $U$ and $Q$ jointly span $\mathbb{R}^K$, we can write 
\begin{equation} \label{eq:span}
\hat{W} = Q \hat{R} + U \hat{S},
\end{equation}
where $\hat{R}$ and $\hat{S}$ are $(K-k)\times (K-k)$, resp.\ $k\times(K-k)$, matrices. 

By plugging (\ref{eq:Ahat}) and (\ref{eq:span}) into (\ref{eq:EVEV}), we get:
\begin{equation} \label{eq:basis}
U D \hat{S} + \hat{\Psi} Q  \hat{R} + \hat{\Psi} U \hat{S} = Q \hat{R} \hat{\Lambda} + U \hat{S} \hat{\Lambda},
\end{equation}
since $U'U$ gives the  identity matrix. 
Pre-multiplying both sides of Equation (\ref{eq:basis}) by $Q'$, we get $Q'\hat{\Psi} Q  \hat{R} + Q'\hat{\Psi} U \hat{S} 
= \hat{R} \hat{\Lambda}$, which yields:
\begin{equation} \label{eq:1}
\hat{\Lambda} = \hat{R}^{-1} ( Q' \hat{\Psi} Q ) \hat{R} + \hat{R}^{-1} ( Q' \hat{\Psi} U) \hat{S}. 
\end{equation}
(We show below that $\hat{R}$ is invertible).
Similarly, by pre-multiplying both sides of Equation (\ref{eq:basis}) by $U'$, we get $D \hat{S} + U' \hat{\Psi} Q \hat{R} + U'\hat{\Psi}U \hat{S} = \hat{S} \hat{\Lambda}$, which yields:
\begin{equation} \label{eq:2}
\hat{S} = D^{-1} \left( - U' \hat{\Psi} Q \hat{R}  - U'\hat{\Psi} U \hat{S} + \hat{S} \hat{\Lambda} \right).
\end{equation}

Let us now derive an expansion for $\hat{\Lambda}$ from Equations (\ref{eq:1}) and (\ref{eq:2}). First, from the Weilandt-Hoffmann inequality (see Tao (2012) p.\ 137), we know $\sum_{j=1}^K \vert \delta_j( A + \hat{\Psi}) - \delta_j(A) \vert^2 \leq \Vert \hat{\Psi} \Vert^2$, which implies 
\begin{equation*}
\Vert\hat{\Lambda}\Vert = \sqrt{ \sum_{j=k+1}^K \delta_j( \hat{A} )^2} \leq \Vert \hat{\Psi} \Vert.
\end{equation*}
Second, from Equation (\ref{eq:span}), we get $I_{K-k} = \hat{W}'\hat{W} = \hat{R}'\hat{R} + \hat{S}'\hat{S}$, which implies $\Vert \hat{R} \Vert = \sqrt{ Tr (\hat{R}'\hat{R})} \leq \sqrt{K-k}$. Third, using the above bounds and $\Vert U \Vert = \sqrt{k}$ and $\Vert Q \Vert =  \sqrt{K-k}$, Equation (\ref{eq:2}) yields:
\begin{equation*}
\Vert \hat{S} \Vert \leq \Vert D^{-1} \Vert \left( K^{3/2} \Vert \hat{\Psi} \Vert + (K+1) \Vert \hat{\Psi} \Vert \Vert \hat{S} \Vert \right).
\end{equation*}
Thus, if $\Vert \hat{\Psi} \Vert \leq \frac{1}{2 \Vert D^{-1} \Vert (K+1)}$, then
\begin{equation} \label{eq:3}
\Vert \hat{S} \Vert \leq 2 \Vert D^{-1} \Vert K^{3/2} \Vert \hat{\Psi} \Vert. 
\end{equation}
From Equation (\ref{eq:2}), we get that, if $\Vert \hat{\Psi} \Vert \leq \frac{1}{2 \Vert D^{-1} \Vert (K+1)}$, then
\begin{equation} \label{eq:4}
\hat{S} = - D^{-1} U' \hat{\Psi} Q \hat{R}  + O\left( \Vert D^{-1} \Vert^2 K^{5/2} \Vert \hat{\Psi} \Vert^2 \right),
\end{equation}
where the bound $O\left( \Vert D^{-1} \Vert^2 K^{5/2} \Vert \hat{\Psi} \Vert^2 \right)$ is uniform \footnote{By this, we mean that $O\left( \Vert D^{-1} \Vert^2 K^{5/2} \Vert \hat{\Psi} \Vert^2 \right) \leq C \Vert D^{-1} \Vert^2 K^{5/2} \Vert \hat{\Psi} \Vert^2$ for a universal constant $C$ that is independent of $A$, $\hat{\Psi}$, and $K$.} . Moreover, using $\Vert \hat{\Psi} \Vert \leq \frac{1}{3 \Vert D^{-1} \Vert (K+1)^{3/2}}$, we get from  Inequality (\ref{eq:3}) that $\Vert \hat{S} \Vert \leq 2/3$, and thus $\Vert I_{K-k} - \hat{R}'\hat{R} \Vert \leq (2/3)^2$ and $\hat{R}$ is invertible. 

We now plug (\ref{eq:4}) into the RHS of Equation (\ref{eq:1}) to get an expansion for $\hat{\Lambda}$. In order to control the remainder term, we need a bound on $\Vert \hat{R}^{-1} \Vert$. We have:
\begin{eqnarray}
\Vert \hat{R}^{-1} \Vert^2 &=& Tr \left( (\hat{R}^{-1})'(\hat{R}^{-1}) \right) = Tr \left( ( \hat{R}'\hat{R})^{-1} \right) \nonumber \\
&=& \sum_{j=1}^{K-k} \delta_j \left( ( \hat{R}'\hat{R})^{-1} \right) \leq 
(K-k) \delta_1 \left( ( \hat{R}'\hat{R})^{-1} \right) = \frac{K-k}{\delta_{K-k} ( \hat{R}'\hat{R}) }. \label{ineq:5}
\end{eqnarray}
Further, using the equation $\hat{R}'\hat{R} = I_{K-k} - \hat{S}'\hat{S}$ that we derived above, as well as the Courant-Fischer formula, which represents
eigenvalues as solutions of constrained quadratic optimization problems (see Appendix 2 of Gagliardini, Ossola, and Scaillet (2019)),  we have:
\begin{eqnarray*}
\delta_{K-k} ( \hat{R}'\hat{R}) &=& \underset{x \in \mathbb{R}^{K-k}: \Vert x \Vert=1}{\min} ~x' ( \hat{R}'\hat{R} ) x 
= 1 - \underset{x \in \mathbb{R}^{K-k}: \Vert x \Vert=1}{\max} ~ x' ( \hat{S}'\hat{S} ) x
 \\
&=&  1 - \delta_1 ( \hat{S}'\hat{S}) \geq 1 -  \Vert \hat{S} \Vert^2 \geq 1 - 4 \Vert D^{-1} \Vert^2 K^3 \Vert\hat{\Psi}\Vert^2,
\end{eqnarray*}
if $\Vert \hat{\Psi} \Vert \leq \frac{1}{2 \Vert D^{-1} \Vert (K+1)}$, where we use  (\ref{eq:3}) for obtaining the last inequality. Hence, if $\Vert \hat{\Psi} \Vert \leq \frac{1}{3 \Vert D^{-1} \Vert (K+1)^{3/2}}$, then $\delta_{K-k} ( \hat{R}'\hat{R}) \geq  1 - \frac{4}{9} \frac{K^3}{(K+1)^3} \geq 1/2$, which yields $\Vert \hat{R}^{-1} \Vert \leq \sqrt{2 (K-k)}$ from (\ref{ineq:5}). Equipped with the last inequality, we plug (\ref{eq:4}) into the RHS of Equation (\ref{eq:1}) and get:
\begin{eqnarray} \label{eq:6}
\hat{\Lambda} &=& \hat{R}^{-1} \left( Q ' \hat{\Psi} Q - (Q'\hat{\Psi} U) D^{-1} (U' \hat{\Psi} Q) \right) \hat{R} + O \left(    \Vert D^{-1} \Vert^2 K^{4} \Vert \hat{\Psi} \Vert^3  \right),
\end{eqnarray}
where the remainder term is uniform, if $\Vert \hat{\Psi} \Vert \leq \frac{1}{3 \Vert D^{-1} \Vert (K+1)^{3/2}}$.

Now we use $\delta_{k+j}(\hat{A}) = \delta_j(\hat{\Lambda})$, equation (\ref{eq:6}) and the Weilandt-Hoffmann inequality to get:
\begin{eqnarray*}
\delta_{k+j}(\hat{A}) &=& \delta_j \left( \hat{R}^{-1} \left( Q ' \hat{\Psi} Q - (Q'\hat{\Psi} U) D^{-1} (U' \hat{\Psi} Q) \right) \hat{R} \right) + O \left(    \Vert D^{-1} \Vert^2 K^{4} \Vert \hat{\Psi} \Vert^3  \right) \\
&=& \delta_j  \left( Q ' \hat{\Psi} Q - (Q'\hat{\Psi} U) D^{-1} (U' \hat{\Psi} Q) \right) + O \left(    \Vert D^{-1} \Vert^2 K^{4} \Vert \hat{\Psi} \Vert^3  \right),
\end{eqnarray*}
where the second equality holds because matrices $A$ and $R^{-1} A R$ have the same eigenvalues. 
The conclusion follows. 

\newpage

\clearpage

\newpage

\clearpage

\newpage

\clearpage

\newpage

%
%
		%
%
%


\clearpage

\newpage

\clearpage

\newpage

\clearpage

\newpage

\clearpage

\newpage

\clearpage

\newpage

\clearpage

\newpage

\clearpage

\newpage

\clearpage

\newpage

\newpage

\begin{table}  
\begin{center}
\title{\textbf{Size and power under DGP1}} 

\bigskip

\begin{tabular}{|cc|ccc|ccc|}
				\hline
				&          & \multicolumn{3}{c|}{Size (\%)} & \multicolumn{3}{c|}{Power (\%)} \\
				& $T$      & 6       & 12     & 24     & 6       & 12      & 24     \\ \hline
				$\sqrt{n}\mathscr{S}(k)$ & $n=500$  & 4.4   & 5.6  & 6.2  & 92   & 100    & 100   \\
				&          & (0.81)  & (0.22) & (0.24) & (16.1)  & (0.0)   & (0.0)  \\
				& $n=1000$ & 4.4   & 5.4  & 5.7  & 92    & 100    & 100   \\
				&          & (0.82)  & (0.24) & (0.24) & (18.9)  & (0.0)   & (0.0)  \\
				& $n=5000$ & 4.7   & 5.3  & 5.2  & 99    & 100    & 100   \\
				&          & (0.39)  & (0.21) & (0.21) & (6.9)   & (0.0)   & (0.0)  \\ \hline
				$\mathscr{S}^*(k)$       & $n=500$  & 5.9   & 5.1  & 5.2  & 0.59    & 69    & 97   \\
				&          & (0.36)  & (0.21) & (0.23) & (29.6)  & (24.9)  & (4.4)  \\
				& $n=1000$ & 5.7   & 5.0  & 5.1  & 69    & 89    & 100   \\
				&          & (0.35)  & (0.21) & (0.21) & (32.0)  & (14.8)  & (1.2)  \\
				& $n=5000$ & 5.5   & 4.8  & 5.0  & 92    & 99    & 100   \\
				&          & (0.27)  & (0.21) & (0.21) & (20.1)  & (6.2)   & (0.0)  \\ \hline
			\end{tabular}
\end{center}
\caption{For each statistic and sample size combination $(n,T)$, we provide the size and power in \% under DGP1. Nominal size is $5\%$. Power refers to rejection frequencies for statistics $\sqrt{n}\mathscr{S}(2)$ and $\mathscr{S}^*(2)$. In parentheses, we report the standard deviations for size and power across $100$ different draws of the factor path.}
\label{MC:Table1}
\end{table}

\newpage

\clearpage

\begin{table}

\begin{center}
\title{\textbf{Rejection rates of $\sqrt{n} \mathscr{S}(k)$, $k=2$, for DGP2}} 

\bigskip

\begin{tabular}{|cccccccc|}
\hline
     $\kappa$ & $0$ & $0.25$  & $0.40$ &  $0.50$ & $0.60$  & $0.75$  &  $1$  \\ 
		& strong & \multicolumn{2}{c}{semi-strong} & weak & \multicolumn{3}{c|}{vanishing} \\ \hline
	\multicolumn{1}{|c}{\%}   &   \multicolumn{7}{c|}{$c=0.1$}   \\
  $n=500$ &  $5.5$  & $6.9$   & $5.6$  & $5.2$  & $5.3$  & $5.1$   & $5.0$  \\
  $n=1000$ & $86$  & $6.6$ & $5.4$ & $5.1$  &  $5.4$  & $5.2$ & $5.4$   \\
  $n=5000$ & $100$  & $10$  & $5.6$  & $5.3$  & $5.0$   & $5.1$ & $5.4$  \\  \hline
	\multicolumn{1}{|c}{\%}   &   \multicolumn{7}{c|}{$c=1$}  \\
  $n=500$ &  $100$  & $100$   & $40$   & $13$   & $7.3$  & $5.5$ & $5.0$ \\
  $n=1000$ &  $100$ & $100$ & $44$ & $13$ & $6.6$  & $5.5$  &  $5.0$  \\
  $n=5000$ & $100$  & $100$  & $59$ & $13$  & $6.2$  & $5.2$ & $5.0$  \\ \hline
	\multicolumn{1}{|c}{\%}   &   \multicolumn{7}{c|}{$c=10$}  \\
  $n=500$ & $100$   & $100$   & $100$  & $100$  & $100$  & $50$ & $7.0$ \\
  $n=1000$ & $100$  & $100$  & $100$  & $100$  & $100$  & $34$ & $5.8$ \\
  $n=5000$ & $100$  & $100$  & $100$  & $100$  & $99$  & $16$ & $5.0$ \\  \hline
\end{tabular}
\end{center}
\caption{We report the rejection frequency in \% of statistic $\sqrt{n}\mathscr{S}(2)$ for each combination of constants $c$ and $\kappa$ in the parameterization of the beta variance $\sigma_{\beta,3}^2 = c n^{-\kappa}$ of the third factor, and cross-sectional size $n$. The time series dimension is $T=6$.}
\label{MC:Table2}
\end{table}

\newpage


\clearpage

\begin{table}

\begin{center}
\title{\textbf{Rejection rates of $\mathscr{S}^*(k)$, $k=2$, for DGP2}} \label{MC:Table3}

\bigskip

\begin{tabular}{|cccccccc|}
\hline
     $\kappa$ & $0$ & $0.25$  & $0.40$ &  $0.50$ & $0.60$  & $0.75$  &  $1$  \\
		& strong & \multicolumn{2}{c}{semi-strong} & weak & \multicolumn{3}{c|}{vanishing} \\ \hline
	\multicolumn{1}{|c}{\%}   &   \multicolumn{7}{c|}{$c=0.1$}   \\
  $n=500$ &  $9$  & $5.8$   & $5.4$  & $5.9$  & $5.8$   & $5.8$   & $5.5$  \\
  $n=1000$ & $14$  & $5.5$  & $5.0$  & $5.5$ & $5.1$  & $5.3$ & $4.7$  \\
  $n=5000$ & $47$   & $5.3$  & $5.0$  & $5.4$  & $4.9$   & $5.0$ & $5.3$  \\  \hline
	\multicolumn{1}{|c}{\%}   &   \multicolumn{7}{c|}{$c=1$}   \\
  $n=500$ &  $98$  & $25$   & $7.9$   & $6.7$   & $5.6$  & $5.8$ & $5.5$ \\
  $n=1000$ & $100$  & $32$  & $8.7$  & $6.0$  & $5.3$  & $5.3$ & $5.2$ \\
  $n=5000$ & $100$  & $58$  & $9.9$  & $6.0$  & $5.5$  & $4.9$ & $4.8$  \\ \hline
	\multicolumn{1}{|c}{\%}   &   \multicolumn{7}{c|}{$c=10$}   \\
  $n=500$ &  $99$  & $99$   & $96$   & $64$   & $29$  & $10$ & $5.5$ \\
  $n=1000$ & $100$  & $100$ & $98$ & $66$  & $26$ & $8$ & $5.2$ \\
  $n=5000$ & $100$  & $100$  & $100$  & $71$  & $23$ & $6$ & $5.4$ \\  \hline
\end{tabular}
\end{center}
\caption{We report the rejection frequency in \% of statistic $\mathscr{S}^*(2)$ for each combination of constants $c$ and $\kappa$ in the parameterization of the beta variance $\sigma_{\beta,3}^2 = c n^{-\kappa}$ of the third factor, and cross-sectional size $n$. The time series dimension is $T=6$.}
\label{MC:Table3}
\end{table}

	\begin{table}

\begin{center}
		\title{\textbf{Rejection rates of $\sqrt{n}\, \mathscr{S}(k)$, $k=2$, for DGP3}}
		
		\bigskip 
		
			\begin{tabular}{|llllllll|}
				\hline
				\multicolumn{1}{|c}{$\kappa$} &
				\multicolumn{1}{c}{$0$} &
				\multicolumn{1}{c}{$0.25$} &
				\multicolumn{1}{c}{$0.4$} &
				\multicolumn{1}{c}{$0.5$} &
				\multicolumn{1}{c}{$0.6$} &
				\multicolumn{1}{c}{$0.75$} &
				\multicolumn{1}{c|}{$1$} \\
				\multicolumn{1}{|c}{} &
				\multicolumn{1}{c}{strong} &
				\multicolumn{2}{c}{semi-strong} &
				\multicolumn{1}{c}{weak} &
				\multicolumn{3}{c|}{vanishing} \\ \hline
				\multicolumn{1}{|c}{\%}   &   \multicolumn{7}{c|}{$c=0.1$}                                 \\
				$n=500$  & $85$ & $6.8$ & $6.4$ & $5.9$ & $6.1$ & $6.0$ & $5.7$ \\
				$n=1000$ & $100$ & $8.7$ & $6.3$ & $5.8$ & $6.1$ & $5.9$ & $5.3$ \\
				$n=5000$ & $100$ & $14$  & $5.7$ & $5.3$ & $5.4$ & $5.3$ & $4.7$ \\ \hline
				\multicolumn{1}{|c}{\%}   &   \multicolumn{7}{c|}{$c=1$}                                \\
				$n=500$  & $100$ & $100$  & $79$  & $20$  & $9.2$ & $6.7$ & $6.0$ \\
				$n=1000$ & $100$ & $100$  & $78$  & $22$  & $8.1$ & $5.6$ & $5.2$ \\
				$n=5000$ & $100$ & $100$  & $96$  & $20$  & $6.9$ & $5.1$ & $5.3$ \\ \hline
				\multicolumn{1}{|c}{\%}   &   \multicolumn{7}{c|}{$c=10$}                                \\
				$n=500$  & $100$ & $100$  & $100$  & $100$  & $100$  & $85$  & $8.0$ \\
				$n=1000$ & $100$ & $100$  & $100$  & $100$  & $100$  & $68$  & $6.1$ \\
				$n=5000$ & $100$ & $100$  & $100$  & $100$  & $100$  & $30$  & $5.5$ \\ \hline
			\end{tabular}%
			\end{center}
			\caption{We report the rejection frequency in \%  of statistic $\sqrt{n} \mathscr{S}(2)$ for each combination of constants $c$ and $\kappa$ in the parameterization of the beta variance $\sigma_{\beta,3}^2 = c n^{-\kappa}$ of the third factor, and cross-sectional size $n$. The time series dimension is $T=6$. The errors follow individual ARCH(1) processes. }
		\label{MC:Table4}
	\end{table}

\clearpage

\begin{table}
	\begin{center}
	
\title{\textbf{Rejection rates of $ \mathscr{S}^*(k)$, $k=2$, for DGP3}}

\bigskip

		\begin{tabular}{|llllllll|}
			\hline
			\multicolumn{1}{|c}{$\kappa$} &
			\multicolumn{1}{c}{$0$} &
			\multicolumn{1}{c}{$0.25$} &
			\multicolumn{1}{c}{$0.4$} &
			\multicolumn{1}{c}{$0.5$} &
			\multicolumn{1}{c}{$0.6$} &
			\multicolumn{1}{c}{$0.75$} &
			\multicolumn{1}{c|}{$1$} \\
			\multicolumn{1}{|c}{} &
			\multicolumn{1}{c}{strong} &
			\multicolumn{2}{c}{semi-strong} &
			\multicolumn{1}{c}{weak} &
			\multicolumn{3}{c|}{vanishing} \\ \hline
			\multicolumn{1}{|c}{\%}   &   \multicolumn{7}{c|}{$c=0.1$}                                    \\
			$n=500$  & $8.8$ & $5.4$ & $5.6$ & $5.1$ & $5.7$ & $5.3$ & $6.1$ \\
			$n=1000$ & $15$  & $5.4$ & $5.5$ & $5.0$ & $4.8$ & $5.1$ & $5.1$ \\
			$n=5000$ & $53$  & $5.3$ & $5.3$ & $5.4$ & $5.0$ & $5.0$ & $5.6$ \\ \hline
			\multicolumn{1}{|c}{\%}   &   \multicolumn{7}{c|}{$c=1$}                                    \\
			$n=500$  & $98$  & $26$  & $9.1$ & $5.6$ & $5.4$ & $5.9$ & $5.4$ \\
			$n=1000$ & $100$  & $33$  & $9.0$ & $6.1$ & $5.1$ & $5.6$ & $5.2$ \\
			$n=5000$ & $100$  & $65$  & $9.4$ & $5.8$ & $5.2$ & $5.3$ & $6.1$ \\ \hline
			\multicolumn{1}{|c}{\%}   &   \multicolumn{7}{c|}{$c=10$}                                      \\
			$n=500$  & $98$  & $98$  & $98$  & $73$  & $30$  & $9.2$ & $5.6$ \\
			$n=1000$ & $100$  & $100$  & $99$  & $78$  & $34$  & $8.1$ & $5.4$ \\
			$n=5000$ & $100$  & $100$  & $100$  & $76$  & $24$  & $6.4$ & $5.3$ \\ \hline
		\end{tabular}
		\end{center}
		\caption{We report the rejection frequency in \%  of statistic $\mathscr{S}^*(2)$ for each combination of constants $c$ and $\kappa$ in the parameterization of the beta variance $\sigma_{\beta,3}^2 = c n^{-\kappa}$ of the third factor, and cross-sectional size $n$. The time series dimension is $T=12$. The errors follow individual ARCH(1) processes. }
		\label{MC:Table5}
\end{table}
	
\clearpage

\begin{table}
	\begin{center}
	
\title{\textbf{Size and power of $ n\mathscr{T}(k)$ under DGP4}}

\bigskip
		
			\begin{tabular}{|cc|ccc|ccc|}
				\hline
				&
				&
				\multicolumn{3}{c|}{Size ($\%$)} &
				\multicolumn{3}{c|}{Power ($\%$)} \\
				&
				$T$ &
				6 &
				12 &
				24 &
				6 &
				12 &
				24 \\ \hline
				$\hat{\Sigma}_{U,1}$ &
				$n=500$ &
				4.58 &
				4.78 &
				4.83 &
				99 &
				100 &
				100 \\
				&
				&
				(0.54) &
				(0.19) &
				(0.22) &
				(7.0) &
				(0.0) &
				(0.0) \\
				&
				$n=1000$ &
				4.83 &
				4.91 &
				4.94 &
				100 &
				100 &
				100 \\
				&
				&
				(0.23) &
				(0.22) &
				(0.19) &
				(4.8) &
				(0.0) &
				(0.0) \\
				&
				$n=5000$ &
				4.97 &
				5.00 &
				5.03 &
				100 &
				100 &
				100 \\
				&
				&
				(0.23) &
				(0.21) &
				(0.21) &
				(0.0) &
				(0.0) &
				(0.0) \\ \hline
				$\hat{\Sigma}_{U,2}$ &
				$n=500$ &
				\multicolumn{1}{l}{3.94} &
				\multicolumn{1}{l}{3.37} &
				\multicolumn{1}{l|}{2.30} &
				\multicolumn{1}{l}{99} &
				\multicolumn{1}{l}{100} &
				\multicolumn{1}{l|}{100} \\
				&
				&
				\multicolumn{1}{l}{(0.42)} &
				\multicolumn{1}{l}{(0.19)} &
				\multicolumn{1}{l|}{(0.16)} &
				\multicolumn{1}{l}{(7.0)} &
				\multicolumn{1}{l}{(0.0)} &
				\multicolumn{1}{l|}{(0.0)} \\
				&
				$n=1000$ &
				\multicolumn{1}{l}{4.49} &
				\multicolumn{1}{l}{4.10} &
				\multicolumn{1}{l|}{3.36} &
				\multicolumn{1}{l}{100} &
				\multicolumn{1}{l}{100} &
				\multicolumn{1}{l|}{100} \\
				&
				&
				\multicolumn{1}{l}{(0.23)} &
				\multicolumn{1}{l}{(0.19)} &
				\multicolumn{1}{l|}{(0.18)} &
				\multicolumn{1}{l}{(0.0)} &
				\multicolumn{1}{l}{(0.0)} &
				\multicolumn{1}{l|}{(0.0)} \\
				&
				$n=5000$ &
				\multicolumn{1}{l}{4.95} &
				\multicolumn{1}{l}{4.80} &
				\multicolumn{1}{l|}{4.64} &
				\multicolumn{1}{l}{100} &
				\multicolumn{1}{l}{100} &
				\multicolumn{1}{l|}{100} \\
				&
				&
				\multicolumn{1}{l}{(0.23)} &
				\multicolumn{1}{l}{(0.21)} &
				\multicolumn{1}{l|}{(0.18)} &
				\multicolumn{1}{l}{(0.0)} &
				\multicolumn{1}{l}{(0.0)} &
				\multicolumn{1}{l|}{(0.0)} \\ \hline
			\end{tabular}%
		
	\end{center}
		\caption{For each sample size combination $(n,T)$, we provide the size and power in $\%$ under DGP4 for the test statistic based on instruments. Nominal size is $5\%$. Size and power refer to rejection frequencies for statistics ${n}\mathscr{T}(k)$ with $k=3$ and $k=2$, respectively. In parentheses, we report the standard deviations for size and power across $100$ different draws of the factor path $F$ and the matrix $\Gamma$ linking betas to instruments. The upper panel uses the estimator $\hat{\Sigma}_{U,1}=\hat{\sigma}^2 [I_T\otimes \hat{Q}_{zz}]$ for simulating the critical values, and the lower panel uses $\hat{\Sigma}_{U,2}=\frac{1}{n}\sum_{i=1}^n [(\hat{\varepsilon}_{i}\hat{\varepsilon}'_{i})\otimes (z_iz'_i)]$.}
		\label{MC:Table6}
		
		\end{table}

\end{document}